\documentclass[fleqn,usenatbib]{mnras}
\usepackage{epsfig}
\usepackage{amsmath}
\usepackage{amssymb}
\usepackage{graphicx}
\usepackage{bm}
\usepackage{amsfonts,epsf}
\usepackage{color}
\usepackage{subfigure}
\usepackage{appendix}
\usepackage{soul}
\newcommand{\fig}[1]{Fig.~\ref{#1}}

\newcommand{\eq}[1]{equation~(\ref{#1})}
\newcommand{\eqs}[2]{equations~(\ref{#1}--\ref{#2})}

\newcommand{\be}{\begin{equation}}
\newcommand{\ee}{\end{equation}}

\newcommand{\bem}{\begin{multline}}
\newcommand{\eem}{\end{multline}}

\newcommand\bea{\begin{eqnarray}}
\newcommand\eea{\end{eqnarray}}

\makeatletter
\renewcommand*{\thesection}{\arabic{section}}
\renewcommand*{\thesubsection}{\thesection.\arabic{subsection}}
\renewcommand*{\p@subsection}{}
\makeatother

\title[High-energy flares from radiative reconnection]{High-energy synchrotron flares powered by strongly radiative relativistic magnetic reconnection: 2D and 3D PIC simulations}

\author[K. M. Schoeffler et al.]{K. M. Schoeffler,$^{1,2}$\thanks{E-mail: kevin.schoeffler@tecnico.ulisboa.pt (KMS)}
T. Grismayer,$^{1}$
D. Uzdensky,$^{3}$
L. O. Silva$^{1}$\\
$^{1}$GoLP/Instituto de Plasmas e Fus\~ao Nuclear,
Instituto Superior T\'ecnico,\\ Universidade de Lisboa, 1049-001 Lisboa, Portugal\\
$^{2}$Institut f\"ur Theoretische Physik, Ruhr-Universit\"at Bochum, Bochum, Germany\\
$^{3}$Center for Integrated Plasma Studies, Physics Department, \\ University of Colorado, Boulder CO 80309, USA\\
}
\date{\today}

\begin{document}
\setcounter{secnumdepth}{2}
\label{firstpage}
\pagerange{\pageref{firstpage}--\pageref{lastpage}}
\maketitle

\begin{abstract}
The time evolution of high-energy synchrotron radiation generated in a
relativistic pair plasma energized by reconnection of strong magnetic fields is
investigated with two- and three-dimensional (2D and 3D) particle-in-cell (PIC)
simulations.  The simulations in this 2D/3D comparison study are conducted with
the radiative PIC code OSIRIS, which self-consistently accounts for the
synchrotron radiation reaction on the emitting particles, and enables us to explore
the effects of synchrotron cooling.  Magnetic reconnection causes
compression of the plasma and magnetic field deep inside magnetic islands
(plasmoids), leading to an enhancement of the flaring emission, which may help
explain some astrophysical gamma-ray flare observations.  Although radiative
cooling weakens the emission from plasmoid cores, it facilitates additional
compression there, further amplifying the magnetic field $B$ and plasma
density~$n$, and thus partially mitigating this effect. Novel simulation
diagnostics utilizing 2D histograms in the $n\mbox{-}B$ space are developed and
used to visualize and quantify the effects of compression. The $n\mbox{-}B$
histograms are observed to be bounded by relatively sharp power-law boundaries
marking clear limits on compression. Theoretical explanations for some of these
compression limits are developed, rooted in radiative resistivity or
3D kinking instabilities.  Systematic parameter-space studies with respect to
guide magnetic field, system size, and upstream magnetization are conducted and
suggest that stronger compression, brighter high-energy radiation, and perhaps
significant quantum electrodynamic (QED) effects such as pair production, may
occur in environments with larger reconnection-region sizes and higher
magnetization, particularly when magnetic field strengths approach the critical
(Schwinger) field, as found in magnetar magnetospheres.

\end{abstract}



\section{Introduction}
\label{sec-intro} 

Bright, rapid gamma-ray flares occur throughout the cosmos, coming from sources associated with relativistic compact objects --- neutron stars and black holes --- both in our own Galaxy and beyond. 
Among extra-galactic flaring gamma-ray sources, perhaps the most spectacular ones are gamma-ray bursts (GRBs) observed at cosmological distances: both long (several seconds) GRBs resulting from violent, explosive deaths of very massive stars, and short ($\leq$2 seconds) GRBs from neutron-star mergers \citep{Piran2005, Meszaros2006, Berger2014}, including the recently observed relatively weak short GRB associated with the gravitational-wave event GW-170817 detected by LIGO \citep{Abbott2017, Goldstein2017, Davanzo2018}. Another important class of powerful extragalactic sources flaring violently in the gamma-ray band is coronae and relativistic jets of active galactic nuclei (AGN) powered by accreting supermassive black holes residing at the centers of many galaxies, such as~M87. For example, ultra-rapid ($\sim$ 10 minutes) Very-High-Energy (VHE) TeV flares are observed by ground-based Cerenkov telescopes from M87 \citep{Abramowski2012}
and from many blazars (relativistic AGN jets pointing directly along our line of sight) \citep{Albert2007b, Aharonian2007, Aleksic2011, Madejski2016}; blazars are also observed to have simultaneous GeV flares on 1-day time-scales \citep{Tanaka2011}.
Some of the most notable manifestations of variable gamma-ray activity from Milky Way sources include 
pulsed broad-band high-energy emission (peaking in the GeV range) from young pulsars such as Crab and Vela~\citep[see, e.g.,][for a recent review]{Philippov2022}; 
the enigmatic day-long 100MeV--1GeV flares from the Crab pulsar wind nebula (PWN)~\citep{Abdo2011,Tavani2011,Buehler2014}; 
very short and intense hard-X-ray and soft gamma-ray flares from magnetars 
(e.g., \citealt{Mazets1999}, \citealt{Palmer2005}; see \citealt{Kaspi2017} for a recent review); 
and nonthermal high-energy emission extending at least up to MeV energies from accreting stellar-mass black holes in X-ray Binaries (XRBs) such as Cyg X-1 \citep{Remillard2006, Zdziarski2012}, which also sometimes exhibit VHE ($\geq$100GeV) hour-long flares \citep{Albert2007a}. 

The leading radiation mechanisms responsible for these flares can be either synchrotron or inverse-Compton (IC), depending on the source. Thus, in neutron-star systems, the magnetic fields are strong and the radiation is often dominated by synchrotron emission, even in the gamma-ray range. 
For sufficiently strong fields, the radiation emission takes place in the discrete, quantum-electrodynamic (QED) regime, where the emission of a single photon causes a significant drop in the emitting particle's energy. Moreover, the interaction of the emitted energetic gamma-ray photons with the ambient strong magnetic field can lead to electron-positron pair production, thus providing an important source of pair plasma populating the neutron-star magnetosphere. These QED processes are especially important for magnetars --- young neutron stars with ultra-strong magnetic fields exceeding the QED (Schwinger) field, $B_Q \equiv m_e^2 c^3/e\hbar \simeq 4.4 \times 10^{13}\, {\rm G}$ $=E_Q$ (in Gaussian units)~\citep[e.g.,][]{Duncan1992}. 
In contrast, in environments with weaker magnetic fields, e.g., those around rapidly accreting black holes (e.g., in coronae of XRBs and quasars), radiative cooling is often dominated by IC scattering \citep{Albert2007a}, which may also sometimes happen in the QED Klein-Nishina regime and power  prodigious pair production \citep{Beloborodov-2017, Mehlhaff2020}.

In all of these cases, magnetic reconnection provides an attractive mechanism for explaining the high-energy flares~\citep{Romanova1992,Lyubarsky-1996,DiMatteo1998,Lyutikov-2003, Jaroschek_etal-2004, Giannios-2008,Giannios2009,Giannios2010, Giannios-2013,Nalewajko2011, Nalewajko-2012, McKinney2012, Uzdensky2011,Uzdensky_etal-2011,Cerutti2012,Cerutti2013, Uzdensky_Spitkovsky-2014,Sironi-2015, Cerutti_etal-2016, Beloborodov-2017, Philippov_Spitkovsky-2018, Lyutikov_etal-2018, Werner2018,Werner2019,Giannios_Uzdensky-2019,Mehlhaff2020, Hakobyan2023, Hakobyan2023-pulsar, Chen2023}.
During reconnection, free energy contained in oppositely directed magnetic fields is rapidly converted to bulk flows, plasma heating, and nonthermal particle acceleration; moreover, in strongly radiative cases much of this energy is promptly converted into radiation. Furthermore, reconnecting current sheets are unstable to the secondary tearing instability leading to the generation of magnetic islands (plasmoids), or flux ropes in three dimensions~(3D) \citep{Loureiro2007, Bhattacharjee2009, Uzdensky2010}.
As the freshly energized plasma tends to accumulate inside these islands, bursts of radiation are expected to be emitted from there \citep{Giannios-2013,Cerutti2013, Sironi2016, Petropoulou2016, Beloborodov-2017,Schoeffler2019,Sironi2020}

Magnetic reconnection is therefore a potential cause of observed gamma-ray and X-ray flares.  Several previous radiative-PIC studies have investigated reconnection with radiative cooling due to inverse Compton scattering, where energetic particles upscatter soft photons from an ambient radiation bath \citep{Werner2019, Mehlhaff2020, Sironi2020, Sridhar2021}. However, in reconnection regimes with strong magnetic fields, especially found near pulsars and magnetars, the radiation cooling is predominantly caused by synchrotron emission~\citep{Lyubarsky-1996, Uzdensky_Spitkovsky-2014, Cerutti_etal-2016}. 
Relativistic collisionless reconnection with synchrotron cooling has been studied with radiative-PIC simulations, mostly in two dimensions~(2D), in a number of previous works \citep{Jaroschek2009, Cerutti2013, Cerutti2014, Nalewajko2018, Schoeffler2019, Hakobyan2019, Hakobyan2023}.  
It will also be the focus of the present paper, which will be devoted to studying the interplay between 3D and radiative cooling effects.

In our previous 2D computational study~\citep{Schoeffler2019}, reconnection was shown to cause a sudden jump in the radiation emission.  
The reconnection process leads to plasma heating and nonthermal particle acceleration, both directly by the reconnecting electric field, and by the evolution and merging processes of the plasmoids. Increased plasma density, magnetic field, and temperature, caused by the compression of islands in~2D, leads to stronger emission of radiation. 
Radiative cooling was shown to further enhance the compression and subsequent radiation at the cores of magnetic islands.
In a strong magnetic field, the enhanced radiation can reach into the gamma-ray band, potentially inducing QED effects such as pair production~\citep{Schoeffler2019}.

The intriguing results of our previous 2D computational study~\citep{Schoeffler2019} naturally lead to an important question of whether the observed very strong compression effects will still occur in a more realistic 3D system. 
Building up on that study, in this paper we will show that enhanced compression is indeed possible in~3D at some level, and hence 3D relativistic magnetic reconnection in strong magnetic fields could still explain the occurrence of gamma-ray flares in astrophysical systems. However, the maximum degree of compression achievable in 3D remains rather modest, as compressing flux ropes tend to get disrupted by the kink instabilities. A moderate out-of-plane (so-called ``guide") magnetic field can stabilize the kink and helps keep the plasma from escaping the flux ropes. However, at the same time, the magnetic pressure of this same guide field resists and limits the compression. It turns out that the compression is maximized for moderate values of the guide field, comparable to the upstream reconnecting field.

In this paper, we conduct a large, comprehensive study using 2D and 3D particle-in-cell (PIC) simulations using the OSIRIS framework~\citep{OSIRIS}. 
First and foremost, we look at the importance of 3D effects on these reconnecting systems with strong fields, which has not yet been thoroughly investigated in dedicated radiative-PIC simulation studies \citep[see, however,][]{Cerutti2014}.
Furthermore, we have developed novel numerical diagnostic tools to characterize and understand in detail the plasma and magnetic field compression in magnetic islands and the emission of radiation in these reconnection regimes. This includes 2D histograms characterizing the spatial correlations between plasma density, magnetic field strength, and plasma temperature, which help us elucidate the degrees of compression that enhance the radiation emission.  
Extensive exploration of various broad parameter spaces elucidates the conditions under which gamma-ray flares can be expected.

This paper is organized as follows. In Section~\ref{sec-setup} we will introduce the numerical setup for our 2D and 3D radiative PIC simulations that will be presented throughout this paper. 
In Section~\ref{sec-diags} we will introduce the new diagnostics used to examine the emission of radiation, divided into (a) the
different estimates of the total radiated power and of the local emissivity as a function of space and (b) 2D correlation histograms of the density, magnetic field, and temperature, which help quantify the degree of compression and spatial correlations between these quantities.
In Section~\ref{sec-2D} we will examine 2D simulations utilizing these new diagnostics considering different synchrotron cooling strengths characterized by different values of the normalized reconnecting magnetic field~$B_0/B_Q$. In Section~\ref{sec-3D} we will present and analyze the results of full 3D simulations. In Section~\ref{sec-param} we will present a broad parameter-space study exploring the effects of several important system parameters, such as the guide magnetic field, the system size and aspect ratio, and the upstream plasma magnetization.
Finally, in Section~\ref{sec-concl} we will summarize the conclusions found in this work, and discuss how magnetic reconnection in strong fields may power radiation observed in astrophysical gamma-ray flares. We also include appendices with a more detailed description of the setup in Appendix~\ref{sec-append-setup}, a more developed explanation of the theoretical boundaries of density in the density-magnetic field histograms in Appendix~\ref{sec-append-nbound}, a derivation of an effective resistivity due to synchrotron radiation in Appendix~\ref{sec-append-raddiss}, and an associated theoretical boundary of magnetic fields in the density-magnetic field histograms in Appendix~\ref{sec-append-Bbound}.

\section{Numerical setup}
\label{sec-setup}

We conducted both 2D and 3D PIC studies of
relativistic  reconnection in a pair plasma, taking advantage of the OSIRIS framework~\citep{OSIRIS}. 
OSIRIS self-consistently includes synchrotron radiation and the QED process of pair production by a single gamma-ray photon propagating across a strong electromagnetic field~\citep{GrismayerPOP,GrismayerPRE}.
However, in this study, we look at a regime where, although the back-reaction caused by the synchrotron radiation plays an important role, the QED processes are not relevant.
In these simulations, we track the total amount of radiated energy emitted by each particle at every time step and use the local estimation for the emissivity~$\epsilon_{\rm est}$ (described in Section~\ref{subsec-emissivity}) to track emission as a function of space and time.  

We simulate an initial double relativistic Harris current-sheet equilibrium~\citep{Harris1962, KirkHarris} with periodic boundary conditions, which is explained in more detail in Appendix~\ref{sec-append-setup}.  For our simulations presented in Sections~\ref{sec-2D} and \ref{sec-3D}, we focus on the fiducial values of the key system parameters described in this section, while we will vary some of them in the parameter scans in Section~\ref{sec-param}. 

The computational domain is initially filled with a relativistically hot Maxwell-J\"uttner background electron-positron plasma with uniform density (of each species)~$n_b$ and temperature~$T_b =4 m_e c^2$. These parameters are chosen to yield a high upstream ``hot" plasma magnetization
\be
\sigma_h \equiv \frac{B_0^2}{4 \pi (2n_b) h_b} = 25.76,
\ee
where $B_0$ is the reconnecting magnetic field oriented along the $\hat{x}$ direction and $h_b$ is the relativistic enthalpy per particle in the upstream background ($h\approx 4T$ for ultrarelativistic temperatures). This corresponds to an upstream plasma beta $\beta_{\rm up} \equiv {8 \pi (2n_b) T_b/B_0^2} = 1/2\sigma_h = 0.0202$. 
Note that the value of magnetization adopted in this paper is greater than the value $\sigma_h=6.44$ of our previous work \citep{Schoeffler2019}.
We also include an out-of-plane ($\hat{z}$) uniform guide magnetic field~$B_G = 0.4 B_0$.
The cold magnetization, discussed in Appendix~\ref{sec-append-setup}, is $\sigma_c\equiv B_0^2/4\pi (2n_b) m_e c^2=412$. 

In addition to the uniform background, we
include two anti-parallel initial Harris current layers, each lying in a $y={\rm const}$ plane and carrying electric current in the $\pm \hat{z}$ direction. 
The layers are composed of drifting Maxwell-J\"uttner distributions of counter-streaming electrons and positrons with central density (of each species) $n_0 = 37 n_b$, rest-frame temperature~$T_0=6.92 m_e c^2$, an initial half-thickness~$\delta = 2.55 \rho_L$, and a drift velocity~$v_d/c=0.56$ for each species (Lorentz factor $\gamma_d=1.21$, proper velocity $u_d = \gamma_d v_d/c = 0.68$).

Here, our main fiducial normalizing length-scale
\be
\rho_L \equiv
\gamma_T m_e c^2/eB_0 = \gamma_T c/\Omega_c
\ee
is defined as the Larmor radius of a background particle with a Lorentz factor corresponding to the peak of the initial upstream relativistic Maxwell-J\"uttner distribution, $\gamma_T \equiv
2T_b/m_e c^2$.
Here, $\Omega_c \equiv eB_0/m_ec$ is the classical (nonrelativistic) gyrofrequency.
Other important length-scales include the respective background (nonrelativistic) skin depth $d_e \equiv \left[m_e c^2/4 \pi (2 n) e^2\right]^{1/2}$ and
Debye length $\lambda_D \equiv \left[T/4 \pi (2 n) e^2\right]^{1/2}$, defined by the initial background plasma parameters ($n=n_b$ and $T=T_b$):
\be
d_e = \rho_L \left(\sigma_h m_e c^2/ T_b\right)^{1/2} \approx 2.53 \rho_L,
\ee
\be
\lambda_D = \rho_L \sigma_h^{1/2} \approx 5.08 \rho_L,
\ee
both of which are larger than $\rho_L$, scaling as $\sigma_h^{1/2}\rho_L$ for relativistic temperatures.
We can also introduce the values of these length-scales in the Harris sheet where $n = n_0$ and $T = T_0$:
\be
\rho_{L,H} = \rho_L \left(T_0/T_b\right) \approx 1.73 \rho_L,
\ee
\be
d_{e,H} = \rho_L \left(\sigma_h n_b m_e c^2/ n_0 T_b\right)^{1/2}  \approx 0.42 \rho_L,
\ee
\be
\lambda_{D,H} = \rho_L \left(\sigma_h n_b T_0/ n_0 T_b\right)^{1/2}  \approx 1.10 \rho_L.
\ee
The Larmor radius and Debye length in the reconnection regions increase as time progresses, due to the heating of the plasma.
Our fiducial simulation domain size is $2 L_x \times
2 L_y (\times 2 L_z) = 628.8\rho_L\times 628.8\rho_L (\times 117.2\rho_L)$ in 2D~(3D), and the simulations are run for about $3.16$ light crossing times~$L_y/c$ ($t_{\rm max} = 7948~\Omega_c^{-1}$).

Our typical 2D (3D) simulation domain size consists of $1280 \times 1280 (\times 240$) computational grid cells of size $\Delta x = \Delta y ( = \Delta z) = 0.49\rho_L$, initially with $16~(8)$ particles per species in each cell, with a total of about $9.0 \times 10^7~(1.0 \times 10^{10})$ particles. 
There are thus about $1700~(8900)$ initial macroparticles per Debye cube in the background plasma.
Although initially there are only $\sim 80~(90)$ macroparticles per Debye cube in the Harris sheet, once the background plasma enters the reconnection region, this number becomes much larger.  
The simulations are typically run with a
time step of $\Delta t = 0.5 \Delta x/c =\ 0.25 \rho_L/c = 0.25 \gamma_T
\Omega_c^{-1}$.

A novel feature of our simulations is the self-consistent inclusion of optically thin radiation emission by relativistic particles due to strong magnetic fields.  Depending on the importance of QED effects, OSIRIS can treat radiation emission with two alternative implementations: a continuous description of classical radiation reaction, and a quantized description that includes the QED processes.

To determine whether QED (discrete-emission) effects are important for a given emitting particle, we calculate the relativistic invariant for an electron (or positron) of energy $\gamma m_e c^2$ and momentum~$\bm{p}$ moving in an electromagnetic field
\begin{equation}
	\label{chi_3vectordef}
	\chi_e=\frac{1}{B_Q}\sqrt{\left(\gamma \bm {E}+\frac{\bm {p}}{m_ec}\times \bm {B} \right)^2 - \left( \frac{\bm {p}}{m_ec}\cdot \bm {E} \right)^2}, 
\end{equation}
which in our parameter regimes, where usually $B \gg E$, can be approximated by
\begin{equation}
\label{chi_apprxdef}
\chi_e \approx \frac{\gamma B}{B_Q}.
\end{equation}
As this parameter increases, the particle will emit higher-energy photons, and, once $\chi_e$ approaches $1$, QED effects including discrete gamma-ray emission, and, for even higher~$\chi_e$, pair production, can start playing an important role.
However, in the simulations presented in this paper,
the $\chi_e$ parameter does not usually reach
significantly high values even for very energetic particles (i.e., $\chi_e \lesssim 1$).  We thus use the
continuous description, with the radiation back-reaction accounted for classically using the Landau-Lifshitz model~\citep{LandauLifshitz} for the radiative drag force, while we keep track of the total radiated energy.

The radiative cooling is significant when the synchrotron cooling time~$\sim \left(\alpha_{\rm fs} \chi_e \Omega_c\right)^{-1}$, where $\alpha_{\rm fs} \equiv e^2/\hbar c \approx 1/137$ is the fine structure constant, is shorter than or comparable to the relevant timescale of the simulation, i.e., a few global light crossing times~($t_{\rm max} \approx 3.16 L_y/c$).
After one light crossing time~$L_y/c$, a relativistic particle with energy $\gamma m_e c^2$ moving in a magnetic field~$B$ experiences significant cooling (i.e., loss of a significant fraction of its energy) when
\begin{equation}
\label{signftcooling}
	\frac{2}{3}\, \alpha_{\rm fs}\chi_e \Omega_c \frac{L_y}{c} = 
    2 \gamma \, \frac{B^2}{B_0^2} \,\ell_B > 1,
\end{equation}
a parameter directly related to the global magnetic compactness
\begin{equation}\label{magcompactness}
	\ell_B \equiv \sigma_T L_y \frac{U_{B0}}{m_ec^2}.
\end{equation}
Here, $\sigma_T$ is the Thomson cross-section, and $U_{B0}=B_0^2/8\pi$ is the initial upstream magnetic energy density.
Note that in our fiducial set of simulations, with fixed values of $L_y/\rho_L=314.4$ and $\gamma_T = 2T_b/m_e c^2 =8$, the compactness $\ell_B$ scales just linearly (instead of quadratically) with~$B_0$, since $\rho_L = \gamma_T m_e c^2/eB_0 \propto B_0^{-1}$.

We assume that synchrotron emission is the dominant radiation mechanism. For the simulation parameters adopted in this study, the Thomson optical depth
is $\tau_T \equiv \sigma_T L_y n_b = 1.35 \times 10^{-4}$, and thus the radiation occurs in an optically thin regime. 
Here we ignore the IC scattering of both the synchrotron photons (i.e., synchrotron self-Compton, SSC) and any possible ambient photons of external origin (i.e., external~IC); we also neglect synchrotron self-absorption; investigating the effect of these additional radiative processes is left for future studies.

For reference, we define the characteristic radiation-reaction limit Lorentz factor~$\gamma_{\rm rad}$ described, e.g., in~\citep{Uzdensky_etal-2011, Uzdensky2016,Werner2019,Sironi2020, Mehlhaff2020,Mehlhaff2021}.  
This factor is defined as the particle Lorentz factor for which the radiation-reaction force is equal to the acceleration force by the reconnecting electric field $E_{\rm rec} = \beta_E B_0$, or, equivalently, the particle's radiative cooling time is approximately equal to the gyro-period.  
For synchrotron radiation, this limit is
\be
\label{synchgammarad}
\gamma_{\rm rad}^2 = \frac{1}{\sin^2\alpha}\frac{3}{2} \frac{\beta_E}{\alpha_{\rm fs}}\frac{B_{Q}}{B_0} = \frac{1}{2\ell_B} \frac{\Omega_c L_y}{c} \frac{\beta_E}{\sin^2\alpha} = 
\frac{4\pi e}{\sigma_T B_0}\, \frac{\beta_E}{\sin^2\alpha}\, , 
\ee
where $\beta_E\sim 0.1$ is the dimensionless reconnection rate (reconnection inflow velocity normalized to the speed of light), and $\alpha$ is the particle's pitch angle with respect to the magnetic field.

In order to investigate the effects of radiative cooling, in this paper we present the results (in both 2D, Section~\ref{sec-2D}, and~3D,
Section~\ref{sec-3D}) from three simulations with different cooling strengths. The cooling strength is controlled by varying the reconnecting magnetic field strength, using the same magnetic field values as those used by~\cite{Schoeffler2019}:
\begin{itemize}
    \item  (1) Classical Case: $B_0/B_Q = 4.53 \times 10^{-6}$ (i.e., $B_0 = 2\times10^8$\,G), where the peak local average value of $\chi_e$ reaches $\chi_e\sim 0.0003 \ll \left(\alpha_{\rm fs}\Omega_c t_{\rm max}\right)^{-1} \approx 0.017$, and $2\gamma (B/B_0)^2 \ell_B  \approx 0.004$ ($\ell_B \approx 2\times 10^{-5}$), and hence cooling is not important;
    \item  (2) Intermediate Case: $B_0/B_Q = 4.53 \times 10^{-4}$ ($B_0 = 2\times10^{10}$\,G), where $\left(\alpha_{\rm fs} \Omega_c t_{\rm max}\right)^{-1}\approx 0.017 < \chi_e \sim 0.03 \ll 1$, and $2\gamma(B/B_0)^2 \ell_B \approx 0.4$ ($\ell_B  \approx 0.002$), and cooling is becoming important;
    \item  (3) Radiative Case: $B_0/B_Q = 4.53 \times 10^{-3}$ ($B_0 = 2\times10^{11}$\,G), where $\chi_e \sim 0.3$, and $2\gamma(B/B_0)^2 \ell_B  \approx 4$ ($\ell_B \approx 0.02$), and cooling is very important.
\end{itemize}
In these estimations, e.g., of~$\chi_e(B,\gamma)$, to evaluate peak local average values inside of magnetic islands/flux ropes, we have assumed that the magnetic field is enhanced by a factor of about~1.5 (i.e., $B \approx 1.5 B_0$) and the temperature by a factor of~5 (i.e., $T \approx 5 T_b$, $\gamma \approx 5\gamma_T$). 
Furthermore, assuming a normalized reconnection rate $\beta_E = 0.1$ and pitch angle $\alpha = 90^o$, these parameters correspond to $\gamma_{\rm rad} \simeq 2000$, 200, and~$60$ respectively. 
For the radiative case, in order to keep the initial upstream plasma from cooling substantially in the course of the simulation, we restrict the degree of radiative cooling based on the background parameters to $2\gamma_T \ell_B \approx 0.44$, 
with a background average $\chi_e(B_0,\gamma_T) \approx 3.6\times 10^{-2}$. However, as the system evolves, regions develop with an average local value $\chi_e \approx 0.3$ (as shown in Section~\ref{subsec-sigmah}), and energetic particles occur with $\chi_e \approx 1$, allowing for significant local cooling.
Note, however, that unlike in Ref.~\citep{Schoeffler2019}, these values of $\chi_e$ are small enough that there are no significant QED effects (such as discrete photon emission and pair creation). Nevertheless, we still expect qualitatively similar results; negligible cooling in the classical case, and a significant radiated fraction of the released magnetic energy, in part due to a strong  enhancement of magnetic island compression, in the radiative case and, to a lesser extent, in the intermediate case.

In Section~\ref{sec-param}, we also explore the parameter space starting with our 3D radiative case $B_0/B_Q = 4.53 \times 10^{-3}$, and varying $B_G/B_0$, $L_z/\rho_L$, $L_y/\rho_L$, and~$\sigma_h$. 

\section{Diagnostics}
\label{sec-diags}
\subsection{Estimated radiated power and emissivity}
\label{subsec-emissivity}

While it is possible to do in-situ measurements in reconnection experiments and even in the Earth's magnetosphere using spacecraft, for phenomena that
take place around remote astrophysical objects like neutron stars, the only data that can be obtained comes from observations of radiation.  We thus pay special attention to diagnostics measuring the radiation emitted in these environments both as a function of time and of space.

Each particle emits radiation at any given moment in time with a power that is a function of $\chi_e$ and, in the classical regime ($\chi_e \ll 1$), can be expressed as 
\be
\label{classicalpower}
P = \frac{2}{3}\, \frac{\alpha_{\rm fs} m_e c^2}{t_C}\, \chi_e^2 \approx \frac{2}{3}\, \frac{e^2}{c}\,\gamma^2 \Omega_c^2 \sin^2\alpha \, ,
\ee
where the second expression, for classical synchrotron radiation, is valid as long as the magnetic field is much stronger than the electric field.
Here $t_C = \hbar/m_e c^2 \approx 1.29 \times 10^{-21} \text{s}$ is the electron Compton time.

While the total power emitted in a given optically thin system is calculated by summing the powers $P$ radiated by each particle, it is also useful to study the location in space where the radiative power is emitted from. Instead of considering a discrete sum of particles, one can consider a 6D distribution of particles~$f(\bm {p},\bm {x})$ over the momentum space~$\bm {p}$ and the coordinate space~$\bm {x}$, and calculate~$P$ as a function of $\bm {p}$, $\bm {B}\left(\bm {x}\right)$, and~$\bm {E}\left(\bm {x}\right)$.  
Then, the total power emitted at a given moment in time can be expressed as:
\be
\label{emissionsum}
P_{{\rm tot}} = \int {\rm d}^3p\, {\rm d}^3x\, f({\bm p},{\bm x}) P({\bm p},{\bm x}) = 
\int {\rm d}^3x \, \epsilon({\bm x}), 
\ee
where the local emissivity
\be
\label{emissivity}
\epsilon({\bm x}) \equiv \int {\rm d}^3p \, f({\bm p},{\bm x})P({\bm p},{\bm x})
\ee
is the power emitted from a unit volume in space.
The total power is thus proportional to the volume-averaged emissivity $P_{\rm tot} \sim \left<\epsilon(\bm {x})\right>$, where the angle brackets represent an average over space.  
In our simulations, we calculate $P_{\rm tot}$ at each time step by summing the radiation from all the particles in the simulation.  
We normalize $P_{\rm tot}$ to its initial value~$P_{{\rm tot},0}$, $\bar{P}_{\rm tot} \equiv P_{\rm tot}/P_{{\rm tot},0}$, in order to emphasize the relative enhancement of radiation due to reconnection. 

In principle, in PIC simulations it is possible to precisely measure both
$P_{\rm tot}$ and $\epsilon$ by summing the power emitted by each particle within
each given small volume element, e.g., in each grid cell. 
However, in MHD simulations, for example, the details of the particle distribution are not available.  Furthermore, only limited data is
available in observations. We will thus explore several fluid-level methods of estimating $P_{\rm tot}$ using various assumptions, and check their fidelity by comparing them with our exact kinetic measurements. 

Although we do not introduce a diagnostic for the precise (particle-based) value of $\epsilon(\bm {x},t)$ in our simulations, we will define a reasonable fluid-based method of estimation where we assume a local isotropic Maxwell-J\"uttner distribution in the comoving frame corresponding to the local drift velocity $\bm{v}_d$ of each species. 

To obtain the effective local temperature from a time-evolving distribution that is not
necessarily Maxwell-J\"uttner, we take the temperature tensor $T_{ij} \equiv
m_e^{-1}\int {\rm d}^3p \, (p_i p_j/\gamma) f({\bm p})/ \int {\rm d}p^3 \, f({\bm p})$ for the background
electron species calculated in its local rest frame [i.e., in the so-called Eckart frame \citep{Eckart1940}, where the local current of that species vanishes].
Here, $p_i$ is the $i$th component ($i=x,y,z$) of the momentum,
$\gamma = \sqrt{1+p^2/m_e^2c^2}$, and $f(\bm{p})$ is the momentum distribution function.
The effective temperature is then defined using the trace of the temperature tensor, $T \equiv
Tr(T_{ij})/3$.  
While this temperature initially only represents the temperature of the background plasma, as the background population mixes with the current-sheet population, this temperature becomes a representative temperature of the system.
We also define a representative density~$n$, which is the total local particle density of one species, (e.g., positrons), including both initially background and Harris current-sheet particles.

Assuming again that $E \ll B$, we can say that $\chi_e \approx \gamma B/B_Q$, and substitute~\eq{classicalpower} into~\eq{emissivity} to get an estimate for the emissivity. Based on our assumption of an isotropic Maxwell-J\"uttner distribution in the local comoving frame, we can integrate over the pitch angles~$\alpha$ and momenta, the result of which is proportional to $B_\perp^2 [1 + u_d^2({\bm x})] + B_\parallel^2$. Here, $\bm B_\parallel$ and $\bm B_\perp$ are defined with respect to the local bulk fluid velocity $\bm v_d({\bm x})$ of the given species, and $u_d({\bm x}) \equiv \gamma_d v_d/c$ is the fluid's local normalized proper velocity, with $\gamma_d$ being the corresponding Lorenz factor.
For simplicity, we will not include in our estimation the Doppler-boosting enhancement in radiated power based on~$u_d$, whose direction may be difficult to determine in observations. We, therefore, find the following estimation only in terms of local, space-dependent parameters~$n(\bm {x}),T(\bm {x}),$ and~$B(\bm {x})$:
\be
\label{emissivityest}
\epsilon_{\rm est}(\bm {x}) \equiv \frac{16}{3}\frac{e^4}{m_e^4c^7}\, [2n(\bm {x})] \,T^2(\bm {x}) B^2(\bm {x})\,.
\ee
Note that while this estimate is calculated in the lab frame, the expression takes the temperature variable calculated in the comoving (Eckart) frame.
The factor of $2$ in front of the density represents the two species, electrons and positrons.  We normalize $\epsilon_{\rm est}$ to the initial background plasma value~$\epsilon_{{\rm est},0}$, evaluated with  $T=T_b$, $B^2=B_0^2 + B_G^2$, and $n=n_b$.  The contribution of the background plasma to the initial total normalized estimated radiation power is $P_{{\rm back},0}\equiv\int {\rm d}^3x \, \epsilon_{{\rm est},0} \approx 0.41 P_{{\rm tot},0}$, with all the simulations performed using the fiducial parameters described in Section~\ref{sec-setup}. (The overall effect of the weaker magnetic field at the center of the current sheets is negligible.) 
We also calculated the initial total normalized estimated radiation power due to the Harris sheet: $P_{{\rm HS},0}\equiv\int {\rm d}^3x
\, \epsilon_{\rm est,HS} \approx 0.59 P_{{\rm tot},0}$, where $\epsilon_{\rm est,HS}$ is
$\epsilon_{\rm est}$ evaluated with $T=T_0$, $B^2=B_x^2(y)(1+3u_d^2) + B_G^2$, and $n=n(y) - n_b$. 
Here, in order to get a more accurate estimate of the initial emissivity, we account for the drifts perpendicular to the magnetic field by including the additional factor of $1+3u_d^2$, where $u_d$ is the proper speed of the drifting particle populations in the initial Harris current layer, as defined in Section~\ref{sec-setup}.  
The radiation from the two populations thus accounts for all of the initial radiation $P_{{\rm tot},0} \approx P_{{\rm back},0} + P_{{\rm HS},0}$.

Our first simplified estimate of the total radiated power $P_{\rm tot}$ at any given moment in time is defined as
\be
\label{Pest}
P_{\rm tot,est} \equiv P_{{\rm back},0} \left<\frac{n}{n_b}\frac{T^2}{T_b^2} \frac{B^2}{B_0^2 + B_G^2}\right>,
\ee
and is calculated by substituting \eq{emissivityest} into~\eq{emissionsum}.
Here, $\left<...\right>$ is the volume-average of the product of $n$, $T^2$, and $B^2$ normalized to the background values: $n_b$, $T_b^2$, and $B_0^2 + B_G^2$, which are used to calculate~$P_{{\rm back},0}$. 
At $t=0$,~$P_{\rm tot,est} \approx 0.49 P_{{\rm tot},0}$. 
While this estimation includes the density from the Harris sheet population, it initially underestimates its radiation because the temperature diagnostic is based only on the background population. The estimation, therefore, takes into account neither the higher temperature of the initial Harris population nor the relativistic enhancement due to the bulk flows of electrons and positrons carrying the electric current.
As described earlier, we have ignored the increased radiation due to the bulk flows, out of simplicity.
Both the currents and, later, reconnection outflows do persist throughout the simulations. However, while the enhanced radiation due to the bulk flows does play a role (e.g., as in the minijet model of \citealt{Giannios2009, Giannios2010}; see also \citealt{Nalewajko2011, Giannios-2013}), as we will show below, the simplified estimation of the total, bolometric radiated power remains qualitatively accurate.
Regions with the highest thermal energy content (namely, large plasmoids) tend to have low bulk-flow velocities, and thus the enhancement of radiation due to the bulk flows is limited.
Although the assumption of a local Maxwell-J\"uttner distribution is initially accurate, this estimate ignores any kinetic effects which can play a role as the distribution evolves. Therefore, while the above estimate is reasonable for relatively steep spectra, for harder, highly nonthermal spectra the kinetic effects play an important role, especially for the high-energy emission, as we discuss in Section~\ref{subsec-kinetic} and at the end of Section~\ref{subsec-2Dcomp}.

We also define an even more basic estimation for~$P_{\rm tot}$, making a connection to situations where one knows only the total (volume-integrated) particle kinetic energy (and hence pressure) and magnetic energy as functions of time:
\be
\label{Pest2}
P_{\rm tot,est2} \equiv P_{{\rm back},0} \left<\frac{nT}{n_bT_b}\right>^2\left<\frac{B^2}{B_0^2 + B_G^2}\right>/\left<\frac{n}{n_b}\right>\, .
\ee
At $t=0$, we find $P_{\rm tot,est2} \approx 0.62\, P_{{\rm tot},0}$.
This initial estimation is somewhat larger than $P_{\rm tot,est}$ because it does not take into account the initial anticorrelation between the magnetic field and the plasma pressure due to the initial pressure balance across the current sheet.  Note that this eventually becomes a positive correlation, as discussed in the next subsection. 
Due to the particle number conservation, as the spatial distribution of $n$ evolves, $\left<n\right>$ remains constant in time, and so only one value is needed for calculating~$P_{\rm tot,est2}$. In cases not studied here, where a significant number of pairs are created, the time evolution of this factor would be important.

The simple estimates $P_{\rm tot,est}$ and $P_{\rm tot,est2}$ for the total radiated power~$P_{\rm tot}$ provide convenient estimations from the often limited measurements available in MHD simulations or from observations. Furthermore, their comparison with the actual exact $P_{\rm tot}$ measured in the radiative PIC simulations helps elucidate and highlight the importance of the spatial correlation between the magnetic field and plasma pressure and of the kinetic effects not included in the estimates. Furthermore, diagnostics showing the spatial distribution of the local estimated emissivity~$\epsilon_{\rm est}(\bm x)$ allow us to understand better how sudden enhancements of radiation occur in the context of the reconnection process.

\subsection{Parameter-space histograms}
\label{subsec-histogram}

In our previous 2D work, we have argued that, due to radiative cooling, the magnetic fields and density of the plasma are strongly compressed in the cores of magnetic islands~\citep{Schoeffler2019}.  Although this effect enhances the radiation in these regions, in this paper we will show that it only mitigates the loss of the emitted power due to the overall cooling of the radiating particles.  
We will argue that the positive correlation of the magnetic fields and density inside islands leads to enhanced radiation compared to the simple estimate~$P_{\rm tot,est2}$, such that $P_{\rm tot} \gg P_{\rm tot,est2}$. A histogram of the gridpoints in the $n\mbox{-}B$ space will both allow us to obtain a quantitative measure for the degree of compression and show that there is in fact a correlation between magnetic fields and density.

\begin{figure}
	\noindent\includegraphics[width=0.5\textwidth]{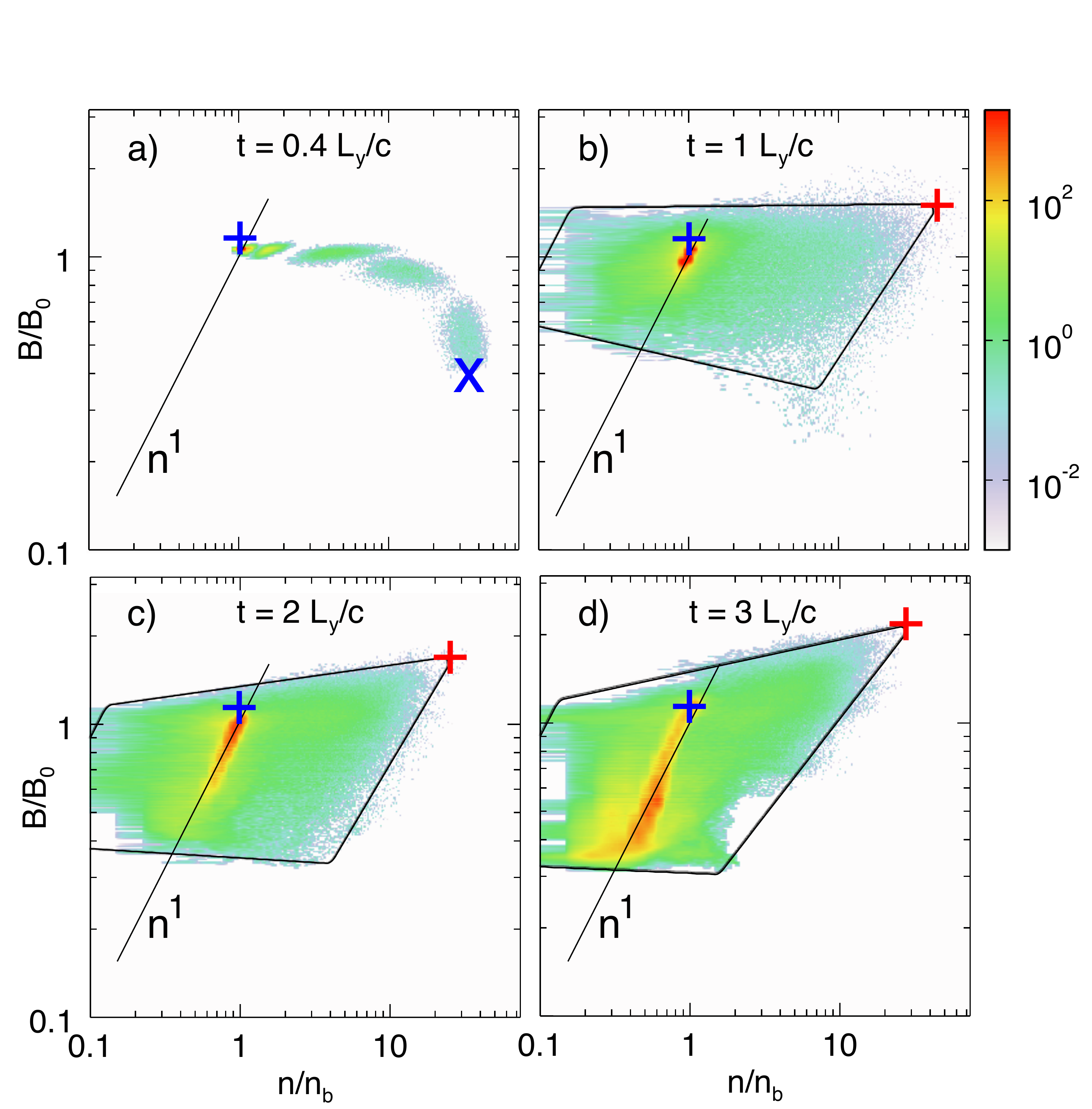}
  \caption{\label{HistBox}
  	Histograms in log-log $n\mbox{-}B$ space of the 3D radiative simulation with $B_G/B_0=0.4$, $L_y/\rho_L= 314.4$, and $\sigma_h = 25.76$, at 4 time-snapshots: (a)  $tc/L_y=0.4$, (b) $tc/L_y=1$, (c) $tc/L_y=2$, and (d) $tc/L_y=3$. At intermediate and late times (panels b-d), the histograms exhibit clear, well-defined power-law boundaries forming 4-sided polygons. 
    The blue plus signs mark the initial conditions of the upstream background plasma, the blue ``X" (in panel~a) represents the initial conditions at the center of the current sheet ($n/n_b = 37, B/B_0 = 0.4$), the red plus signs mark the upper right vertex of the best-fit polygons, and the $B\sim n$ scaling [see \eq{frozenin}] is highlighted.
	}
\end{figure}

Therefore, we visualize this compression and correlation between $n$ and $B$ via the 2D distribution of simulation points in the $n\mbox{-}B$ parameter space. As an illustration, in~\fig{HistBox} we examine the histograms for the 3D radiative case taken at $0.4$, $1$, $2$, and $3$ light crossing times. We show joint 2D histograms of the local values of the normalized $n/n_b$ and $B/B_0$ 
 at each gridpoint. An integral of the histogram over a region of $n\mbox{-}B$ space represents the fraction of the volume with the values of $n$ and $B$ that lie in that region.

The vast majority of the gridpoints are part of the upstream background, where initially $n = n_b$ and $B = \sqrt{B_0^2 + B_G^2} \approx 1.08 B_0$, indicated by the blue plus sign in~\fig{HistBox}. The background plasma is well frozen into the magnetic field. As the upstream, unreconnected magnetic flux is depleted over time via magnetic reconnection, the upstream magnetic field and density drop, keeping the magnetic flux~$B \Delta y$ and number of particles~$n \Delta y L_x$ within a given upstream flux tube of (time-changing) width~$\Delta y$ constant.  The upstream field and density thus follow the simple ideal-MHD relation
\be
\label{frozenin}
\frac{B}{B_0} = \frac{n}{n_b},
\ee
assuming there is not much variation in the $\hat{x}$ direction.  
This simple linear trend can be noted in~\fig{HistBox} for $n/n_b < 1$ where the narrow orange/red band extends over time to lower values of $n$ and $B$ following~\eq{frozenin}.
Aside from this basic observation, after a couple of light-crossing times for the 3D cases, the plasma from the central midplane of the initial Harris current sheet indicated by the blue ``X" in~\fig{HistBox} mixes with the background plasma with the help of a kinking instability described in Section~\ref{subsubsec-nbound}. 

One of the most striking features of the histograms prominently seen at intermediate and late times is that the histograms become bounded above, below, and to the right by clear, distinct limits that can be modeled by power laws. These limits, which constrain the compression of density and magnetic field, will be further discussed in Section~\ref{subsec-hist3D}.

In order to better understand how the compression depends on radiative cooling strength quantified by~$B_0/B_Q$, and on various other parameters in Section~\ref{sec-param}, we design here a novel numerical procedure for measuring the degree of compression. First, we note that, for the 3D simulations after about a light crossing time (starting at $tc/L_y=0.85$), the boundaries of the histogram in log-log $n\mbox{-}B$ space can be approximated with a best-fit of a four-sided polygon.
The parameters describing this polygon are first estimated by hand to match the histogram. A step function with value 1 inside the polygon, with a 20-point smooth, is compared with another step function with value 1 where the histogram is non-zero, with a 10-point smooth.  The parameters of each of the lines are then optimized to a best-fit~\citep{MPFIT}. After each time step, the previous best fit is used as the new initial estimation. 

For example, in~\fig{HistBox}, at $tc/L_y=1, 2,$ and~$3$, the respective slopes of the boundaries (power-law indices $B \sim n^\alpha$) are ($\alpha = 0.005$, $0.075$, and $0.113$) above, and ($0.78$, $0.88$, and $0.67$) to the right of the histogram.  These lines cross at the points  $(n/n_b,B/B_0)=(48.5, 1.53), (24.9, 1.71),$ and $(28.8, 2.18)$ respectively, indicated by red crosses.  
These intersection points mark the upper right corners of the polygons and thus give us an estimation for a maximum level of both $n$ and~$B$.  The final slopes match reasonably well with theoretical predictions that $\alpha = 1/12$, $\alpha = 1/6$, and $\alpha = 1$, which will be discussed in Section~\ref{sec-3D}.

A similar histogram can be constructed for the $n\mbox{-}T$ space instead of the $n\mbox{-}B$ space. This diagnostic furnishes us a convenient visual tool for examining the spatial correlations between $n$ and~$T$.

We will be using these histogram diagnostics extensively in
Sections~\ref{subsec-space},~\ref{subsec-temp},~\ref{subsec-2Dcomp},~\ref{subsec-hist3D}, and throughout Section~\ref{sec-param}, especially in Section~\ref{subsec-Lz}.

\section{2D Results}
\label{sec-2D}
In this section we will explore results from three 2D simulations with varying levels of radiation losses: the classical case $B_0/B_Q = 4.53 \times 10^{-6}$, the intermediate case $B_0/B_Q = 4.53 \times 10^{-4}$, and the radiative case $B_0/B_Q = 4.53 \times 10^{-3}$.  
The rest of the simulation parameters are held fixed here at their fiducial values listed in Section~\ref{sec-setup}.
We show the process of reconnection and the effects that radiation cooling/back-reaction has on it.

\begin{figure*}
	\noindent\includegraphics[width=0.98\textwidth]{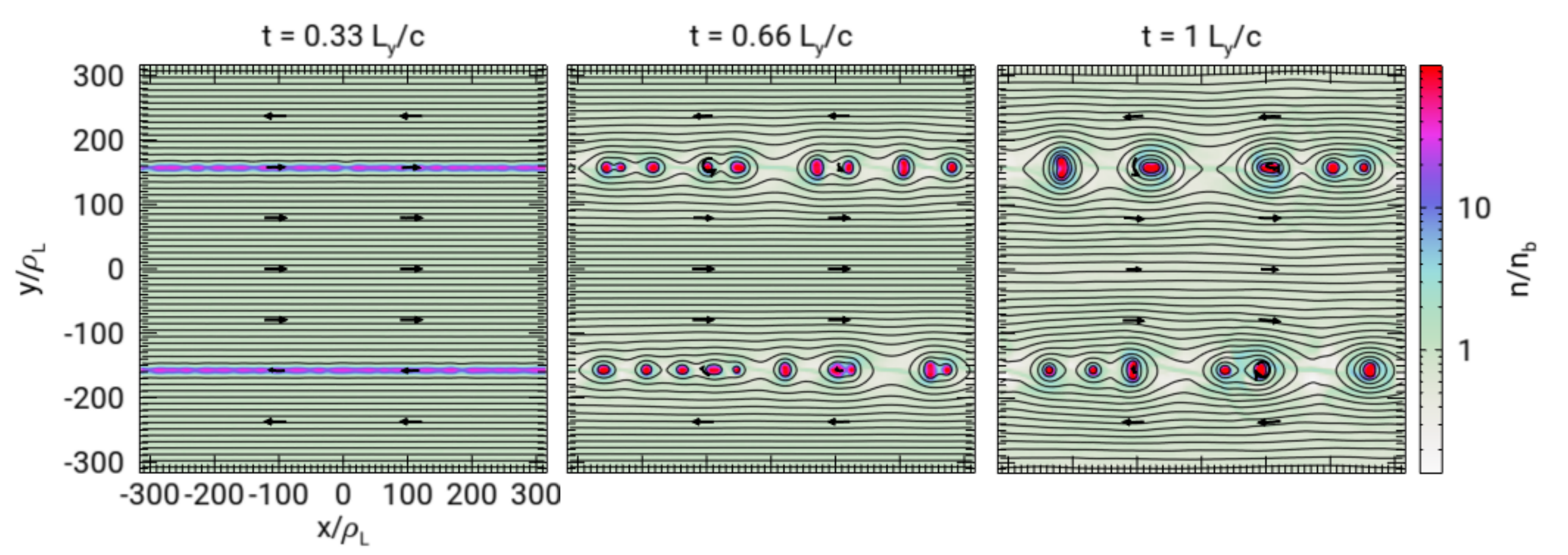}
  \caption{\label{islandformation}
  	Maps of density with in-plane magnetic field lines overlaid for the 2D radiative case
	$B_0/B_Q = 4.5 \times 10^{-3}$, at (a) $t = 0.33 L_y/c$, (b) $t = 0.66 L_y/c$, and (c) $t = 1 L_y/c$. 
	}
\end{figure*}
In all three simulations, the initial current sheet is unstable to the tearing
instability, leading to the formation of multiple magnetic islands separated by
X-points, where magnetic reconnection converts the upstream magnetic energy
into plasma kinetic energy in the form of bulk outflows, heating, and
nonthermal particle acceleration.  The plasma density maps showing the current sheet with super-imposed magnetic field lines (lines of constant magnetic flux), shown in three panels in~\fig{islandformation}, illustrate the generation and merging of magnetic islands during the first light crossing time (up to $t = L_y/c$) of the radiative case, which is qualitatively representative of the other cases as well.

One of the key characteristics of the magnetic reconnection process is the reconnection rate.  To compute it, we first calculate the magnetic flux function $\psi \equiv \hat{z} \cdot \int \bm{B}_{xy}\times {\rm d}\bm{l}$, 
where $\bm{B}_{xy}$ is the in-plane ($xy$) magnetic field, 
and where the integral is taken over the line/contour starting at the bottom left corner, going vertically along the $\hat{y}$ direction, and then horizontally along the $\hat{x}$ direction. 
The reconnection rate measures how fast the difference in $\psi$ between the two current sheets decreases, multiplied by a factor of $1/2$ accounting for the magnetic flux being divided between the two reconnecting current sheets. 
We calculate it using two measures: (i) 
the difference between the major X-points of the two current sheets, corresponding, respectively, to the minimum value of $\psi$ in the upper current sheet and the maximum value of $\psi$ in the lower current sheet 
(defined by the planes $y = \pm L_y/2$, where the current sheets are initially centered), and (ii) the difference between the two values calculated by averaging $\psi$ along each current sheet (i.e., along the previously defined plane). 
The corresponding reconnection rates (defined as the absolute value of the time derivative of the flux) are found to be $0.25$ 
and $0.08~B_0 c_A$, respectively (where $c_A \approx c$ is the upstream Alfv\'en speed), between $tc/L_y = 0.5$ and $1$, after which the rate slows down by a factor of about~$4$.  This is consistent with the predicted reconnection rate for magnetized pair plasmas calculated by~\cite{Goodbred2022}. Although there is a slight trend of decreased reconnection rate for the more radiative cases (stronger~$B_0/B_Q$), the differences are of the same order as the error ($\sim10\%$).

\begin{figure}
	\noindent\includegraphics[width=0.50\textwidth]{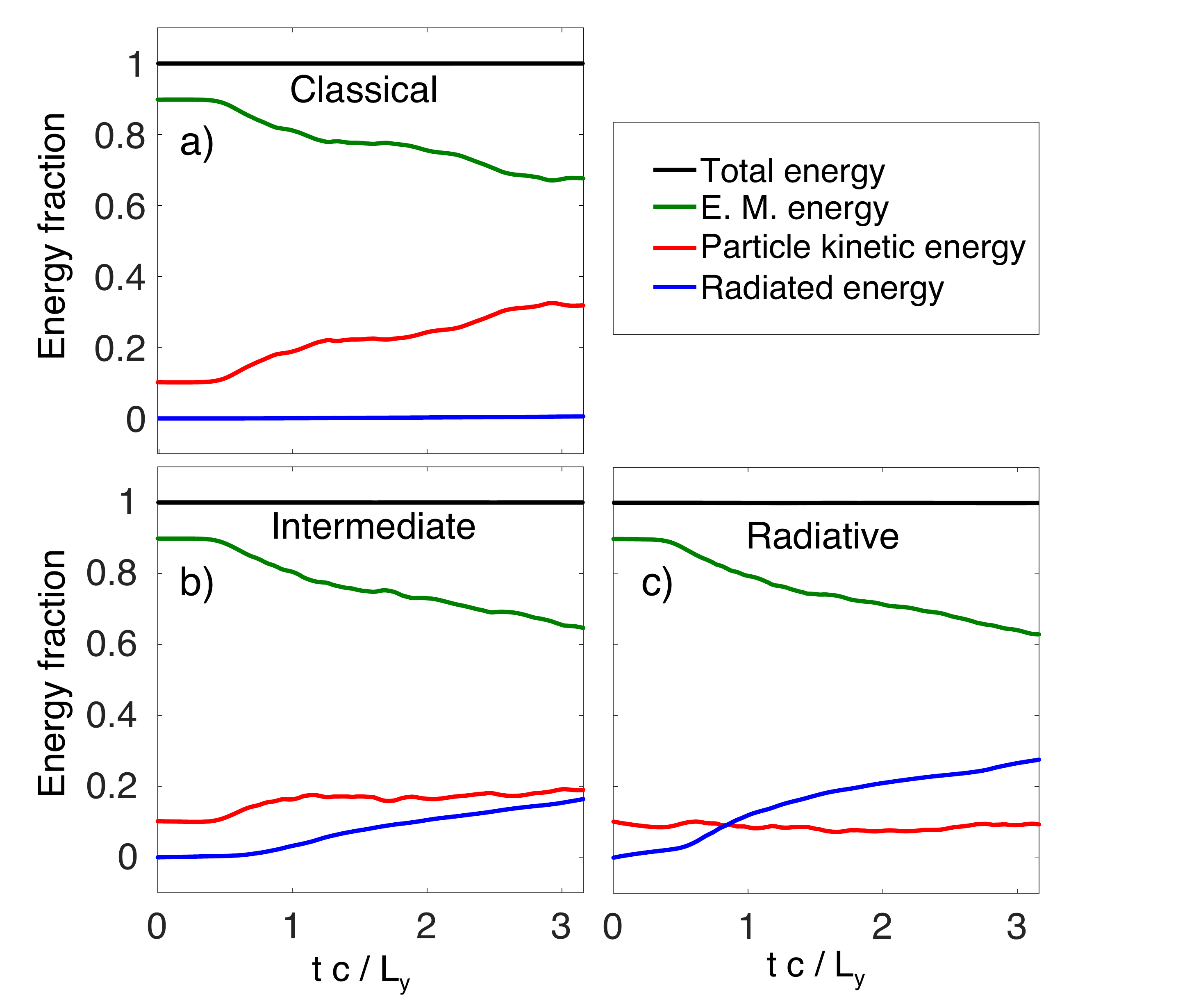}
  \caption{\label{energy2d}
  	 Time-evolution plots of energy partition between electromagnetic (green), particle kinetic (red), and radiated (blue) energies, along with their sum (black), for the 2D simulations with (a) $B_0/B_Q = 4.5 \times 10^{-6}$, (b) $B_0/B_Q = 4.5 \times 10^{-4}$, and (c) $B_0/B_Q = 4.5 \times 10^{-3}$. 
 	}
\end{figure}

The conversion of energy from the magnetic field to the kinetic energy of the plasma particles, and the subsequent conversion to radiation, is shown for all three cases in~\fig{energy2d}.
Unlike the classical case, where a negligible amount of the kinetic energy is radiated away, significant energy goes to radiation in  the intermediate and radiative cases, increasing with the strength of the upstream magnetic field.
In these cases, especially in the radiative case, the particle kinetic energy stays nearly flat throughout most of the evolution from $t \simeq 1\, L_y/c$ onward, while the radiation energy steadily increases; this indicates that the particles act as efficient radiators in this case, promptly converting the energy they receive from magnetic field dissipation into radiation.

Although even in the initial state the thermal particle motion of the plasma leads to some synchrotron radiation energy losses, we will show in Section~\ref{subsec-power} that the radiated power~$P_{\rm tot}$ increases rapidly and significantly during the onset of reconnection, in agreement with all estimates for~$P_{\rm tot}$. 
We will then show in Section~\ref{subsec-space} that, when considering the emissivity as a function of space, a positive correlation between the plasma density~$n$ and magnetic field strength~$B$ leads to an enhanced~$P_{\rm tot}$. 
Although this correlation is more prominent in more radiative cases, we will then show in Section~\ref{subsec-temp} that the correlation between $n$ (or $B$) with the temperature $T$ becomes negative and reduces the normalized~$\bar{P}_{\rm tot}$.  Finally, in Section~\ref{subsec-kinetic} we will address important kinetic effects that affect the emitted power.

\begin{figure}
\noindent\includegraphics[width=0.5\textwidth]{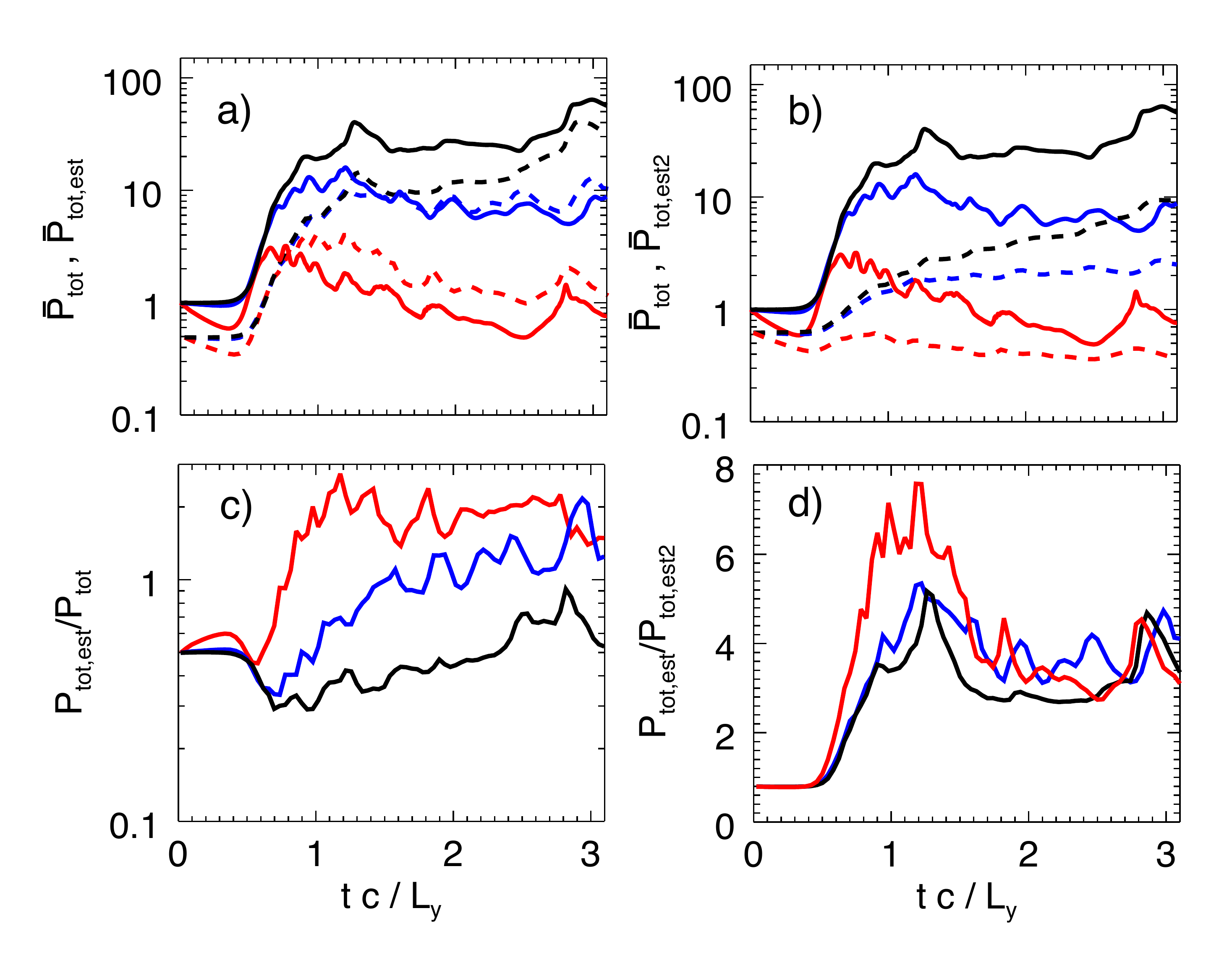}
  \caption{\label{emissionvstime}
  	(a)
	Total normalized radiated power~$\bar{P}_{\rm tot} \equiv P_{\rm tot}/P_{{\rm tot},0}$ (solid lines) for 2D simulations with $B_0/B_Q = 4.5 \times 10^{-6}$ (black, classical case), $4.5 \times 10 ^{-4}$ (blue,
	intermediate case) and $4.5 \times 10 ^{-3}$ (red, radiative case). 
	These colors are used for all panels in this figure. 
    The dashed lines represent the normalized estimated power~$\bar{P}_{\rm tot,est} \sim \left<n T^2 B^2\right>$ [see \eq{Pest}].
	(b)
	Total normalized radiated power~$\bar{P}_{\rm tot}$ (solid lines) and the second normalized power estimation~$\bar{P}_{\rm tot,est2} \sim \left<n T\right>^2 \left<B^2\right>$ [see \eq{Pest2}] in dashed lines.
	(c)
	Ratio~$P_{\rm tot,est}/P_{\rm tot}$.
	(d)
	Ratio of the two estimations of power
        radiated~$P_{\rm tot,est}/P_{\rm tot,est2}$.  
	}
\end{figure}

\subsection{Total emitted power}
\label{subsec-power}
Before discussing the enhancements of the power emitted as a result of magnetic reconnection, we should note the dependence of the radiated power per particle on the magnetic field strength~$B_0/B_Q$ (at fixed $\sigma_h$, etc.). On the one hand, as the radiated power for a given particle is proportional to~$B_0^2$, there is the trivial effect that the most radiative cases (i.e., those with stronger~$B_0/B_Q$) will clearly radiate much more than less radiative ones. 
For the radiative case, $P_{{\rm tot},0}$ is a factor of $10^2$ larger than in the intermediate case and a factor of $10^6$ larger than in the classical case.  
On the other hand, here we are interested in the relative modifications to this trivial scaling due to various factors. Therefore, we will focus the discussion in this paper on the normalized radiated power~$\bar{P}_{\rm tot} \equiv P_{\rm tot}/P_{{\rm tot},0}$. (We will also use this bar notation when plotting estimates $\bar{P}_{\rm tot,est}$ and $\bar{P}_{\rm tot,est2}$, which have the same normalization.)
We will show that this relative enhancement in radiation is weaker for the most radiative cases. That is, the normalized radiation is weaker, but the actual amount of radiation remains much greater.

We first examine the time evolution of the total normalized radiated power~$\bar{P}_{\rm tot}$ for all three 2D cases, which we plot in~\fig{emissionvstime}(a,b) (solid lines). After $t \approx 0.5 L_y/c$, magnetic reconnection gets started and the power of emission abruptly increases by a factor as high as~$30$. 
This is caused by the increase in temperature and a concentration of  magnetic fields inside magnetic islands discussed in Section~\ref{subsec-space}.

The major effect of stronger radiative cooling, quantified by~$B_0/B_Q$, is a drop in~$\bar{P}_{\rm tot}$.  While for the classical case (black lines in~\fig{emissionvstime}), after $t \approx 1 L_y/c$, $P_{\rm tot}$ remains close to a factor of $30$ above the initial state's~$P_{{\rm tot},0}$, for higher $B_0/B_Q$ in the more radiative cases (blue and red lines), the normalized power is limited and even decreases with time for the most radiative case.
This is caused primarily by a decrease in the average particle kinetic energy due to radiative cooling. As the cooling is particularly strong in the densest regions, where the magnetic field is compressed, the effect is enhanced by a loss of the positive correlation between the temperature and density found in the classical case discussed in Section~\ref{subsec-temp}. 

In our previous work~\citep{Schoeffler2019}, we showed that in 2D simulations radiative cooling led to significant additional compression of the magnetic field and density inside magnetic islands (most pronounced for the highest~$B_0/B_Q$), caused by the necessity to maintain a magnetostatic equilibrium. 
The relatively weak guide field $B_G = 0.05\,B_0$ adopted in that study was not able to prevent this compression, and this resulted in a concentration of much stronger radiative losses at the cores of the magnetic islands.  One might then conjecture that this could lead to an enhanced overall normalized power in the more radiative cases, in an apparent contrast with our results presented here in~\fig{emissionvstime}(a).  However, after performing a similar analysis to the data of that previous study, we find the results are qualitatively similar to those presented here.  There was an initial sudden spike in $\bar{P}_{\rm tot}$ once reconnection got started, but the enhancement was weaker for higher $B_0/B_Q$ (more radiative cases) and it decayed with time [similar to~\fig{emissionvstime}(a)].  
The localized enhancement of $\epsilon_{\rm est}$ was not strong enough to counterbalance the overall cooling-driven decrease in $\bar{P}_{\rm tot}$ for stronger~$B_0/B_Q$. In fact, the decrease in relative power was even more pronounced than in the simulations of the present work.

As shown in~\fig{emissionvstime}(a) [see also~\fig{emissionvstime}(c)] the estimated power~$P_{\rm tot,est}$ [see \eq{Pest}], plotted with a dashed line in \fig{emissionvstime}(a), is a qualitatively good predictor of~$P_{\rm tot}$ and, in particular, qualitatively captures the dependence of $\bar{P}_{\rm tot}$ on~$B_0/B_Q$.
In the classical case, $P_{\rm tot,est}$ moderately underestimates~$P_{\rm tot}$, because it does not include the enhancement of radiation due to bulk flows and kinetic effects.
For the intermediate case, it provides an excellent approximation.
However, $P_{\rm tot,est}$ overestimates $P_{\rm tot}$ somewhat for the radiative case. This overestimation is caused by kinetic effects that we will discuss in Section~\ref{subsec-kinetic}.  As shown in~\fig{emissionvstime}(c), the ratio~$P_{\rm tot,est}/P_{\rm tot}$ typically
reaches as high as $\sim 2$ for the most radiative case.

The simpler normalized estimate of power~$P_{\rm tot,est2}/P_{{\rm tot},0}$ [see \eq{Pest2}] is shown as dashed lines in~\fig{emissionvstime}(b). 
During the active reconnection stage [$t \simeq (0.5-2)L_y/c$], it strongly underestimates the emitted power, by a factor as high as~$\sim10$; a significant under-estimation, although not as dramatic, is observed at later times as well.
The reason for this is that $P_{\rm tot,est2}$ does not take into account the positive spatial correlation between strong magnetic field and large kinetic energy density (i.e., plasma pressure), which enhances the radiated power. 
This correlation will be discussed in more detail in Section~\ref{subsec-space} and Section~\ref{subsec-temp}. We highlight the importance of the correlation
in~\fig{emissionvstime}(d) by taking the ratio of the estimated emission~$P_{\rm tot,est}$, which takes into account these correlations, to the estimation from~$P_{\rm tot,est2}$, which does not.  This ratio can reach values as high as~$7$.

\begin{figure*}
  \noindent\includegraphics[width=0.98\textwidth]{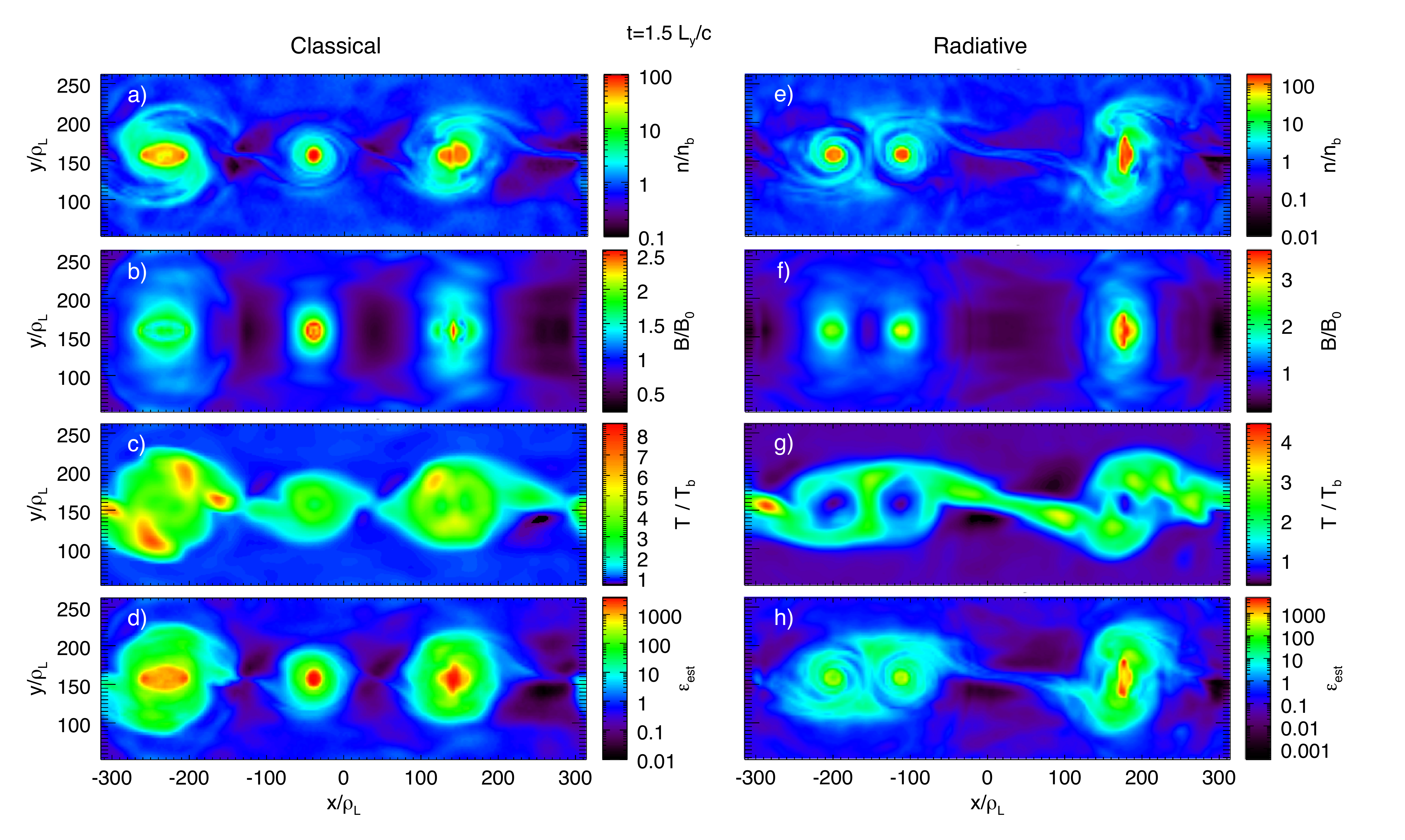}
  \caption{\label{2Dmaps}
	{\it Left column:} Maps of (a) electron density, (b) total magnetic field, (c) effective temperature, and (d) normalized estimated local synchrotron emissivity~$\epsilon_{\rm est}/\epsilon_{{\rm est},0} \sim n T^2 B^2$  for the 2D classical case $B_0/B_Q = 4.5 \times 10^{-6}$ at $t = 1.5 L_y/c$. 
    {\it Right column:} Respective maps (e,f,g,h) for the 2D radiative case $B_0/B_Q = 4.5 \times 10^{-3}$ at the same time $t = 1.5 L_y/c$. 
	}
\end{figure*}

\subsection{Spatial correlation between plasma density and the magnetic field}
\label{subsec-space}

Compression of magnetic fields and density near the centers of magnetic islands~leads to enhancements of the local emissivity~$\epsilon(\bm{x})$, and the total emitted power~$P_{\rm tot}$. At about $t = 1 L_y/c$, when the ratio $P_{\rm tot,est}/P_{\rm tot,est2}$, which quantifies
the importance of the correlation, shown in~\fig{emissionvstime}(d), is highest, reaching a factor of about~$5$, the power in~\fig{emissionvstime}(a-b) is significantly enhanced. During the next light-crossing time the enhancement drops down to about~$3-4$.

At the time $t=1.5 L_y/c$, the compression of $n/n_b$ and $B/B_0$ is illustrated for both the classical case in~\fig{2Dmaps}(a-b) and the radiative case in~\fig{2Dmaps}(e-f).  
The corresponding plasma temperature maps are shown in panels (c) and (g) of \fig{2Dmaps} and will be discussed in more detail in Section~\ref{subsec-temp}.
The maximum density and magnetic field are both found near the centers of the magnetic islands,
indicating a clear correlation between the magnetic field energy and plasma densities.%
\footnote{Note that the apparent extremely strong (reaching~$\gtrsim 100 n_b$!) peak density enhancement inside island cores is mostly explained by the very high density of the plasma in the initial Harris layer, $n_0=37n_b$, which quickly collects in plasmoid cores and subsequently undergoes only a moderate  compression.}
We provide evidence of the enhancement of local emissivity by examining the estimated emissivity $\epsilon_{\rm est}$ as a function of space in~\fig{2Dmaps}(d,h), which shows the strongest emission exceeding the background levels by factors of more than $1000$ in the centers of the magnetic islands.

In the radiative case, there is a noticeable decrease, throughout most of the volume, in the normalized $\epsilon_{\rm est}/\epsilon_{{\rm est},0}$ compared to the classical case [see~\fig{2Dmaps}(d,h)], consistent with the drop in $P_{\rm tot,est}$ shown in~\fig{emissionvstime}(a). This can be explained by the reduction in the effective temperature caused by the radiative cooling. Interestingly, however, at the specific time $t=1.5 L_y/c$ shown in \fig{2Dmaps}, the peak values of $\epsilon_{\rm est}/\epsilon_{{\rm est},0}$ for the radiative and the classical cases are about the same. This is because the negative effect of cooling on the emissivity is compensated, at this particular time, by the stronger peak compression of the magnetic field in plasmoid cores in the radiative case [see~\fig{2Dmaps}(b,f)].    

Indeed, in our previous paper~\citep{Schoeffler2019} we showed that the potential loss of pressure support inside the islands due to radiative cooling (most pronounced for the highest~$B_0/B_Q$) is prevented by the enhanced compression of the plasma density, which in turn drives the compression of the magnetic field (see below). This compression, in principle, should lead to higher synchrotron emissivity. 
The compression is not as pronounced in the simulations presented here due to the stronger guide magnetic field~$B_G/B_0=0.4$ instead of $B_G/B_0=0.05$ adopted in~\citep{Schoeffler2019}. 
However, it still counteracts the direct suppression of the emissivity by radiative cooling and hence may explain why the peak~$\epsilon_{\rm est}$ in~\fig{2Dmaps}(d,h) was not strongly affected by the cooling. 

\begin{figure}
  \noindent\includegraphics[width=0.5\textwidth]{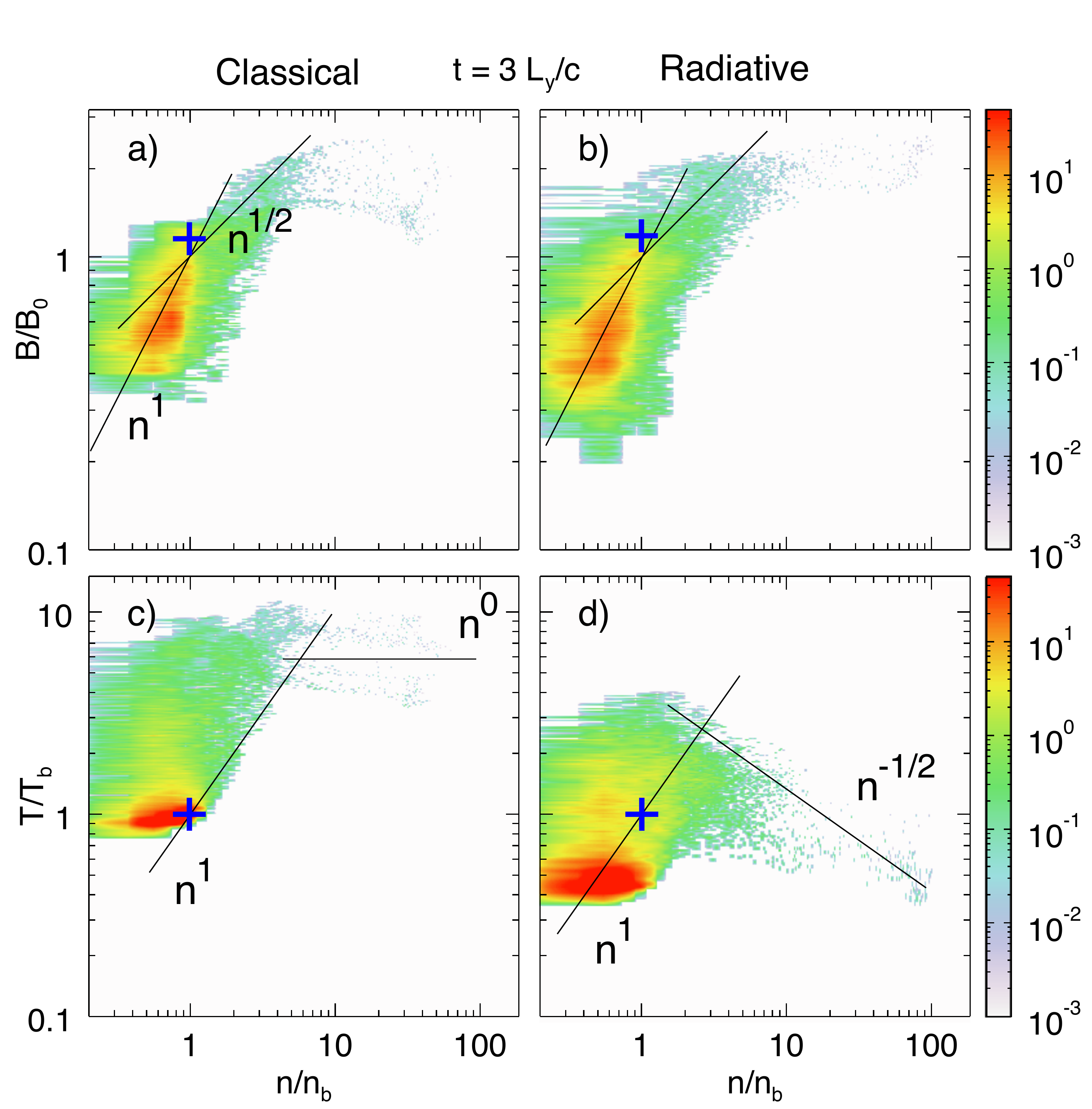}
  \caption{\label{Histogram2D}
  	{\it Top row:} Histograms in $n\mbox{-}B$ space  
	at $t = 3 L_y/c$ in terms of the local density~$n/n_b$ and magnetic field~$B/B_0$,
	for (a) the 2D classical simulation 
	and (b) the 2D radiative simulation. 
    {\it Bottom row:}
	Similar histograms in $n\mbox{-}T$ space in terms of the
	local density~$n/n_b$ and temperature~$T/T_b$,
	for (c) the 2D classical simulation,
	(d) the 2D radiative simulation.
	The blue plus signs represent the initial conditions of the ambient background. 
	The $B\sim n$ [\eq{frozenin}] and $B\sim n^{1/2}$ [\eq{frozenin2}] scalings are shown with thin solid lines in the top panels. 
	Similarly, the scalings $T\sim n$, $T\sim n^0$, and $T\sim n^{-1/2}$ are shown in the $n\mbox{-}T$ space histograms in the bottom panels for reference.
	}
\end{figure}

The enhanced compression can be seen more clearly when examining the $n\mbox{-}B$ histogram shown for the 2D simulations in~\fig{Histogram2D} for $t=3L_y/c$. 
However, before looking at the most compressed regions, let us examine the general features of this histogram.
The basic expected feature, discussed in Section~\ref{subsec-histogram}, is that most of the points start in the background at $n= n_b$, $B = \sqrt{B_0^2 + B_G^2}$, and follow the frozen-in scaling of~\eq{frozenin}. 
In the higher-density region of the histogram, where $n/n_b > 1$,
corresponding to the magnetic islands, a new scaling can be determined, also based on the flux-freezing law. 

First, one should note that the plasma that was initially located deep inside the Harris current sheet, where $n=n_0=37n_b$, as was indicated by the blue X in~\fig{HistBox} at $tc/L_y=0.4$, has moved at later times to the centers of the magnetic islands. This population is represented by a very small number of very-high-density points in~\fig{Histogram2D}, extending up to $n/n_b \simeq 40$ in the classical case and up to $n/n_b \simeq 100$ in the radiative case. On the other hand, the new scaling under the discussion here corresponds to the outer parts of the magnetic islands containing newly reconnected magnetic flux and filled with background plasma, with density $n/n_b \gtrsim 1$.

Let us consider a moderately dense ($n \sim (2-5)\, n_b$), thin annular flux ribbon somewhere inside an island, encircling, but lying outside of, the island's dense, guide-field-dominated inner core. 
Let us examine the self-similar evolution of this ribbon, assuming that its radial thickness $\Delta r$ and its radius~$r$ decrease in unison, in proportion to each other, as the island compresses over time, i.e., $\Delta r(t) \propto r(t)$.
The number of particles~$2\pi n r \Delta r$ and the in-plane magnetic flux~$B_{xy} \Delta r$ enclosed within this flux ribbon should both be conserved as the radius shrinks (neglecting the decay of the magnetic flux due to radiative resistivity, see Appendices~\ref{sec-append-raddiss}
and~\ref{sec-append-Bbound}). 
One can then obtain the following relationship, based on the characteristic values of $n$ and~$B$ inside this flux ribbon, assuming that the in-plane magnetic field $B_{xy}$ dominates over the guide field~$B_z$, so that $B\simeq B_{xy}$:
\be
\label{frozenin2}
\frac{B}{B_0} \sim \left(\frac{n}{n_b}\right)^{1/2}.
\ee
Strictly speaking, this relation should be followed only if the magnetic fields can be well described by a 2D model, since, in a real 3D situation, the compressed plasma could in principle escape the island in the out-of-plane direction.

We can see in the $n\mbox{-}B$ histogram shown in~\fig{Histogram2D} (a-b) that at $t = 3 L_y/c$, the expected correlations~(\ref{frozenin}) and~(\ref{frozenin2}) between $n$ and $B$ due to the frozen-in condition are followed for both the classical and radiative cases.  
For $n/n_b < 1$, it is clear that $B/B_0 \sim n/n_b$, while for $1 < n/n_b \lesssim 10$, $B/B_0 \sim (n/n_b)^{1/2}$ provides a good fit.  For $n/n_b > 10$ (a few points outside of the bulk of the histogram, corresponding to the centers of the primary magnetic islands filled primarily with the initial dense current-sheet plasma), neither of the scalings \eqs{frozenin}{frozenin2} based on frozen-in flux hold, most likely because the compressed guide field $B_z$ dominates here. 
However, we also observe a significant difference between the classical and radiative cases.  
For the classical case, the initial Harris sheet structure is retained; i.e., $B$ decreases with $n$ for very large densities. The initial current sheet was in pressure balance, and thus the magnetic pressure initially decreased along gradients of increasing density. As plasma moves towards the high-density centers of the islands during reconnection, the histogram retains this trend [the magnetic field $B/B_0$ slightly decreases with $n/n_b$ in regions of $n\mbox{-}B$ space where $n/n_b > 10$, shown in~\fig{Histogram2D}(a)]. In contrast, the radiative cooling and subsequent compression present in the radiative case lead to a continued positive correlation between the magnetic field and the density, which results in a somewhat increased magnetic field compression in the radiative case [$B/B_0$ slightly increases with $n/n_b$ above $n/n_b > 10$, shown in~\fig{Histogram2D}(b)].

\subsection{Spatial correlation/anticorrelation between plasma temperature and density}
\label{subsec-temp}

While we briefly considered the importance of the correlation between the particle kinetic energy and the magnetic field energy in Section~\ref{subsec-power}, we mostly focused on the correlation between the plasma density and magnetic field strength in Section~\ref{subsec-space} ignoring any dependence on temperature. 
The correlations of $B$ and $n$ with the temperature $T$ are, however, important because the temperature strongly affects the local emissivity, $\epsilon \sim T^2$.
In fact, in the classical case, there is a positive correlation between the temperature and compressing magnetic fields and density (albeit slightly less pronounced), leading to an even stronger enhancement of the local emissivity $\epsilon(\bm {x})$ and of~$P_{\rm tot}$.  
The enhanced temperature is caused both by heating and particle acceleration via reconnection and by the adiabatic compression of magnetic islands. One can see in~\fig{2Dmaps}(c) that the temperature is increased inside the islands, although it reaches its peak closer to the Y-point region where the reconnection outflows collide with the islands.
 
In contrast, in the radiative case, as seen in~\fig{2Dmaps}(g), there is a general reduction of temperature due to radiative cooling. In particular, the temperature becomes much lower at the centers of the islands, reaching a local minimum. This results in a negative correlation between $n$ and~$T$, which, along with the general cooling, helps explain the clear reduction in~$\bar{P}_{\rm tot}$ shown in~\fig{emissionvstime}(a) for the more radiative cases.

These correlations are also clearly visible in the $n\mbox{-}T$ histograms. In the classical case shown in~\fig{Histogram2D}(c), there is a clear positive correlation between $n$ and~$T$, particularly visible on the right (high-$n$) border of the histogram (with a scaling around $T \sim n$), while the temperatures in the highly compressed ($n\gtrsim 5 n_b$) regions, including island cores, seem to be weakly dependent on~$n$.  
However, in the radiative case~\fig{Histogram2D}(d), one observes  an inverse correlation between $n$ and $T$ when these reach their maximum values (with a scaling around $T \sim n^{-1/2}$). This is expected due to the enhanced cooling at higher $B$ which corresponds to higher densities [as seen in~\fig{Histogram2D}(b)].

\begin{figure}
  \noindent\includegraphics[width=0.5\textwidth]{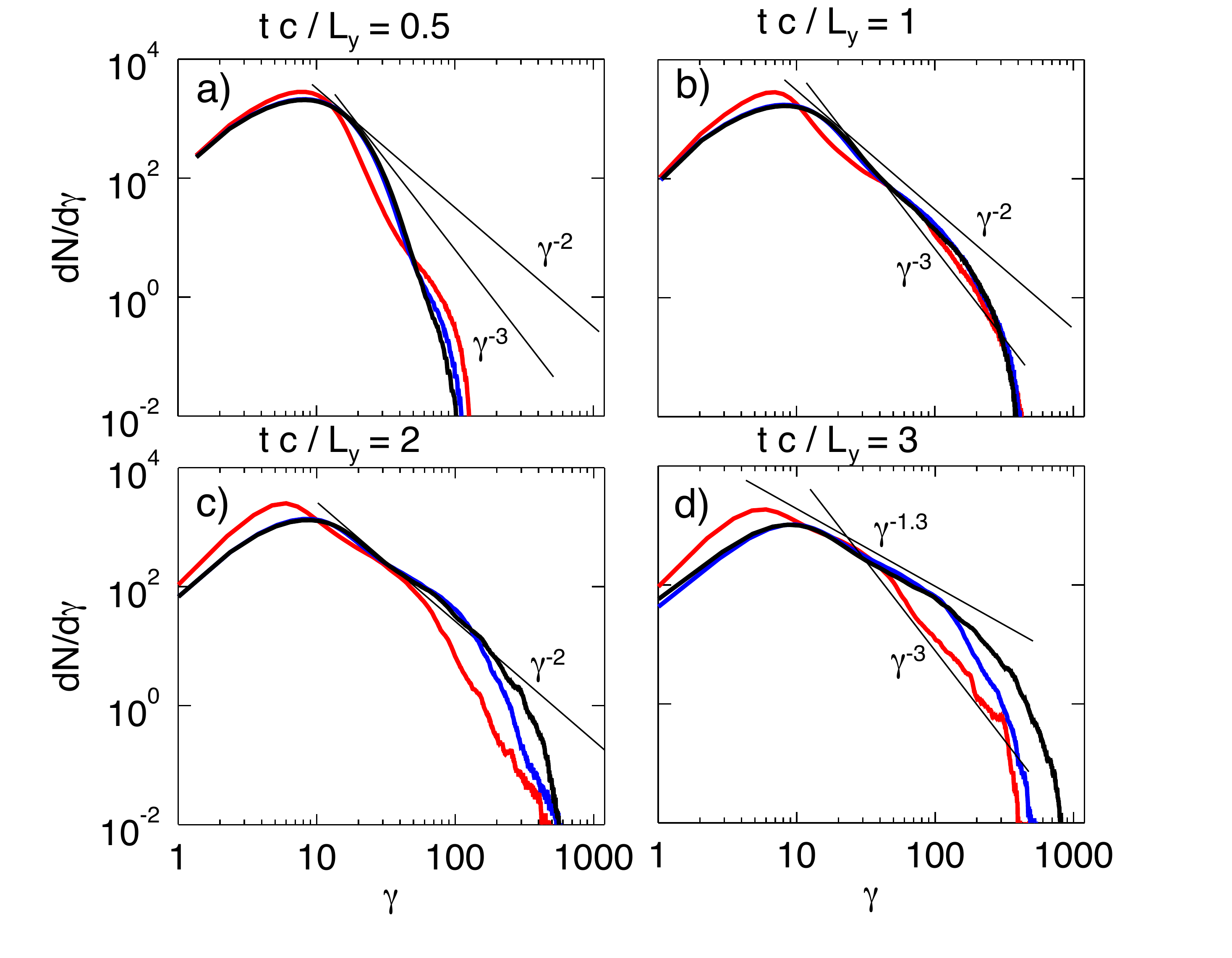}
  \caption{\label{gammadist}
	Electron energy distributions for the 2D classical case
	$B_0/B_Q = 4.5 \times 10^{-6}$ (black), 
	the intermediate case $B_0/B_Q = 4.5 \times 10^{-4}$ (blue), 
	and the radiative case $B_0/B_Q = 4.5 \times 10^{-3}$ (red),
	at $tc/L_y =$ (a) $0.5$, (b)~$1$, (c)~$2$, and (d)~$3$.
	}
\end{figure}

\subsection{Kinetic effects}
\label{subsec-kinetic}
The estimate~$P_{\rm tot,est}$ is based on fluid quantities, assuming that an isotropic Maxwell-J\"uttner distribution in the given species' comoving frame is maintained and thus ignores kinetic effects. 
However, the particle momentum distribution does not in fact remain Maxwellian or isotropic, and the average particle energy alone no longer suffices to determine the emissivity. 
Radiation is dominated by more energetic particles and particles with velocities making large angles with respect to the magnetic field; it is thus affected by features like, respectively, super-Gaussian energy distributions and anisotropic pitch-angle distributions.  
Pitch-angle distribution anisotropy, e.g., caused by the predominant synchrotron cooling of high pitch-angle particles, reduces the emission relative to the level predicted by~$P_{\rm tot,est}$, which may explain the increase in the ratio~$P_{\rm tot,est}/P_{\rm tot}$ seen in~\fig{emissionvstime}(c) for the more radiative simulations. 
On the other hand, nonthermal high-energy particles accelerated during reconnection, which do not always provide a significant contribution to the effective temperature and hence to~$P_{\rm tot,est}$, are expected to radiate significantly more, making $P_{\rm tot,est}$ an under-estimate; this may explain the drop in $P_{\rm tot,est}/P_{\rm tot}$ occurring at the onset of magnetic reconnection around $tc/L_y \approx 0.5$, also visible
in~\fig{emissionvstime}(c) for all three cases.

The energy distributions of the background electrons, shown in~\fig{gammadist} at several different times, display the formation of a nonthermal population after $t = 0.5 L_y/c$ [\fig{gammadist}(a)]. 
By $t = 3 L_y/c$ [shown in \fig{gammadist}(d)], the distribution for the non-radiative case (black) evolves to a hard power law with an index~$\alpha \approx 1.3$, whereas in the radiative case (red) there is a spectral break to a steeper high-energy power law with an index $\alpha \gtrsim 3$, consistent with previous results by
\cite{Werner2016,Werner2019,Hakobyan2019}.  

The moments of the distribution, $\left<\gamma\right>$ (temperature) and  $\left<\gamma^2\right>$ (power radiated), can help us understand the drop in $P_{\rm tot,est}/P_{\rm tot}$ seen in~\fig{emissionvstime}(c) starting at $t \sim 0.5 L_y/c$.  
This drop corresponds to situations where the power-law index of the nonthermal part of the particle energy distribution falls between 2 and~3. Indeed, such a power law has a peculiar property that the first moment of the distribution function (and hence the effective temperature) is dominated by the lower-energy particles with $\gamma$ near the peak of the distribution, while the second moment (and hence the radiated power) is dominated by the highest energy particles. That is, different particle sub-populations are responsible for the temperature, which enters into~$P_{\rm tot,est}$, and for the actual emissivity, which enters into the directly measured~$P_{\rm tot}$; this leads to an underestimation of the emitted power by~$P_{\rm tot,est}$.
We can see in \fig{gammadist}(b) that at $t \sim 1 L_y/c$ the developing power law has become hard enough so that its spectral index is between the critical values $2$ and~$3$ (for the classical and intermediate cases), and this corresponds to a dip in the $P_{\rm tot,est}/P_{\rm tot}$ ratio in~\fig{emissionvstime}(c).
By $t \sim 2 L_y/c$ [see \fig{gammadist}(c)], however, the nonthermal spectra in the classical and intermediate cases, as well as the moderate-energy uncooled part of the spectrum in the radiative case, have hardened even further and their power-law indices start to drop below~$2$.
Both the temperature (the first moment) and the radiative emissivity (the second moment) are now dominated by the same,  highest-energy, particle populations and hence (ignoring the effects of radiative cooling on the pitch-angle distribution which allow $P_{\rm tot,est}/P_{\rm tot}$ to exceed unity in the radiative case) $P_{\rm tot,est}$ becomes a better estimation of the emitted power.

Kinetic effects are therefore expected to enhance the radiated power compared to the average-energy-based estimations ($P_{\rm tot} > P_{\rm tot,est}$) during the early stages of magnetic reconnection, and diminish it ($P_{\rm tot} < P_{\rm tot,est}$) as time progresses for more strongly radiative systems.\\

In summary, in this section, we have shown that, in 2D relativistic radiative reconnection, the total radiated power~$P_{\rm tot}$ is increased at the onset of magnetic reconnection due to the heating and acceleration of particles in the plasma by reconnection, enhanced by the compression and correlation of magnetic fields and plasma density at the centers of magnetic islands,  which can be reasonably well captured by~$P_{\rm tot,est}$, but not by~$P_{\rm tot,est2}$.  In addition, the kinetic, nonthermal effects, which are ignored by~$P_{\rm tot,est}$, can further enhance the radiated power at these early times. 
However, we have also shown that, in the most radiative cases, radiative cooling leads to a pronounced anti-correlation of temperature with density and magnetic field; this causes a decrease in the normalized
radiated power~$\bar{P}_{\rm tot}$. As a result, at late times the enhancements in radiation can be canceled out, and both $P_{\rm tot,est}$ and $P_{\rm tot,est2}$ become better predictors.

\section{3D Results}
\label{sec-3D}

\begin{figure*}
	\noindent\includegraphics[width=0.98\textwidth]{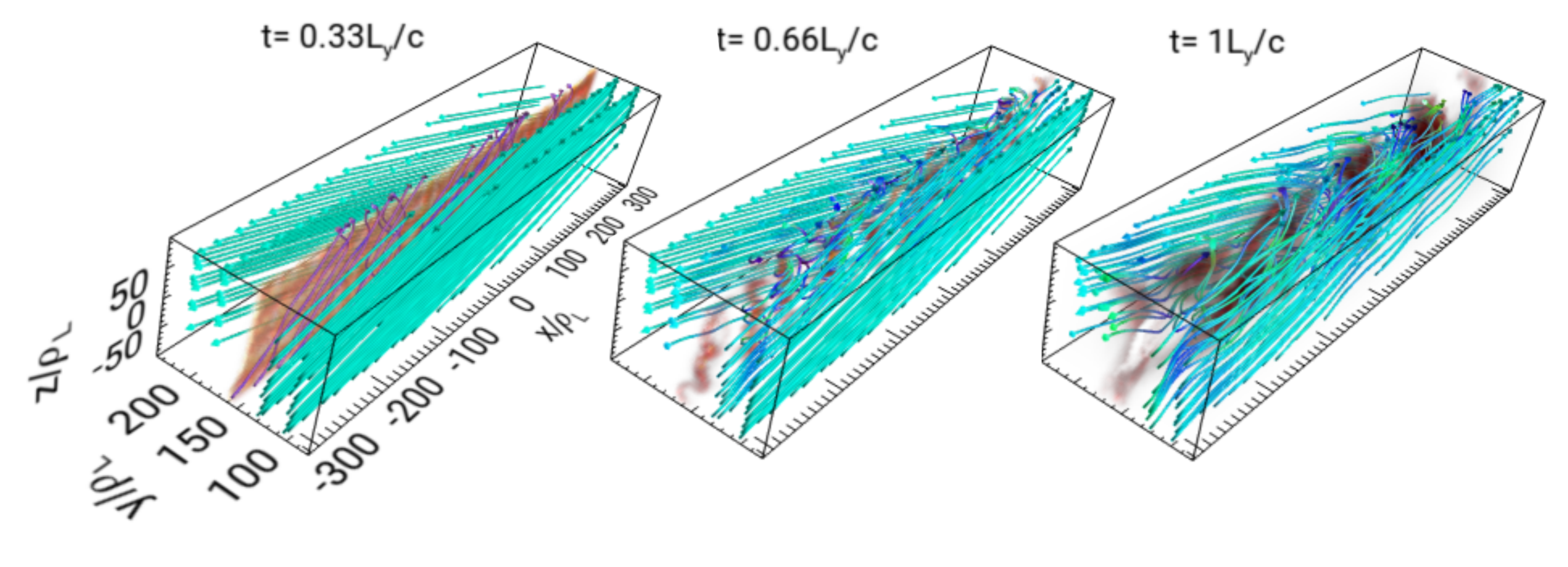}
  \caption{\label{islandformation3d}
  	Zoom-in on the upper current sheet, showing magnetic field lines (green) and a volume rendering of the plasma density (brown),	
	for the 3D radiative case
	$B_0/B_Q = 4.5 \times 10^{-3}$, at (a) $t = 0.33 L_y/c$, 
	(b) $t = 0.66 L_y/c$, and (c) $t = 1 L_y/c$. 
	}
\end{figure*}
\begin{figure}
	\noindent\includegraphics[width=0.5\textwidth]{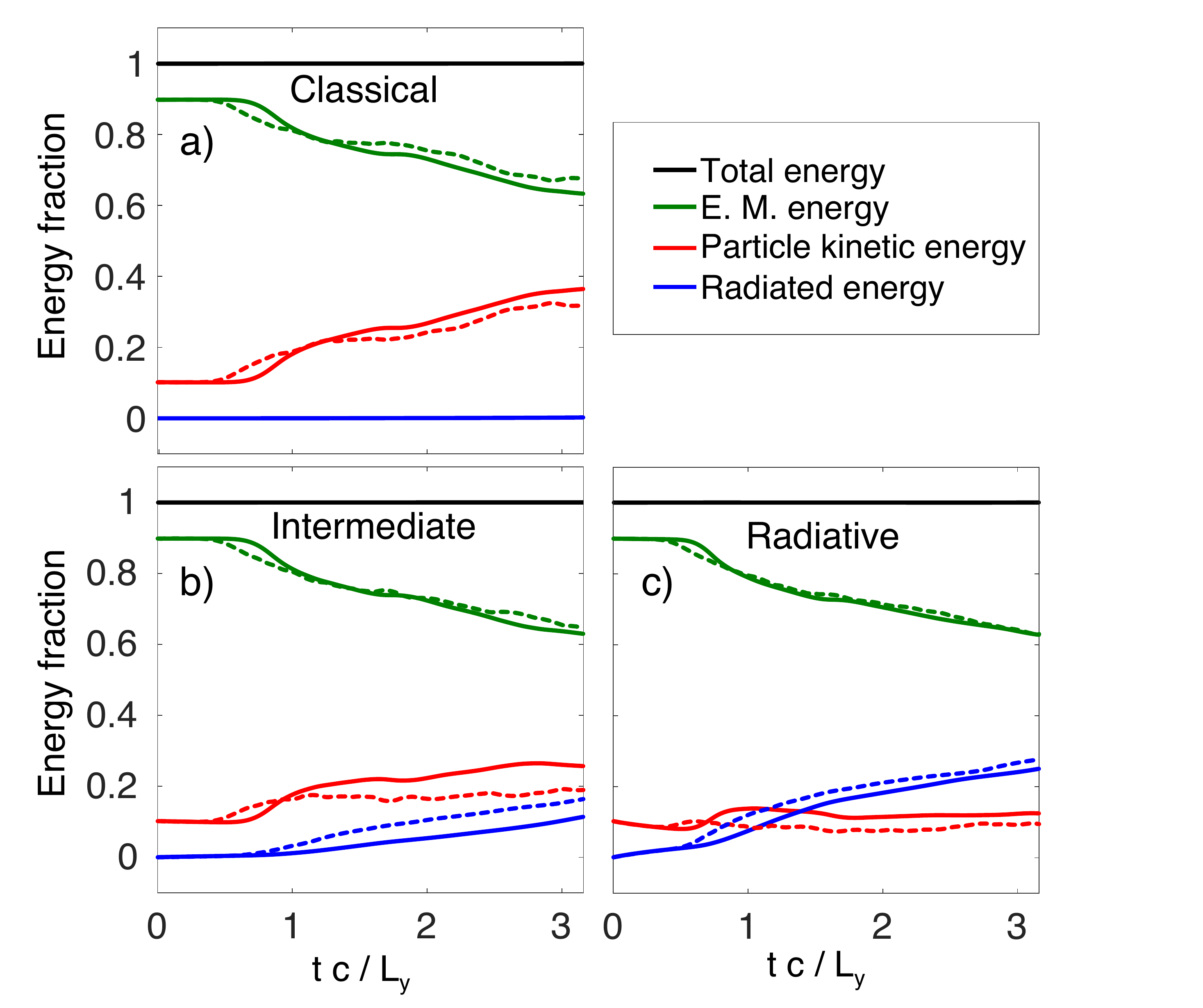}
  \caption{\label{energy3d}
    Time-evolution plots of energy partition between electromagnetic (green), particle kinetic (red), and radiated (blue) energies, along with their sum (black), for the 3D simulations with (a) $B_0/B_Q = 4.5 \times 10^{-6}$, (b) $B_0/B_Q = 4.5 \times 10^{-4}$, 
	and (c) $B_0/B_Q = 4.5 \times 10^{-3}$. 
    For reference, the dotted lines represent the same quantities for the 2D cases.
	}
\end{figure}
As in the 2D study of Section~\ref{sec-2D}, in this section we will explore results from three simulations using the fiducial parameters from Section~\ref{sec-setup} ($\sigma_h = 25.76, B_g/B_0 = 0.4, L_x/\rho_L = L_y/\rho_L = 314.4$, etc.) with varying levels of radiation strength: the classical case $B_0/B_Q = 4.53 \times 10^{-6}$, the intermediate case $B_0/B_Q = 4.53 \times 10^{-4}$, and the radiative case $B_0/B_Q = 4.53 \times 10^{-3}$.  
In these 3D simulations, we adopt the system size in the third dimension to be $L_z/\rho_L=58.6$ ($L_z/L_y = 0.19$). Although this value of $L_z$ is rather small, given our guide field $B_G/B_0 = 0.4$, it still allows the system to exhibit important dynamics in the $\hat{z}$ direction including the development of a kinking instability. We did conduct a parameter-space study varying the guide field in Section~\ref{subsec-BG} and $L_z$ in Section~\ref{subsec-Lz} to justify this choice.  We again show the process of reconnection, the effects that radiation has on it, and now how 3D results differ from~2D. As we will show below, while the guide field keeps the dynamics similar to the 2D case, and many of the standard predictions of reconnection do not differ strongly, the development of a kink mode significantly limits the density compression compared to that found in~2D.

Again, in all cases, the initial current sheet is unstable to the tearing instability, and multiple magnetic islands (plasmoids; flux ropes in~3D) form, driven by magnetic reconnection that converts the upstream magnetic energy into the particle kinetic energy in the form of bulk outflows, heating, and nonthermal particle acceleration.  The plasma density map in the current sheet, with superimposed magnetic field lines, shown in~\fig{islandformation3d}, illustrates the generation and merging of 3D plasmoids during the first light crossing time in the radiative case.  Like in~2D, these dynamics are representative and similar to the other two cases.

The conversion of energy from the magnetic fields to the kinetic energy of the plasma particles (both heating and bulk flows) as a function of time is shown for the three cases in~\fig{energy3d}, comparing the 3D simulations to the 2D ones.  Like in~2D, the particle kinetic energy is rapidly converted into radiation for the more radiative cases, where the radiated energy fraction increases with the strength of radiative cooling characterized by~$B_0/B_Q$. The onset of reconnection, and thus energy transformations occur somewhat later in~3D, but the decay in magnetic energy eventually follows similar curves. Furthermore, for the 3D intermediate and radiative cases, there is slightly less radiation and therefore more particle kinetic energy at late times.

Again we calculate the reconnection rate by looking at the difference in magnetic flux between the two current sheets (using two measures, the difference between the average flux along the planes of the initial current sheets $y = \pm L_y/2$, and between the maximum and minimum values in each of these planes).  
Although in 3D a magnetic flux function is difficult to define in a unique way, we estimate one after averaging the magnetic fields along~$\hat{z}$.
The rate at which the flux decreases, for the radiative case, gives us a normalized reconnection rate of $0.04$ and $0.125 B_0 c_A$ using the two respective measures of flux, a factor of $2$ slower than the equivalent measures in~2D; however, this rate persists for the whole duration of the simulation in agreement with~\citep{Werner_Uzdensky-2021}. 
Also, as in~2D, we do not find a significant dependence of the reconnection rate on radiative cooling strength.

\subsection{Comparisons of radiation, field maps, and their correlations between 3D and 2D simulations}
\label{subsec-2Dcomp}
\begin{figure}
\noindent\includegraphics[width=0.5\textwidth]{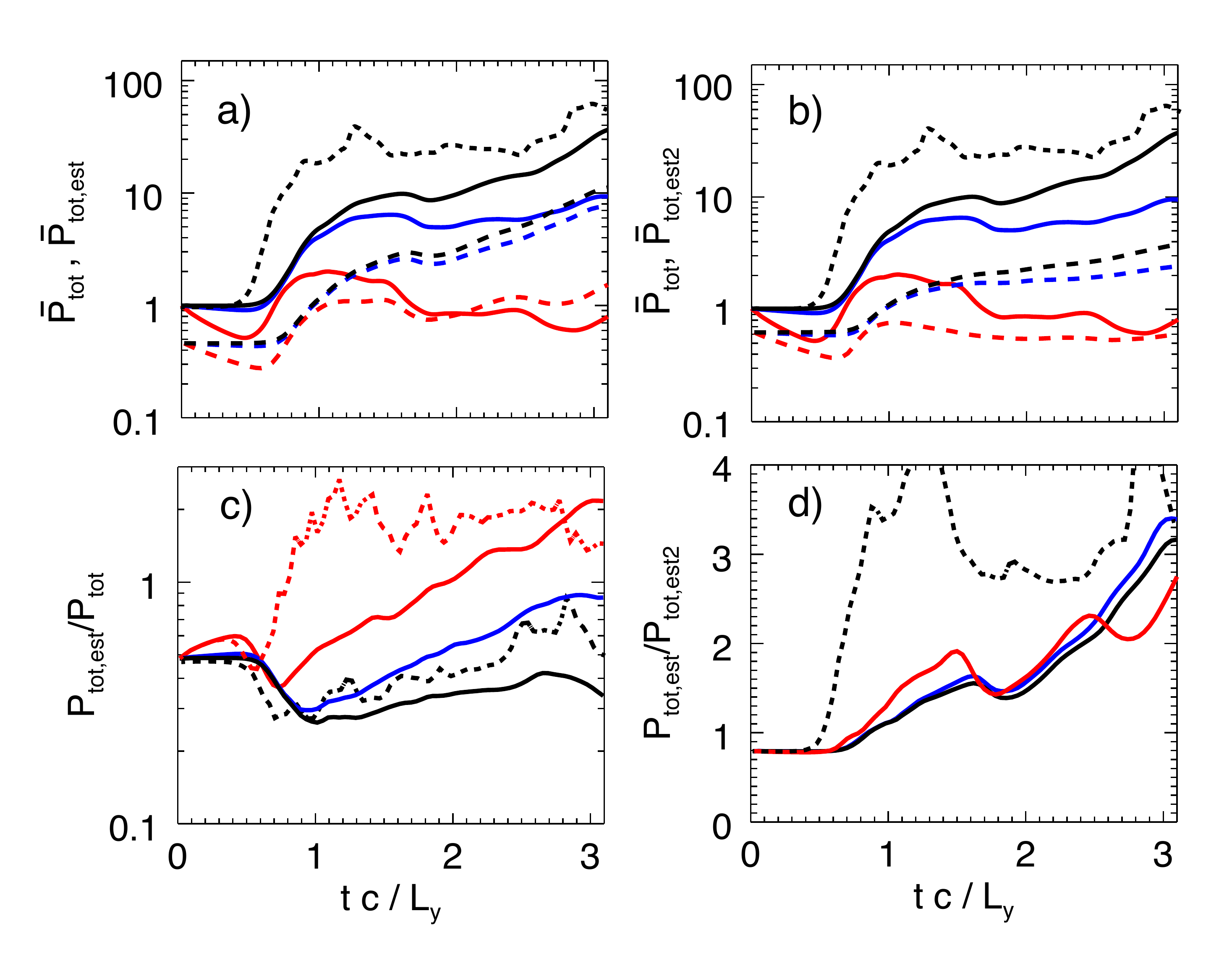}
  \caption{\label{emissionvstime3D}
   	(a)
	Total normalized radiated power~$\bar{P}_{\rm tot} \equiv P_{\rm tot}/P_{{\rm tot},0}$ (solid lines) for 3D simulations with $B_0/B_Q = 4.5 \times 10^{-6}$ (black, classical case), $4.5 \times 10 ^{-4}$ (blue, intermediate case) and $4.5 \times 10 ^{-3}$ (red, radiative case). 
	These colors are used for all panels in this figure. 
    For reference, the black dotted lines represent the same quantities as the solid lines but for the 2D classical case (radiative case shown in red in panel~c). 
    The dashed lines represent the normalized estimated power~$\bar{P}_{\rm tot,est} \sim \left<n T^2 B^2\right>$ [see \eq{Pest}].
	(b)
	Total normalized radiated power~$\bar{P}_{\rm tot}$ (solid lines) and the second normalized power estimation~$\bar{P}_{\rm tot,est2} \sim \left<n T\right>^2 \left<B^2\right>$ [see \eq{Pest2}] in dashed lines.
	(c)
	Ratio~$P_{\rm tot,est}/P_{\rm tot}$.
	(d)
	Ratio of the two estimations of power radiated~$P_{\rm tot,est}/P_{\rm tot,est2}$. 
	}
\end{figure}

For the most part, the power emitted (including its spectra) and its estimates based on the spatial distributions of density, magnetic field, and temperature from the 3D simulations are qualitatively the same as in~2D.
Generally, diagnostics differ only by factors of about~$2$, and we will note some of these modest differences.  However, we will highlight one significant difference: 3D effects tend to disrupt the dense concentrated regions with significantly higher local emissivity at the centers of plasmoids that were found in~2D.

Firstly, the total power and its estimates are qualitatively similar in 2D and~3D.  This can be seen in~\fig{emissionvstime3D}(a,b), where the actual emitted power~$P_{\rm tot}$ and its estimates $P_{\rm tot,est}$ and $P_{\rm tot,est2}$ are roughly comparable to those shown in~\fig{emissionvstime}(a,b) (the dotted line in~\fig{emissionvstime3D} shows the 2D classical result for reference).
However, there are still substantial quantitative differences. The emitted power grows more slowly in~3D, although eventually it reaches magnitudes that are fairly similar to (but slightly less than) those found in~2D. The normalized emitted power $\bar{P}_{\rm tot}$ begins to increase at $t = 0.6-0.7 L_y/c$, about a factor of 1.3 later than in~2D. 
In addition, whereas in the 2D non-radiative case $\bar{P}_{\rm tot}$ stays nearly flat for $t \gtrsim 1 L_y/c$, in 3D it undergoes a steady rise after about $t \simeq 2 L_y/c$, so that the total 2D and 3D radiative powers become very close at late times.
The 3D radiative case differs substantially from its 2D counterpart in terms of the time behavior of $P_{\rm tot,est}/P_{\rm tot}$. In~2D, this ratio [the red curve in~\fig{emissionvstime}(c), also shown in~\fig{emissionvstime3D}(c) as the dotted red curve] quickly rises and then saturates at a level corresponding to $P_{\rm tot,est}$ overestimating~$P_{\rm tot}$ by a factor of about~2. 
For the 3D case, shown in~\fig{emissionvstime3D}(c) with a solid red line, the ratio $P_{\rm tot,est}/P_{\rm tot}$ grows slowly with time throughout the whole simulation, so that $P_{\rm tot,est}$ underestimates $P_{\rm tot}$ until about $t c/L_y = 2$ and reaches the levels of overestimation comparable to the 2D case only by $t c/L_y = 3$.  
Like in~2D, energy is predominantly radiated by the high-energy electrons and positrons moving roughly perpendicular to the magnetic field, leading to deviations from a Maxwellian distribution. Unlike in 2D, the time evolution of these deviations spans the full duration of the simulation.

The enhancement of radiation due to the correlation between the magnetic~$B^2/8\pi$ and thermal~$nT$ energies is similarly present in 2D and~3D.  
In Section~\ref{subsec-power} we showed that the importance of this correlation can be quantified by the ratio $P_{\rm tot,est}/P_{\rm tot,est2}$, shown in~\fig{emissionvstime}(d). 
In 2D this ratio, shown with a dotted black line for the non-radiative case,  rises rapidly during the onset of magnetic reconnection and reaches a saturated value. This differs in 3D [see~\fig{emissionvstime3D}(d)], where the ratio $P_{\rm tot,est}/P_{\rm tot,est2}$, continues to grow slowly and steadily without reaching saturation, and consequently so does the normalized emitted power $\bar{P}_{\rm tot}$ [in \fig{emissionvstime3D}(a,b)] (except for the radiative case, where radiative cooling causes a decrease in the normalized power).  

\begin{figure*}
  \noindent\includegraphics[width=0.98\textwidth]{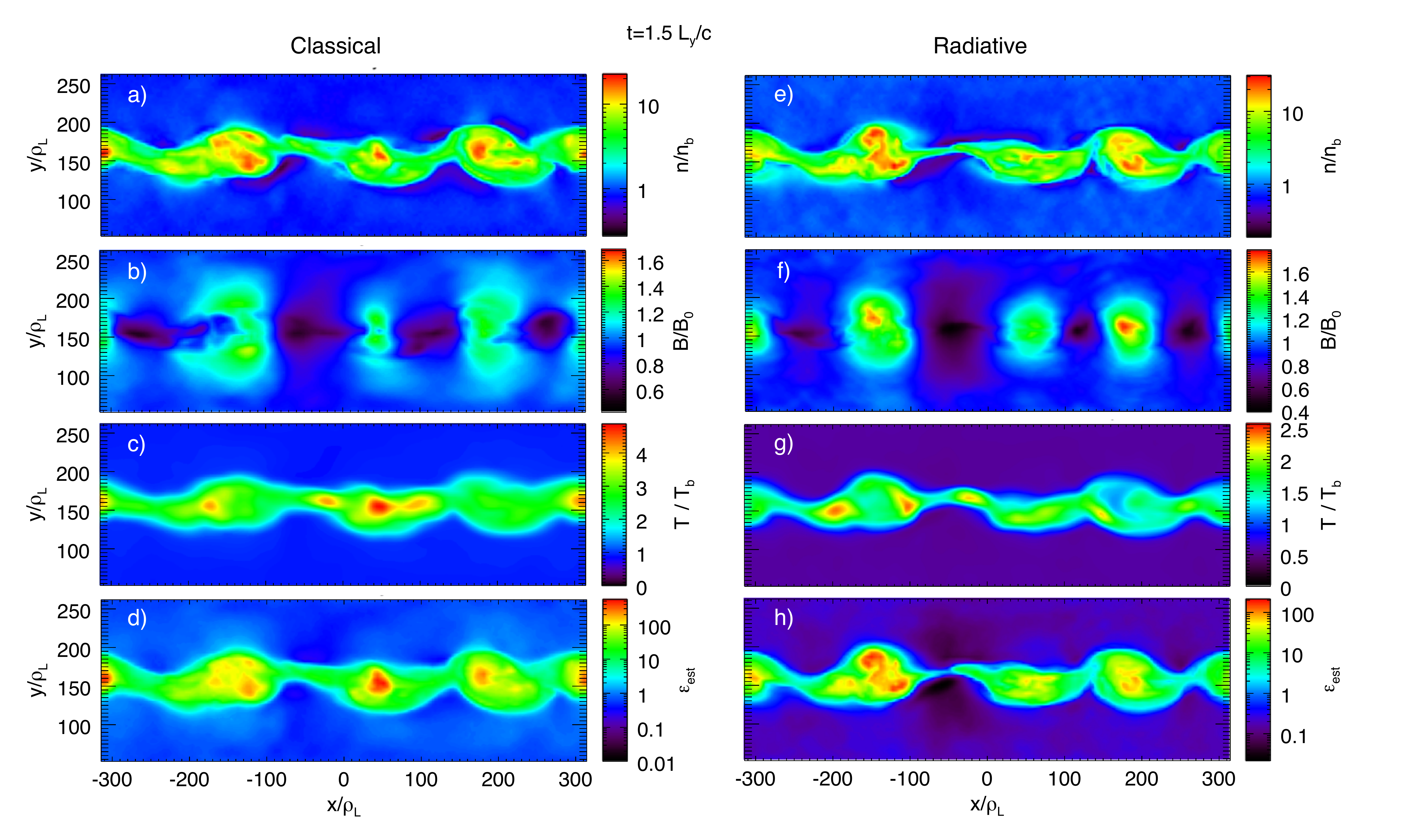}
  \caption{\label{3Dmaps}
	{\it Left column:} Maps of (a) electron density, (b) total magnetic field, (c) effective temperature, and (d) normalized estimated local synchrotron emissivity $\epsilon_{\rm est}/\epsilon_{{\rm est},0} \sim n T^2 B^2$ at a representative slice $z=-4/5\, L_z$ for the 3D classical case $B_0/B_Q = 4.5 \times 10^{-6}$ at $t = 1.5 L_y/c$. 
    {\it Right column:} Respective maps (e,f,g,h) for the 2D radiative case $B_0/B_Q = 4.5 \times 10^{-3}$ at the same time $t = 1.5 L_y/c$. 
	}
\end{figure*}

The most notable difference between the 3D and 2D simulations is that there is significantly less compression of the density in~3D. The respective enhancements of density, shown in~\fig{3Dmaps}(a,e), and of the magnetic field, shown in~\fig{3Dmaps}(b,f), reach values of $\sim 20 n_b$ and $\sim (1.5-2) B_0$, compared to the $\sim 100 n_b$ or $\sim 3 B_0$ in the 2D case.  The density concentration is thus almost an order of magnitude weaker in~3D, while the enhancements of the magnetic field and the temperature, shown in~\fig{3Dmaps}(c,g), are only about a factor of $2$ smaller.
Similar to the 2D case shown in~\fig{2Dmaps}, the spatial correlations between the peak density, magnetic field, and, for the classical case, temperature in~\fig{3Dmaps}, are visible. 

All else being equal, the weaker density compression leads to significantly weaker emissivity at the centers of plasmoids in~3D. The estimated local emissivity $\epsilon_{\rm est}$ reaches peak values that are significantly lower (by a factor of about~$10$) in 3D than in~2D.  
This means that in~3D, regions with significant radiation emission are less concentrated and are spread over a larger volume.  Note that, despite this strong difference in the peak~$\epsilon_{\rm est}$, the total emitted power~$P_{\rm tot}$ remains roughly the same in both 2D and 3D simulations (only a factor of about $2$ higher in~2D). 

\begin{figure}
  \noindent\includegraphics[width=0.5\textwidth]{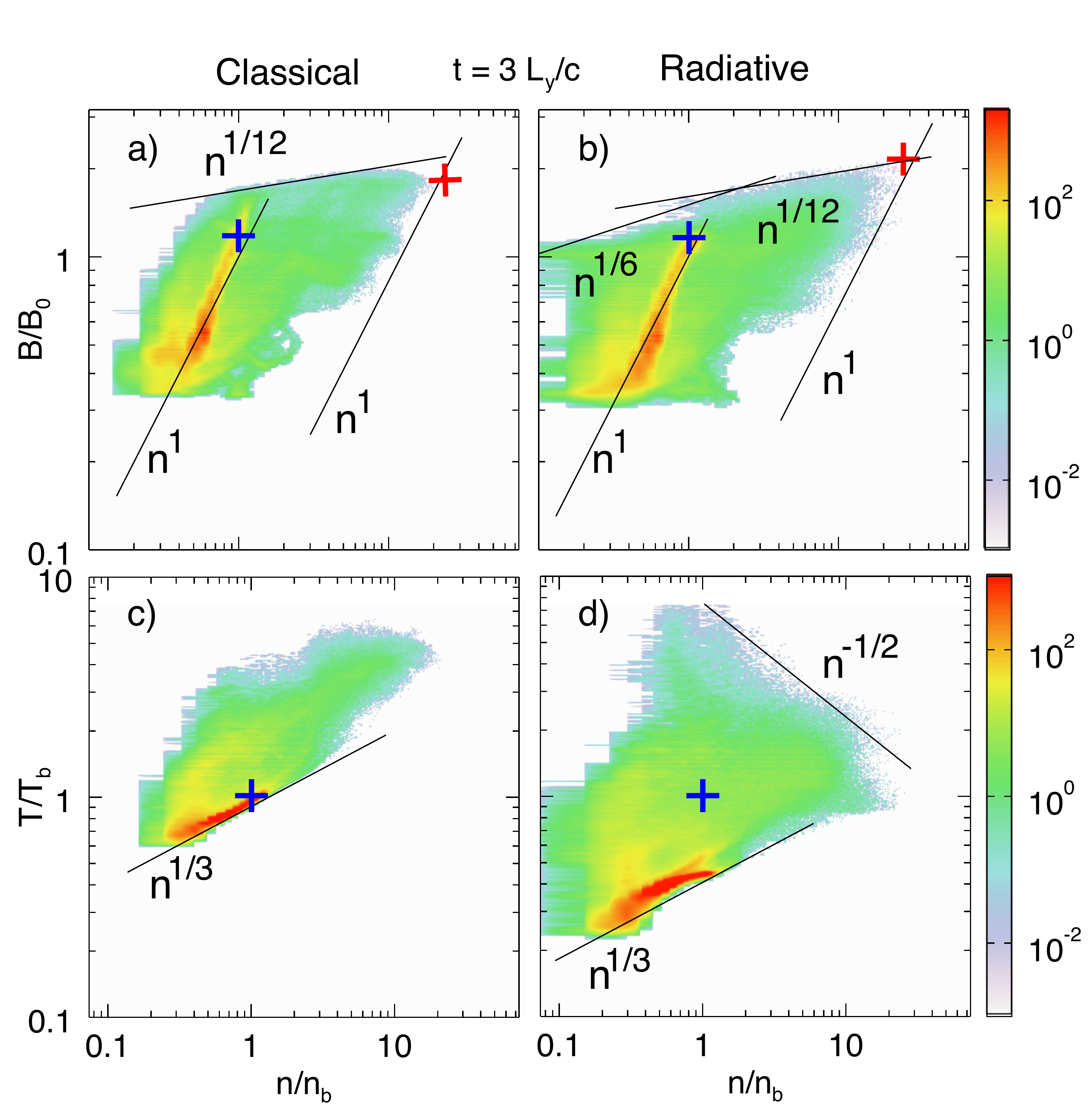}
  \caption{\label{Histogram3D}
  	{\it Top row:} Histograms in $n\mbox{-}B$ space at $t = 3 L_y/c$ in terms of the local density~$n/n_b$ and magnetic field~$B/B_0$
	for (a) the 3D classical case 
	and (b) the 3D radiative case [same as in~\fig{HistBox}(d)]. 
    {\it Bottom row:}
	Similar histograms in $n\mbox{-}T$ space in terms of the
	local density~$n/n_b$ and temperature~$T/T_b$,
	for (c) the 3D classical case,
	(d) the 3D radiative case.
	The blue plus signs represent the initial conditions of the ambient background, while the red plus signs in the $n\mbox{-}B$ histograms mark the upper right vertex of the best-fit polygon boundary of the histogram, as explained in Section~\ref{subsec-histogram}; the $n\mbox{-}B$ coordinates of this vertex are plotted vs. time in \fig{B0scan}.
	The scalings $B\sim n$ [\eq{frozenin} and \eq{nlimit}], $B\sim n^{1/6}$ [\eq{Blimit}], and $B\sim n^{1/12}$ are shown with thin solid black lines in the top panels, while the $T\sim n^{1/3}$ and $T\sim n^{-1/2}$ scalings are shown in the $n\mbox{-}T$ space histograms in the bottom panels for reference.
	}
\end{figure}

The limit on the density compression can be seen in the $n\mbox{-}B$ histogram shown in~\fig{Histogram3D}.  
Like in 2D, in 3D the correlations between $n$ and $B$ due to the frozen-in condition are followed according to \eq{frozenin}
for $n/n_b < 1$.  However, in 3D, as magnetic tension squeezes the plasma to a higher density in a magnetic island, the plasma is free to move out along the $\hat{z}$ direction to regions with a weaker magnetic field. Therefore, variations along the $\hat{z}$ direction caused by, for example, the kink instability, prevent the $n\mbox{-}B$ distribution from following \eq{frozenin2} for $n/n_b > 1$ as found in~2D.
Furthermore, compression of the density is also limited, and the maximum $n/n_b$ drops from $\sim 100$ to close to $30$ (i.e., less than the initial current-sheet density $n_0/n_b = 37$). Not only is the density enhancement limited, but the plasma is also allowed to spread broadly across $n\mbox{-}B$ space, eventually revealing power-law limits that will be described further in Section~\ref{subsec-hist3D}.  

In 3D, the plasma is not as easily trapped and compressed at the centers of plasmoids, where it can be strongly cooled, as occurs in~2D.  Therefore, the anticorrelation between the magnetic field (density) and the temperature, found in the radiative case, is not as strongly pronounced in~3D. This can be seen by comparing \fig{3Dmaps} and \fig{2Dmaps}. However, the cooling still leads to an anticorrelation in~3D.

The similarity of these (anti)correlations between 2D and~3D cases can be also noted from the $n\mbox{-}T$ histograms.  In the classical case shown in~\fig{Histogram3D}(c), there remains a clear positive correlation between $n$ and~$T$, roughly consistent with the relativistic adiabatic scaling $T\sim n^{1/3}$, for the whole range of~$n$.  In contrast, in the radiative case, see \fig{Histogram3D}(d), an inverse correlation between $n$ and $T$ is visible near the maximum values, with a power-law slope close to~$n^{-1/2}$. 

\begin{figure}
  \noindent\includegraphics[width=0.5\textwidth]{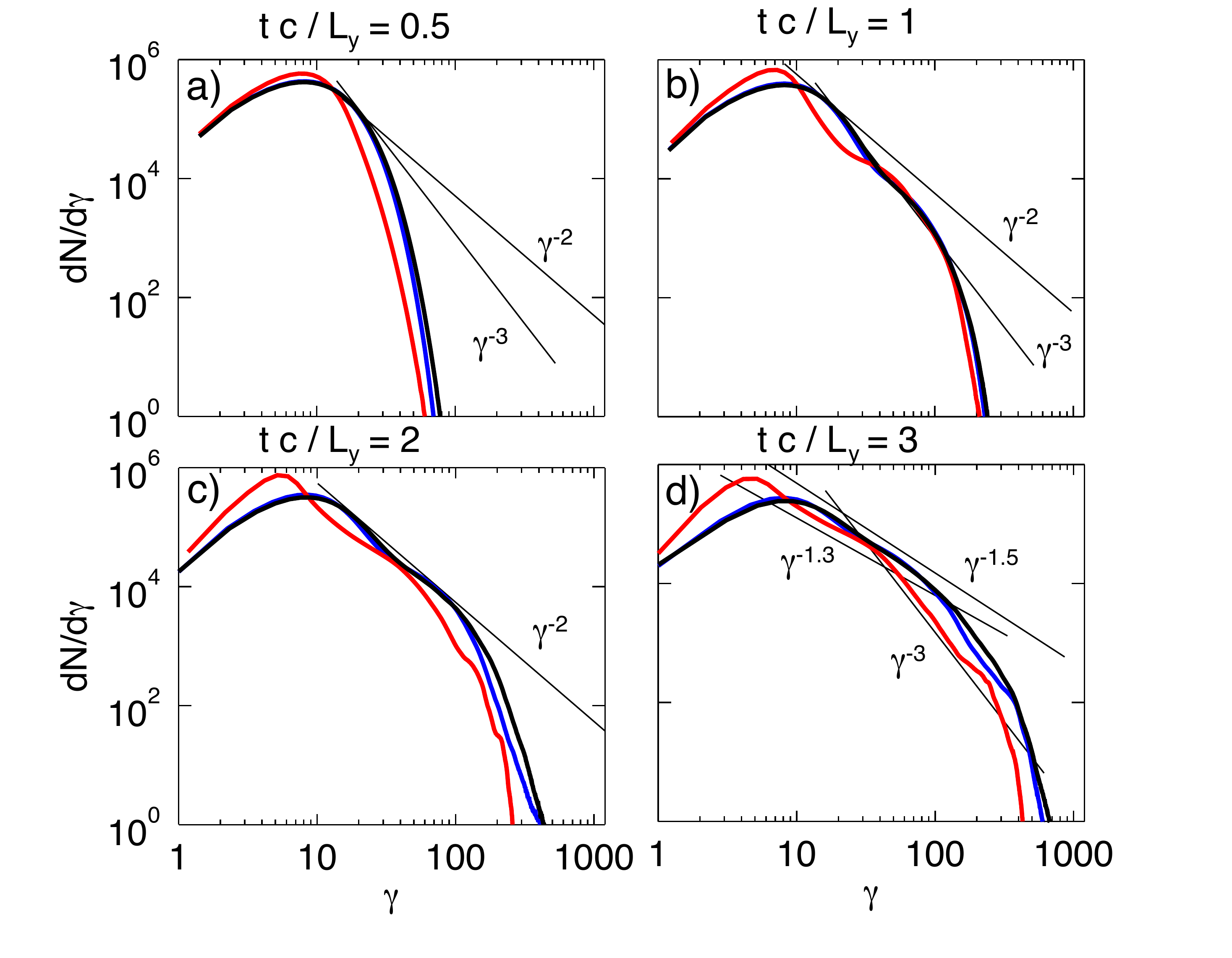}
  \caption{\label{gammadist3D}
    Electron energy distributions for the 3D classical case $B_0/B_Q = 4.5 \times 10^{-6}$ (black), the intermediate case $B_0/B_Q = 4.5 \times 10^{-4}$ (blue), and the radiative case $B_0/B_Q = 4.5 \times 10^{-3}$ (red), at $tc/L_y =$ (a) $0.5$, (b)~$1$, (c)~$2$, and (d)~$3$.
	}
\end{figure}

Finally, the particle energy spectra are almost the same in 3D as in~2D, in agreement with previous studies \citep{Werner_Uzdensky-2017}.
In 3D, shown in~\fig{gammadist3D}, the nonthermal electron population in the particle energy distribution forms more slowly than in the 2D case (shown in ~\fig{gammadist}), and is not yet present by $t = 0.5 L_y/c$.  
However, similar to~2D, at late times ($tc/L_y = 2-3$) a power-law tail is fully formed in 3D runs, with the index reaching ~$\alpha \approx 1.3$ for the radiative case (at moderate energies) and $\approx 1.5$ for the other cases.
Once again, in the 3D radiative case, there is a spectral break to a steeper power law with $\alpha \gtrsim 3$ at higher energies.
These results remain consistent with the results of 2D radiative reconnection PIC simulation studies by  \citep{Werner2019,Hakobyan2019}.  
The limit on the maximum energy of the most energetic electrons is stricter in 3D than in~2D.  As in 2D, kinetic effects influence the accuracy of the estimated power. The spectral index~$2<\alpha<3$ occurs between $t= (1-2) L_y/c$ and may help explain the under-estimation of~$P_{\rm est}$, particularly in the classical case, in~\fig{emissionvstime3D}(c).

We should also remark that, based on the radiative case's background magnetic field strength of $B_0/B_Q = 4.53\times10^{-3}$, all particles with Lorentz factors $\gamma \gtrsim 15$ are expected to emit synchrotron radiation in the gamma-ray regime (i.e., with $\hbar \omega > m_e c^2$), thus potentially feeding powerful pair creation.  While for simplicity we have excluded pair-production effects and other QED physics from the present study, incorporating them self-consistently in PIC studies and examining their back-reaction on the reconnection process itself constitutes a particularly interesting and exciting frontier of extreme plasma astrophysics \citep{Uzdensky2011, Beloborodov-2017, Schoeffler2019, Mehlhaff2021, Hakobyan2019, Hakobyan2023, Chen2023}.

\subsection{Histogram boundaries in 3D}
\label{subsec-hist3D}
As mentioned in Section~\ref{subsec-histogram}, one of the most striking features of the histogram diagnostic from the 3D simulations is that the local levels of magnetic field and density compression are bounded by clear and distinct power laws in the $n\mbox{-}B$ space.
The late-time {\bf ($t=3L_y/c$)} histograms for both the classical and radiative cases are shown in~\fig{Histogram3D}. One can see at the top of the histogram for both these cases, in~\fig{Histogram3D}(a,b), an upper bound on $B$ given by the power law $B/B_0 \sim \left(n/n_b\right)^{1/12}$. At the top of the histogram of the radiative case, in~\fig{Histogram3D}(b), an additional upper bound on $B$, given by the power law $B/B_0 \sim \left(n/n_b\right)^{1/6}$, is seen at lower densities.
Also, to the right of the histogram for both cases, in~\fig{Histogram3D}(a,b), an upper bound on $n$ can be described as $B/B_0 \sim \left(n/n_b\right)^{1}$.
It is also worth mentioning that there is a very clear, robust lower boundary of this histogram: $B_{\rm min} \simeq 0.3 B_0$, essentially independent of~$n$.

To put this in context, the best-fit lines of the boundaries over the entire ranges of $n$ and $B$ for the 3D radiative case presented in~\fig{HistBox}(d) correspond to $B/B_0 = 1.5\left(n/n_b\right)^{0.11}$ and $B/B_0 = 0.23\left(n/n_b\right)^{2/3}$. 
The $0.11\simeq 1/9$ slope of the best-fit upper boundary appears to be roughly an average between the $1/6$ and $1/12$ slopes; it is an artifact of fitting with a single power law a function that is better described as a broken power law. Likewise, the discrepancy between the slopes of the right boundaries shown in~\fig{HistBox}(d) and~\fig{Histogram3D}(b) occurs because, at $t c/L_y = 3$, the power-law boundary is also not distinct along the full range in $n\mbox{-}B$ space. The right boundary does not fit a single power law for low values of magnetic field ($B/B_0~\lesssim~0.6$), and therefore the automatic fit, when applied to the entire range of magnetic-field variation, $B/B_0 \simeq 0.3-2$, does not give an accurate measure of the slope of this power-law boundary. 
However, the fit still provides a good measure of the maximum compression of both $B$ and~$n$ via the intersection point between the two limiting lines, indicated by red plus signs in~\fig{Histogram3D}.

Below we describe a couple of theoretical models that may be used to explain these
power laws, and to get an order-of-magnitude estimate of the coefficients in
front, allowing us to determine the maximum compression theoretically. 

\subsubsection{Density boundary}
\label{subsubsec-nbound}
To the right of the histogram in~\fig{Histogram3D}(a,b) there is a power-law boundary limiting the compression of the plasma density.
In Appendix~\ref{sec-append-nbound}, we present a possible explanation for a boundary with a slope $B_{\rm min} \sim n$ [see \eq{nlimit}], based on the marginal condition for the onset of the kink instabilities found in~3D. Initially, the current sheet can become unstable to the relativistic drift-kink instability (RDKI), while later, the current filaments (flux ropes) can be unstable to other modes including MHD kink.
The kinking of the current filaments, which constitute the highest-density regions, allows the plasma to escape to new locations, thereby checking the growth of the density due to compression. Regions to the right of this histogram boundary are subject to instability, while regions to the left are stable.

\begin{figure}
  \noindent\includegraphics[width=0.5\textwidth]{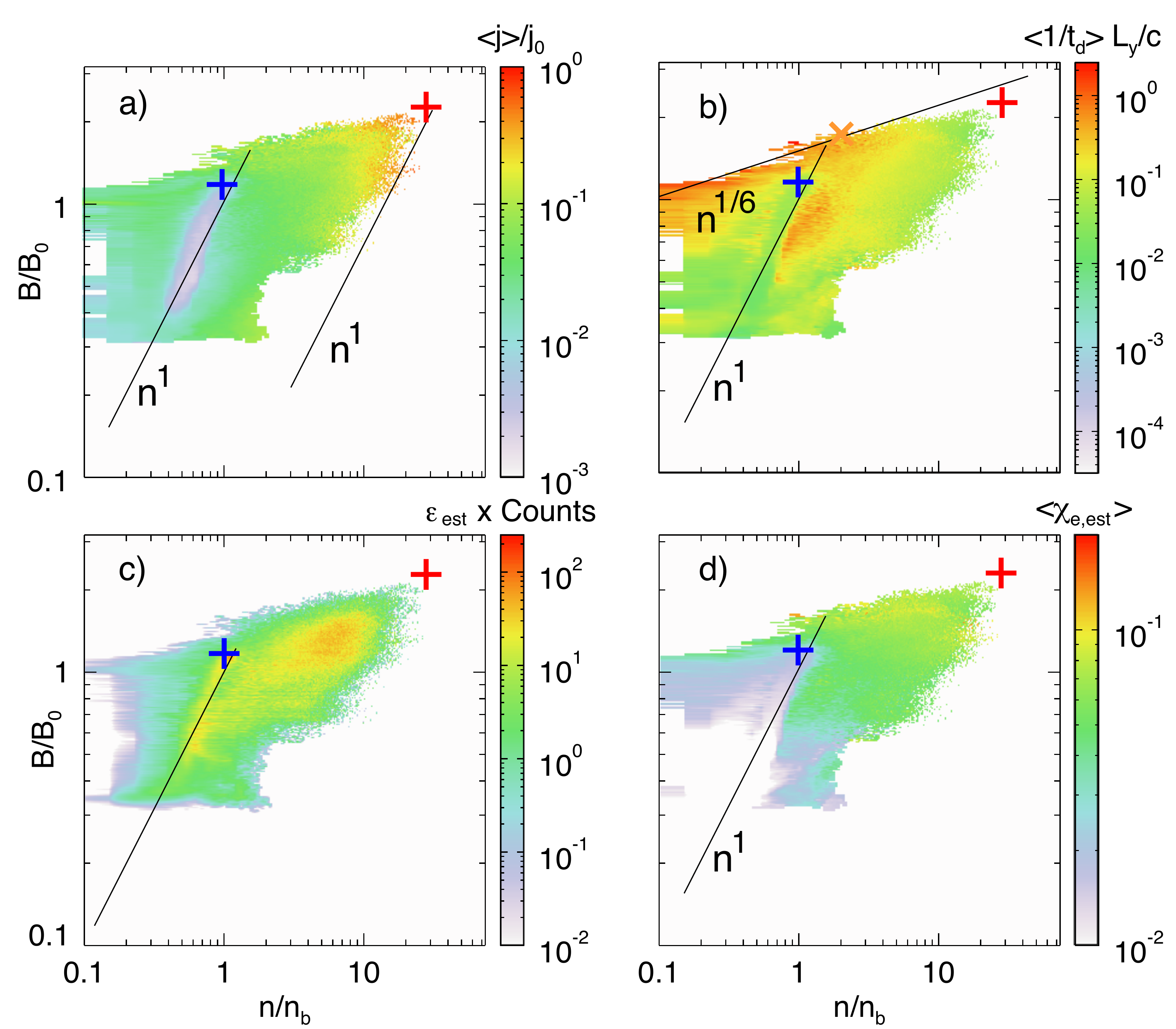}
  \caption{\label{wHistogram}
	2D maps in $n\mbox{-}B$ space of the local average 
    (a) current density $\left<j\right>/j_0$, 
    (b) radiative-resistive magnetic dissipation rate $\left<1/t_d\right>L_y/c$ from~\eq{dissipationrate}, (c) histogram distribution weighted by the local estimated emissivity~$\epsilon_{\rm est}$ from~\eq{emissivityest},
    and 
    (d) $\left<\chi_{e,{\rm est}}\right>$ based on~\eq{chi_apprxdef} with $\gamma = 2 \Theta_e$,
	at $t = 3 L_y/c$ for the 3D radiative case.
	The blue plus signs represent the initial conditions of the ambient background, the red plus signs represent the upper right vertex of the best-fit polygon boundary (see Section~\ref{subsec-histogram}), which is plotted vs. time in \fig{B0scan}. 
	The $B\sim n$ [\eq{frozenin} and \eq{nlimit}] and $B\sim n^{1/6}$ [\eq{Blimit}] scalings are shown with thin solid lines. The orange cross in panel~(b) represents the intercept of the $n^{1/6}$ upper boundary assuming $L_y/c t_d = 0.7$, as explained in the text. 
	}
\end{figure}

\fig{wHistogram}(a) shows that this compression boundary in fact occurs where the electric current density is highest. Instead of the distribution density of the histogram, the average normalized current density $j/j_0$ is shown here for each location in the $n\mbox{-}B$ space.  
The normalization~$j_0 \equiv e n_0 c$ is about equal to the initial peak current density~$1.08 j_0$.  The highest current densities are located at the upper part of the right boundary in the $n\mbox{-}B$ space given by \eq{nlimit} in Appendix~\ref{sec-append-nbound}, suggesting that the location of the boundary is determined by the unstable kinking of current filaments.

The formation of the boundary can be observed as the kink instability evolves.
Initially, the center of the current sheet, marked by an X in~\fig{HistBox}(a), is unstable to the~RDKI.  
After 1 light crossing time, as seen in~\fig{HistBox}(b), the plasma evolves, pushing the histogram into new regions of the $n\mbox{-}B$ space where $n$ tends to be smaller; while lower densities (often with lower current densities) decrease the likelihood of kink instabilities, some of these regions can still be unstable, and kink instabilities continue to grow.  After $2$ or $3$ light crossing times, the nonlinear development of the kink instability is expected to mix high and low-density regions, eventually leaving only regions (confined by a boundary in $n\mbox{-}B$ space) where we predict the kink instabilities to be stable.

In Appendix~\ref{sec-append-nbound}, the $B \sim n$ boundary is predicted to occur where $(B/B_b)/(n/n_B) \approx 0.08$ [see~\eq{eq-kink_boundary-5}]. We test our hypothesis by considering the boundary region, using the local maximum values of compressed
islands, $B/B_0= 2.2$ and $n/n_b=28.8$, from the intersection in~\fig{HistBox} (indicated by red crosses in~Figs.~\ref{Histogram3D} and~\ref{wHistogram}), matching the theoretical predictions remarkably well.

\subsubsection{Magnetic-field boundary}
\label{subsubsec-Bbound}
Above the histogram in~\fig{Histogram3D}, there are power-law boundaries that limit the magnetic field compression. There is an empirically determined boundary with the scaling $B \sim n^{1/12}$, found in both classical and radiative cases. However, there is also evidence for another, somewhat steeper slope, $B \sim n^{1/6}$, for the radiative case at low and moderate plasma densities. Its origin is elucidated in Appendix~\ref{sec-append-Bbound} [see~\eq{Blimit}] based on the radiative dissipation of the magnetic flux, associated with an effective synchrotron resistivity 
\be
\eta_{\rm eff} \simeq \frac{40}{9}\,\frac{e^2B^2}{n m_e^2 c^5}\left(\frac{T}{m_ec^2}\right)^2,
\ee
a function of the local values of $n$, $B$, and~$T$, 
derived in Appendix~\ref{sec-append-raddiss} [see \eq{effressim}].
Therefore, in the radiative case, the combination of this slope and the shallower power law $B~\sim n^{1/12}$ exhibited in the higher-density segment of the boundary [see \fig{Histogram3D}] effectively leads to the intermediate best-fit power law $\alpha=0.11$ over the whole range of~$n$, plotted in~\fig{HistBox}(d).

In the interest of understanding the $1/6$ slope, we look at the 3D radiative case in~\fig{3Dmaps}(h), which shows the emissivity map at~$t=3L_y/c$. It is evident that most of the radiation is produced near the centers of plasmoids where the magnetic field is strongest (the local estimated emissivity $\epsilon_{\rm est}$ is greatest there). As shown in Appendix~\ref{sec-append-Bbound}, the magnetic field dissipation rate via effective radiative resistivity in plasmoid cores is proportional to~$B^2 T^2/n$. Given the parameters of the 3D radiative simulation (described in Appendix~\ref{sec-append-Bbound}), the corresponding magnetic dissipation time-scale $t_d$ is comparable to the radiative cooling time~$t_c$.
Therefore, it is expected that the radiative dissipation has sufficient time to occur and to dominate in these hot, strongly magnetized regions.

To provide firmer evidence that radiative dissipation is most relevant near the upper boundary in $n\mbox{-}B$ space, in~\fig{wHistogram}(b), instead of the distribution of the histogram, we show the average value of the normalized magnetic dissipation rate $\left<1/t_d\right>L_y/c$ [see \eq{dissipationrate} from Appendix~\ref{sec-append-Bbound}] for each location in $n\mbox{-}B$ space. Although we argued earlier that the radiative dissipation is most relevant in regions where the emissivity $\epsilon_{\rm est}$ is greatest, $\left<1/t_d\right>L_y/c$ better determines the relevant regions. 
The picture is, therefore, somewhat nuanced and we need to distinguish two classes of plasmoids. 
First, the cores of primary, first-generation, plasmoids, filled mostly with the dense plasma from the initial Harris current sheet, have the highest emissivity~$\epsilon_{\rm est}$; however, their radiative-resistive magnetic decay rate $t_d^{-1} \propto B^2 T^2/n$ is relatively low because of its inverse scaling with density and because of the cooling-induced anti-correlation between temperature and density.
In contrast, the low-density (and hence relatively low-emissivity) cores of secondary plasmoids, filled with the more tenuous upstream background plasma, have much higher $\left<1/t_d\right>L_y/c$; this is basically because, in order for a smaller number of particles to carry a sufficient current, they must move faster. 
As one can see  in~\fig{wHistogram}(b), the largest values of $\left<1/t_d\right>L_y/c$ are indeed found at the lower-density ($n\lesssim 2n_b$) part of the upper power-law boundary in $n\mbox{-}B$ space.  This is clear evidence that the $B\sim n^{1/6}$ limit on the strength of the magnetic field is indeed related to radiative dissipation.

One can further verify the model by estimating the location of the boundary in $n\mbox{-}B$ space, i.e., the normalization of the power-law scaling.  One can estimate the limit of $B/B_0$ at $n/n_b=2$, near the end of the $n^{1/6}$ scaling,
by solving the expression for $L_y/c t_d$ from Appendix~\ref{sec-append-Bbound} [see \eq{dissipationrate}] with respect to~$B/B_0$, imposing the requirement of significant dissipation during a crossing time, e.g., $L_y/ct_d  \approx 0.7$ (a reasonable number chosen to fit the boundary). 
By taking the parameters of the radiative simulation: $B_0/B_Q =4.53 \times 10^{-3}$, $L_y/\rho_L = 314$, and $\sigma_h = 25.76$, taking the characteristic filament radius from \fig{3Dmaps} to be $r/\rho_L=20$, and setting $\theta_{e,loc} \approx 5 (n/n_b)^{1/3}$, one obtains $B/B_0 \approx 1.7$. 
This is in reasonable agreement with the limits on the histogram shown in \fig{wHistogram}(b), where this point is highlighted with an orange cross.
The $B\sim n^{1/6}$ scaling in \fig{Histogram3D}(b) is valid only for low density $n/n_b \le 2$, and is then replaced by a shallower scaling $B~\sim n^{1/12}$ at higher densities.  
In principle, for more radiative systems, this scaling would be valid for the full range of densities.

\subsubsection{Plasmoids and their compression}
\label{subsubsec-plasmoidcompress}
While discussing the limits on compression, we have focused our attention on the most significant source of radiation, the compressed regions inside plasmoids.  Despite the small area they occupy, the total power they radiate may exceed that from the entire upstream region.
In~\fig{wHistogram}(c), the distribution in $n\mbox{-}B$ space is weighted by the value of $\epsilon_{\rm est}$ for each gridpoint.  This figure illustrates both the significant power radiated from the upstream region, where $B \simeq (B_0^2 + B_G^2)^{1/2}$ and $n \simeq n_b$, and the even greater power radiated from the compressed plasmoid cores, centered around $B = 1.3 B_0$ and  $n= 6.5 n_b$.  Most of these plasmoid regions are located in between (and far from) the two boundaries in $n\mbox{-}B$ space, where neither the density and current are so high that kinking plays a role, nor are the magnetic field and hence $\left<1/t_d\right>L_y/c$ so large that radiative dissipation becomes important.
Thus, the compact, compressed plasmoid-core regions become brightly shining fireballs that contribute significantly to, and perhaps even dominate, the overall emission.

We also wish to highlight the trend that the estimated $\chi_{e,{\rm est}} \equiv (2 T/m_e c^2)\, B/B_Q$ increases in regions of stronger compression (higher~$B/B_0$), as seen in~\fig{wHistogram}(d).
We, therefore, expect that for systems with stronger compression, and hence stronger~$B$, $\chi_{e,{\rm est}}$ could approach or exceed unity, leading to significant discrete hard gamma-ray emission and pair production.

\begin{figure}
  \noindent\includegraphics[width=0.5\textwidth]{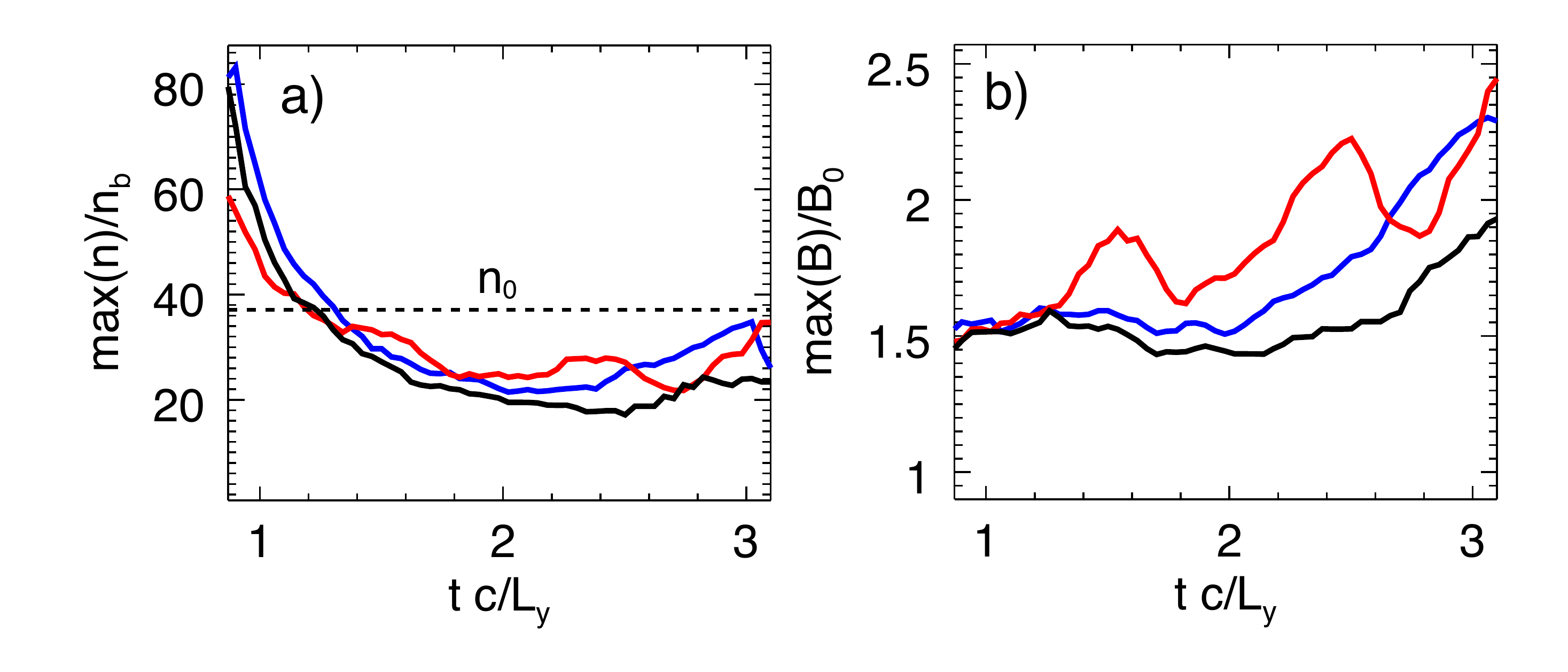}
  \caption{\label{B0scan}
        Peak (a) density and (b) magnetic field corresponding to the upper right vertex of the best-fit polygon boundary of the $n\mbox{-}B$ histogram (see Section~\ref{subsec-histogram}), indicated in \fig{Histogram3D} and \fig{wHistogram}, as functions of time for 3D simulations with a range of magnetic field strengths: $B_0/B_Q= 4.5\times10^{-6}$ (black, classical case), $4.5\times10^{-4}$ (blue, intermediate case), and $4.5\times10^{-3}$ (red, radiative case).
    The dotted line in panel~(a) is the initial density at the center of the Harris current layer.
        }
\end{figure}

In 3D, it is possible to define a useful and simple measure for quantifying the compression of plasmoids using the power-law-boundary fits described
in Section~\ref{subsec-histogram}. 
For times $t > L_y/c$, i.e., after the clear boundaries have developed, a useful measurement of the maximal degree of compression is given by the intersection of the two boundaries in $n\mbox{-}B$ space at the upper right vertex of the histograms. 
This intersection is indicated by the red ``+" signs at $t=3L_y/c$ for the classical case in~\fig{Histogram3D}(a), where $(n/n_b, B/B_0) = (24.0, 1.86)$, and for the radiative case in~\fig{Histogram3D}(b), where $(n/n_b, B/B_0) = (28.8, 2.18)$. 
We present this measurement of compression for both $n$ in~\fig{B0scan}(a) and $B$ in~\fig{B0scan}(b) as functions of time.
Unfortunately, this diagnostic does not work well for the histograms based on our 2D simulations, and thus we only present this diagnostic in~3D.

Although this diagnostic is not yet available before $t \sim 1 L_y/c$ because the clear histogram boundaries have not yet fully formed, before this time, the initial already high density at the center of the current sheet, $n_0 = 37 n_b$, is compressed even more in the centers of magnetic islands, as we have already shown in 2D and~3D in both radiative and non-radiative cases (see Sections~\ref{sec-2D} and~\ref{sec-3D}). By the time the diagnostic becomes available around $t \sim 1 L_y/c$, the density has already compressed to peak values as high as $n/n_b \sim 80$ ($n/n_0 \sim 2$) and the peak density has already begun to decrease.
For all 3D cases the degree of density compression in~\fig{B0scan}(a) drops as a function of time due to kinking (see Section~\ref{subsubsec-nbound} and Appendix~~\ref{sec-append-nbound}). Meanwhile, the magnetic field, which has also already started compressing, continues to grow, as seen in~\fig{B0scan}(b).

We can measure the dependence of the compression on the strength of radiative cooling, controlled by the strength of the upstream reconnecting field~$B_0/B_Q$, by comparing our three simulations.  We find that both $\text{max}\left(n\right)/n_b$ and $\text{max}\left(B\right)/B_0$ have a rather weak dependence on~$B_0/B_Q$ in~3D. 
By $tc/L_y=3$, the respective enhancements of the compression are $\sim 50\%$ and $\sim 30\%$ in the radiative case compared to the classical case. Thus, in contrast with the 2D results from \cite{Schoeffler2019}, the radiative cooling-driven enhancement is relatively modest. Since the out-of-plane magnetic flux due to the initially relatively strong guide-field $B_G/B_0 = 0.4$ is conserved no matter how strongly the plasma cools radiatively, the compression is limited. However, in~3D, as we show in Section~\ref{subsec-BG}, the compression is maximized for this value of the guide field. Although the $B_0/B_Q$-dependence is not so significant (at least for these parameters), we will use these measurements of compression to investigate its dependencies on other parameters in Section~\ref{sec-param} and determine in which regimes more compression may be expected.\\

In summary, in this section, we have shown that, like in~2D, the plasma energization due to 3D relativistic magnetic reconnection leads to a sudden increase in the total radiated power $P_{\rm tot}$ and its simplified fluid-level estimates $P_{\rm tot,est}$ and~$P_{tot,est,2}$.  
This increase is further enhanced by the compression of $n$ and $B$ within the magnetic islands, which we can quantify using theoretical limits in $n\mbox{-}B$ space
(where the limit on density is only found in~3D). The compression enhances the emission of energetic photons and thus may be an important factor in powering gamma-ray flares from various astrophysical systems. 
Simulations performed in 2D give good qualitative agreement with the 3D simulations in overall particle spectra, the total radiation power, and field maps. However, the much stronger density compression, and thus also local emissivity enhancement, found in 2D are disrupted by kinking instabilities that can (and do) develop only in 3D simulations.
Although the density compression is reduced in~3D relative to the 2D case, the sudden increase in radiative power persists.

\section{Parameter scans}
\label{sec-param}
Significant synchrotron gamma-ray emission (i.e., radiation with photon energies $E_\gamma >
m_e c^2$) occurs when the parameter $\chi_e$ of a significant number of
particles gets as large as $\chi_e \sim 1/\gamma$, and thus the synchrotron photon energy, which is about $\hbar (eB/m_ec) \gamma^2$ is of order $m_ec^2$.
Reconnection can cause an enhancement of the $\chi_e$ parameter by particle energization and also by magnetic field compression.

In Sections~\ref{sec-2D} and~\ref{sec-3D}, we have found that, although larger values of $B_0/B_Q$ lead to more intense gamma-ray emission, the enhancement of radiation from the reconnection process becomes less pronounced for stronger magnetic fields, due to radiative cooling of the plasma overall, as well as the fact that locations of most significant radiation also suffer the most radiative cooling.
However, we have also found that radiative cooling leads to an enhanced compression of the magnetic field and plasma density, helping to mitigate these effects. We have thus begun to understand the effect of one parameter, i.e., the normalized strength of the reconnecting magnetic field~$B_0/B_Q$, on gamma-ray emission. 
However, there are several other important parameters to consider which also merit investigation.

To decide which parameters to investigate, we consider some important questions. For example, in what parameter regimes do we expect the strongest compression and the strongest flaring of radiation in the gamma-ray energy range?
In which regimes do we expect 2D models to provide good predictions for the full 3D system? When are kinking instabilities in the $z$ direction important, and how do they affect compression?
Some regimes exist in theory, but are difficult to simulate numerically;
are there regimes with stronger flaring of radiation in the gamma-ray energy range than those we can simulate?
In what regime is significant pair production eventually expected to take place (i.e., typically $\chi_e > 1$)?
Do we expect such regimes to occur in astrophysical environments?

Motivated by these questions, in this section we will explore the effects of several important parameters: 
guide field ($B_G/B_0$), which resists and inhibits compression but can also mitigate 3D effects by suppressing the kinking instabilities; 
system size ($L_y/\rho_L$ and $L_z/\rho_L$), which allows for longer evolution of both relevant 2D and 3D dynamics; 
and upstream plasma magnetization~$\sigma_h$, which quantifies the magnetic energy released during reconnection. 
Each of the subsections that follow presents the findings of an individual parameter scan with respect to one of these parameters. 
We perform these parameter scans by starting with our previous fiducial case, the 3D radiative ($B_0/B_Q = 4.53 \times 10^{-3}$) simulation setup, and individually varying these parameters while keeping the others constant.

\subsection{Parameter scan: guide field \texorpdfstring{$B_G$}{TEXT}}
\label{subsec-BG}

The first step towards finding a regime with significant gamma-ray emission is looking at the dependence on the guide field~$B_G/B_0$.  
A real 3D plasma acts like a 2D simulation only for a sufficiently strong guide field. Therefore, one should view as tentative any conclusions based on 2D simulations with a weak guide field, such as those presented by \cite{Schoeffler2019}, where the compression was unphysically large.  
For the simulations shown in the previous sections of the present paper, we have chosen a guide field of~$B_G/B_0 = 0.4$, which is both strong enough for an order-of-magnitude agreement with 2D simulations, but also weak enough to allow significant compression and thus enhancement of emitted radiation power.

To investigate the dependence of the results on the guide field, we have performed a parameter scan of~$B_G/B_0=0.05$, $0.2$, $0.4$, $0.6$, and~$1.0$, for the 3D radiative case, keeping $B_0/B_Q = 4.53 \times 10^{-3}$, $L_y/\rho_L = L_x/\rho_L = 314.4$, $L_z=0.19\,L_y = 58.6\rho_L$, and $\sigma_h = 25.76$ fixed.
\begin{figure}
  \noindent\includegraphics[width=0.5\textwidth]{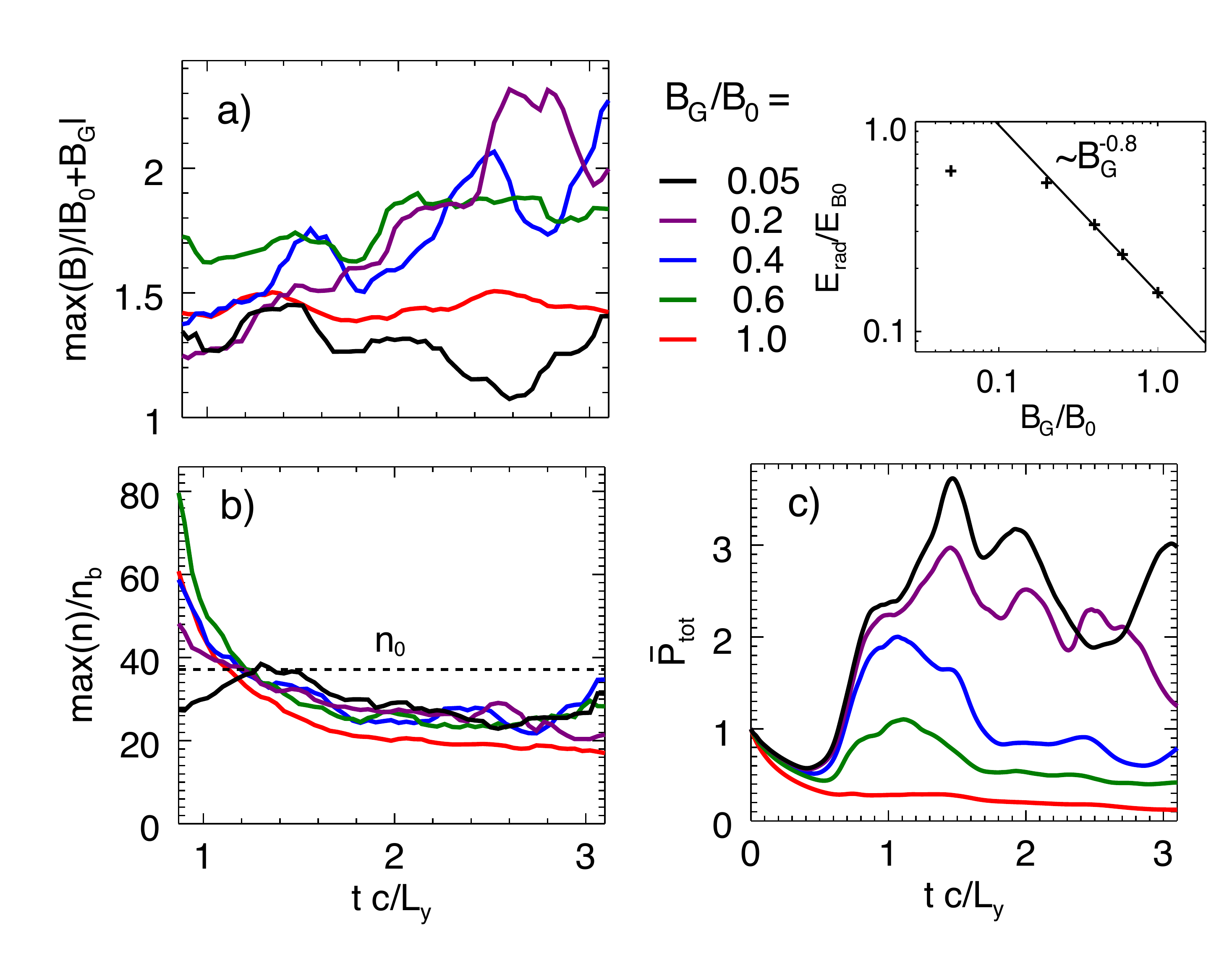}
  \caption{\label{BGscan}
	Peak (a) magnetic field, normalized to the initial upstream field strength $(B_0^2 + B_G^2)^{1/2}$, and (b) plasma density, corresponding to the upper right vertex of the best-fit polygon boundary of the $n\mbox{-}B$ histogram (see Section~\ref{subsec-histogram}), and (c) total normalized
	radiated power~$\bar{P}_{\rm tot}$ (with the time-integrated power as a function of $B_G/B_0$ plotted above), as functions of time for 3D radiative simulations with a range of initial guide magnetic fields: $B_G/B_0= 0.05$ (black), $0.2$ (magenta), $0.4$ (blue), $0.6$ (green), and $1.0$~(red).
    The dotted line in panel~(b) is the initial density at the center of the Harris current layer.
	}
\end{figure}

We again use the histogram diagnostic from Section~\ref{subsec-histogram} to find a good measure of the maximal $n$ and $B$ compression for times $t > L_y/c$ after the clear boundaries have developed.  
We show in~\fig{BGscan}(a) that the compression of the magnetic field has a clear dependence on the strength of the guide field. 
Here we calculate the compression based on the maximal total field compared to the initial upstream total field $|\bm B_0+\bm B_G|$, instead of just~$B_0$, as we are no longer keeping $B_G$ constant.
One should first note that the effect of the guide field on the magnetic compression is non-monotonic. While for strong guide fields e.g., $B_G/B_0 = 1$, the guide-field pressure naturally acts to limit the compression of the plasma and thus the compression of the magnetic field, for a very weak guide field $B_G/B_0 = 0.05$ there is also almost no compression seen. As we will show later in this subsection, the compression, in this case, is disrupted by the development of a kink instability.
We thus find that magnetic-field compression peaks at intermediate guide fields, $B_G/B_0 = 0.2-0.4$.  For these guide fields, the compression fluctuates strongly in time but, on the whole, continues to grow up to $tc/L_y \approx 3$. At this time the compression is strongest for $B_G/B_0 = 0.4$, justifying our choice of this value of $B_G$ for our main fiducial simulations.
A similar trend was found by  \cite{Cerutti2014b}, where nonthermal particle acceleration was also maximized at these moderate guide-field strengths.

As for the compression of the plasma density, we see a similar non-monotonic trend.
\fig{BGscan}(b) shows a clear dependence on $B_G$ at the earliest time that the density compression diagnostic is available, e.g., at $tc/L_y \approx 1$, where it is largest for a guide field~$B_G/B_0 = 0.6$. Soon thereafter, however, the dependence becomes less clear,
with the differences between all the curves except the red one ($B_G/B_0 = 1.0$) being comparable to their fluctuation level. The density compression then rises somewhat for some of the simulations just before the end of the runs, and reaches a maximum at $tc/L_y \approx 3$, occurring at~$B_G/B_0 = 0.4$, similar to the magnetic compression. One can note that the red line ($B_G/B_0 = 1.0$) is consistently below all others starting from about $tc/L_y=1.2$; i.e., that a strong guide field does suppress compression of the plasma. This suppression eventually leads to a compression smaller than in the case with the strongest compression by almost a factor of~2. 

One should note that although the compression is suppressed in weak guide fields, the total radiated power increases as $B_G$ is lowered, see~\fig{BGscan}(c). 
The total energy $E_{\rm rad}$ radiated up to $t_{\rm max} = 3.16 L_y/c$, obtained by integrating the radiated power from ~\fig{BGscan}(c) up to this time, normalized to the initial energy contained in the reconnecting magnetic field~$E_{B0}$, is plotted as a function of~$B_G/B_0$ in the top right panel of \fig{BGscan}, just above panel~(c).
As one might expect, the radiated energy does not depend strongly on the guide field when the guide field is weak, $B_G \lesssim 0.2 B_0$.
For low $B_G/B_0$ approaching $0$, the energy radiated
approaches $E_{\rm rad} \approx 0.6$.
However, there is a clear power-law dependence for stronger guide fields, scaling inversely with~$B_G/B_0$ [as $\sim (B_G/B_0)^{-0.8}$].
This is because the reconnection
rate becomes smaller for higher~$B_G/B_0$, and thus there is less energy dissipation and hence less radiation.  
However, we find that the local average values of $\chi_e$ or $\gamma \chi_e$ increase with the magnetic compression. Therefore, the largest values occur in the simulation with maximal compression ($B_G/B_0= 0.4$).

\begin{figure*}
  \noindent\includegraphics[width=6.0in]{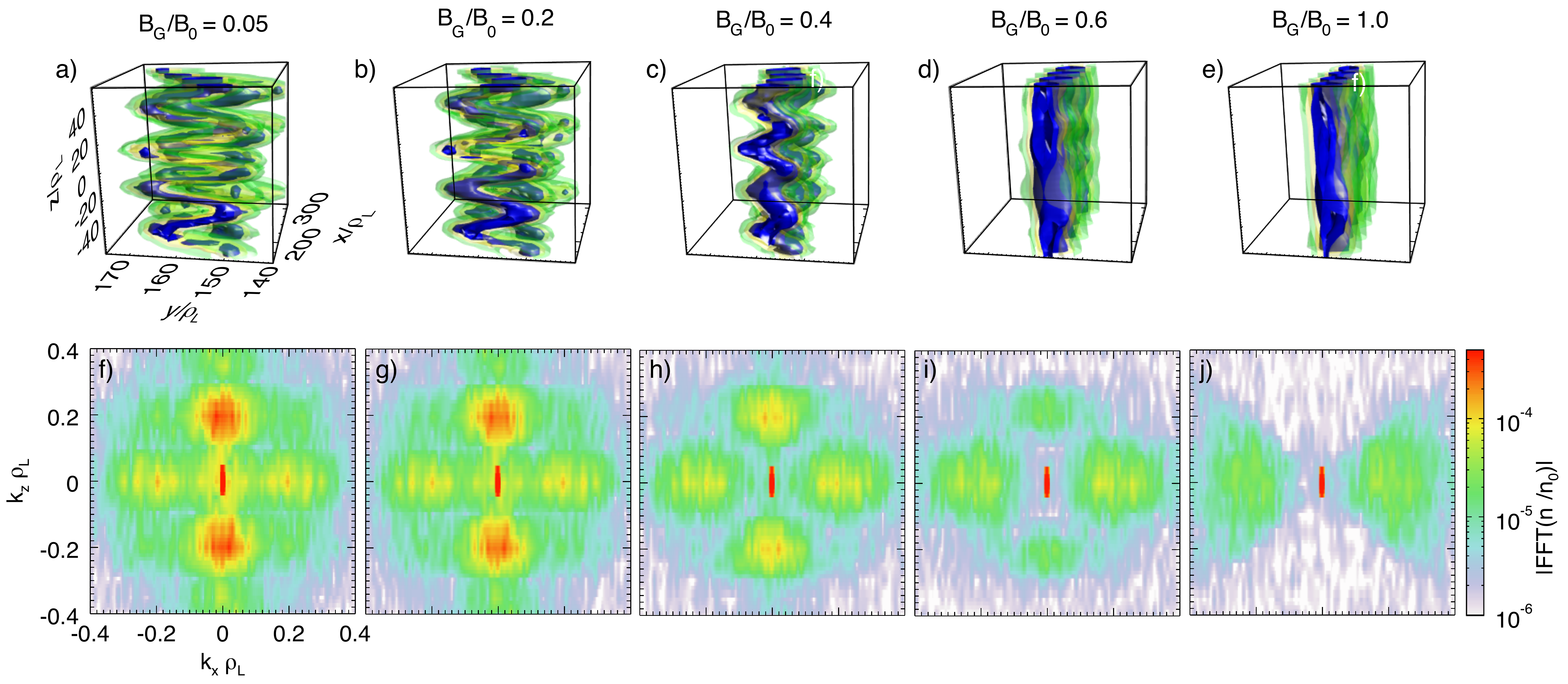}
  \caption{\label{kink}
	3D renderings of density contours (top row: panels a-e) at $tc/L_y = 0.566$ and the spatial Fourier decomposition of the density (bottom row: panels f-j) at $tc/L_y = 0.404$, for 3D radiative-case simulations with different guide-field strengths: from left to right, $B_G/B_0=0.05$, $0.2$, $0.4$, $0.6$, and~$1.0$. Bottom-row panels (f-j) are 2D cuts of the 3D FFT at $k_y\rho_L = 0.34$. 
	}
\end{figure*}

The physical origin of these numerically-observed trends, in particular, the suppression of the compression of plasma and magnetic fields in the weak guide-field regime in~3D, can be traced to the effect that the guide field has on the 3D instabilities developing in the current sheet.
The initial current sheet is unstable to two types of modes; the tearing mode developing primarily in the $\hat{x}$ direction and the RDKI mode primarily in the $\hat{z}$ direction.
To identify the dominant modes, we use fast Fourier transform (FFT) of the density maps from the PIC simulations of this $B_G$ parameter scan at a relatively early time $t c/L_y \approx 0.4$. 
We find the peaks of the Fourier power spectrum at $\bm{k}\rho_L \approx [0.2,0,0]$ ($\bm{k}\delta \approx [0.5,0,0]$) for the tearing mode and $\bm{k}\rho_L \approx [0,0.34,0.2]$ ($\bm{k}\delta = [0,0.87,0.5])$ for the kink mode.
Here the component of the wavenumber directed in the $\hat{y}$ direction just corresponds to the thickness of the unstable current sheet and is not associated with the direction of the unstable mode.

\fig{kink}(a-e) shows early-time, $t \approx 0.6 L_y/c$, 3D renderings of the plasma density contours, which exhibit kinking across a range of guide fields, with decreasing amplitude as the guide field increases. At a slightly earlier time $t c/L_y \approx 0.4$, the calculated FFT of the density in the $(k_x,k_z)$ space at fixed $k_y\rho_L=0.34$ in \fig{kink}(f-j) shows the presence of both tearing and kinking modes (although the tearing mode peaks at $k_y = 0$, it is still visible at $k_y\rho_L=0.34$). 
The tearing mode is slowed down but not fully suppressed by the guide field and is found for all the parameters that we have investigated. It is therefore important to understand the role of the kinking modes which, when significant (for weak guide fields), act to disrupt the compression of the plasma and magnetic fields observed in~2D.
We find that the kinking mode, which is only found in 3D where $k_z \ne 0$, disrupts and limits the compression seen in 2D simulations, and thus explains why the density compression $\max(n)/n_b$ depends strongly on $B_G/B_0$ at $tc/L_y \approx 1$, shown in~\fig{BGscan}(b).  
As one can see in \fig{kink}(f-j), the kink's amplitude is highest for the weakest~$B_G/B_0 = 0.05, 0.2$. 
As $B_G/B_0$ is increased, this mode is suppressed, growing slower and saturating earlier. For $B_G/B_0=1.0$, the kinking mode is completely suppressed for our fiducial value of~$L_z$ [no kinking mode is visible in the Fourier spectrum in \fig{kink}(j)].
While the dominant kinking mode at early stages is the RDKI mode with a fixed wavelength in the $\hat{z}$ direction, $\lambda = 2\pi k_z^{-1} \sim 30 \rho_L$, at later times the MHD kink instability of the flux ropes starts to dominate. The corresponding dominant MHD kink mode's wavelength grows with the guide field and, for our strongest guide-field case $B_G/B_0=1.0$, it can only fit in boxes with $L_z$ larger than simulated in the present parameter scan. %
We will explore the $L_z$-dependence of both of these 3D kinking modes in the next subsection. 

To sum up, the maximum compression in our 3D reconnection simulations occurs at intermediate values of the guide field, e.g., $B_G/B_0 \approx 0.4$, when the compression-disrupting kinking instabilities are somewhat suppressed by the guide field, but, at the same time, the guide-field's pressure $B_G^2/8\pi$ is not strong enough to prevent the compression outright. 

\subsection{Parameter scan: \texorpdfstring{$L_z$}{TEXT}}
\label{subsec-Lz}

As shown in the previous subsection, stronger guide fields suppress variations in the $\hat{z}$ direction, in particular those coming from the kinking modes, making 3D results more like~2D.  In general, 3D simulations become more accurate when the box size $2L_z$ in the guide-field direction (quantified by $L_z/\rho_L$ or~$L_z/L_y$) is increased, allowing for modes with longer wavelengths and more variations to fit in the $\hat{z}$ direction.  
An important question is: for a given guide field, how large does $L_z/\rho_L$ (or~$L_z/L_y$) need to be to capture the relevant 3D physics?  
From the previous subsection, we learned that, for moderate guide fields, at least the initially dominant $k_z\rho_L \approx 0.2$ RDKI mode of the initial Harris current sheet, with wavelength $\lambda = 2\pi/k_z \simeq 30\rho_L$, has to fit in the box of length~$2L_z$. 
Our main fiducial sequence of 3D runs in Section~\ref{sec-3D} adopted the $z$-length $L_z = 58.6 \rho_L = 0.19 L_y$.
This is sufficiently long to resolve 4 initial
wavelengths ($k_z 2 L_z/2\pi \approx 4$), and thus these initial RDKI kinking modes are well captured. 
In the present subsection, we justify our choice for~$L_z$, by comparing simulations with a range of lengths.

We performed a parameter scan of $L_z/\rho_L = 7.325$, $14.65$, $29.3$, $58.6$, $117.2$, and $175.84$, keeping $B_0/B_Q = 4.53 \times 10^{-3}$ (i.e., radiative case), $B_G/B_0=0.4$, and $\sigma_h = 25.76$ fixed.  For computational reasons, these simulations were done using a smaller system size~$L_x/\rho_L = L_y/\rho_L = 157.2$ (i.e., half of our fiducial system size); this allowed us to explore a broad range of aspect ratios from $L_z/L_y=0.047$ to~$1.125$. 
Unfortunately, at this smaller system size, there is not much space and time for significant compression.

\begin{figure}
  \noindent\includegraphics[width=0.45\textwidth]{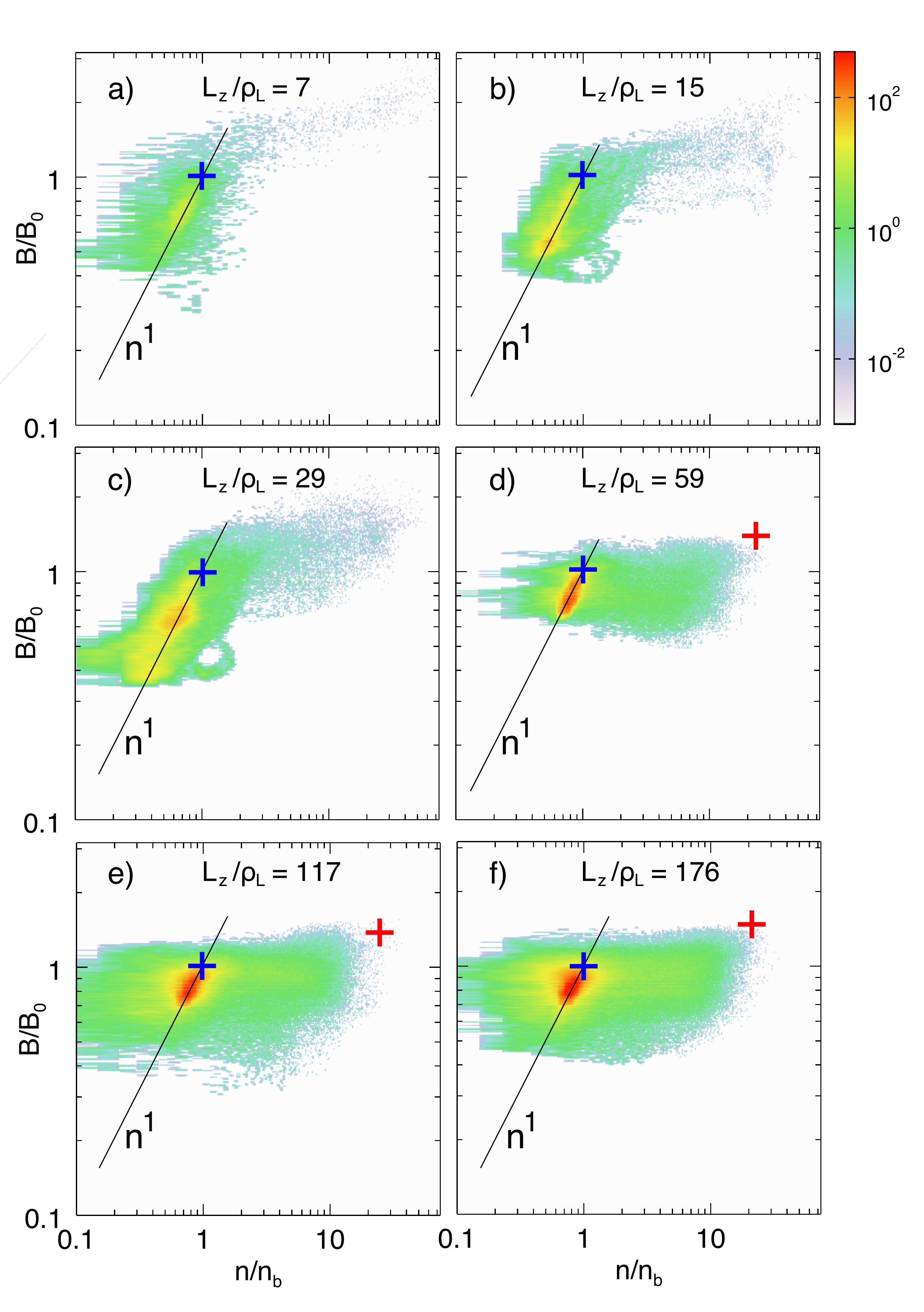}
  \caption{\label{histvsLz}
  	Histograms in $n\mbox{-}B$ space in terms of the local density~$n/n_b$ and magnetic field~$B/B_0$ at $t = 3 L_y/c$
    for the 3D radiative-case simulations with different~$L_z$: $L_z/\rho_L = (a)~7.3$, (b)~$14.7$, (c)~$29.3$, (d)~$58.6$, (e)~$117$, and (f)~$176$.  The blue plus signs represent the initial conditions of the upstream background, the red
	plus signs represent the upper right vertex of the best-fit polygon boundary (see Section~\ref{subsec-histogram}), which is plotted vs. time in \fig{Lzscan}. The $B\sim n$ scaling [\eq{frozenin}] is shown with thin solid lines.
	}
\end{figure}

We find that only the biggest-$L_z/\rho_L$ runs have $n\mbox{-}B$ histograms that resemble those for the 3D runs presented in Section~\ref{sec-3D} and allow for the use of the compression diagnostic from Section~\ref{subsec-histogram}.  
As we see in~\fig{histvsLz}(a-c), the histograms of the runs with the smallest values of~$L_z/\rho_L \in [7.3, 14.7, 29.3]$ resemble those from the 2D simulations [see~\fig{Histogram2D}(b)], where $B/B_0$ follows the frozen-in scaling of \eq{frozenin} for $n/n_b < 1$ and continues to increase with $n$ for $n/n_b > 1$, with a weaker slope.  
On the other hand, in larger-$L_z$ simulations ($L_z/\rho_L > 30$; $L_z/L_y > 0.19$), shown in in~\fig{histvsLz}(d-f), the histograms more strongly resemble those from the fiducial 3D simulations of Section~\ref{sec-3D} [see~\fig{Histogram3D}(b)]. 
In particular, while the bulk of the background plasma still follows the frozen-in scaling of \eq{frozenin}, there are clear and distinct power-law boundaries above and to the right of the distributions. This allows us to employ our standard measures of maximal density and magnetic-field compression (see Section~\ref{subsec-histogram})in terms of the intersection point of these histogram boundaries [red plus signs in \fig{histvsLz}(d-f)].

\begin{figure}
  \noindent\includegraphics[width=0.5\textwidth]{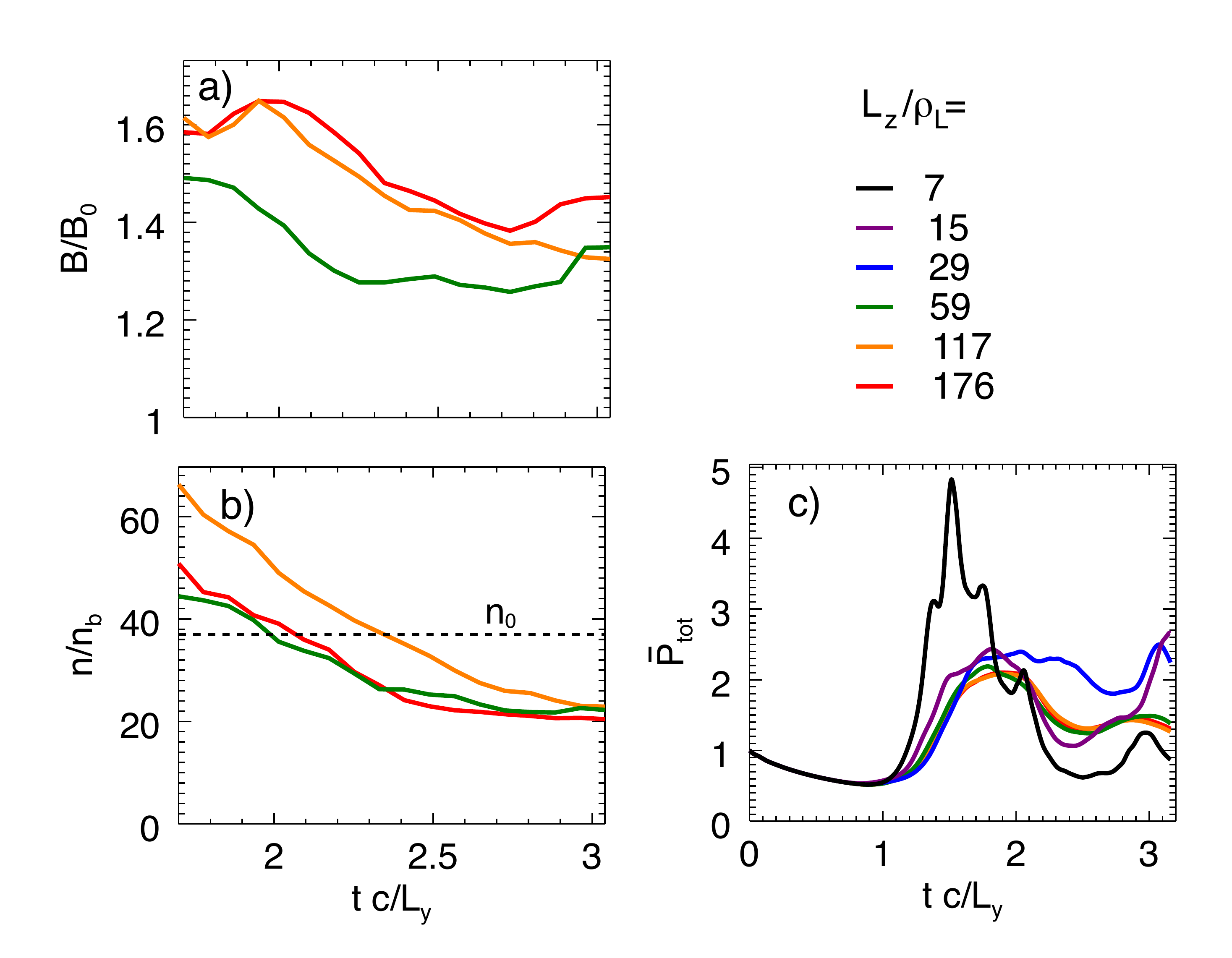}
  \caption{\label{Lzscan}
	Peak (a) magnetic field and (b) plasma density at the upper right vertex of the best-fit polygon boundary (see Section~\ref{subsec-histogram}), indicated with the red plus signs in \fig{histvsLz},
	along with (c) total normalized emitted power~$\bar{P}_{\rm tot}$,  
	as functions of time for the 3D radiative-case simulations with $L_x=L_y = 157.2\,\rho_L$ and with a range of system sizes in the $\hat{z}$ direction: $L_z/\rho_L = 7.3$ (black), $14.7$ (magenta), $29.3$ (blue), $58.6$ (green), $117$ (orange), and $176$ (red).
    The dotted line in panel~(b) is the initial density at the center of the Harris current layer.
	}
\end{figure}

As seen in~\fig{Lzscan}(a-b), the compression in both $n$ and $B$ (from the histogram diagnostic in Section~\ref{subsec-histogram}) is not as strong as that found in the simulations with
larger~$L_y$. However, we can conclude that while the dependence on~$L_z$ is not strong, the compression may be still slightly greater for larger~$L_z$. The normalized emitted power $\bar{P}_{\rm tot}$, shown in \fig{Lzscan}(c), is greater for the simulations with the smallest values of~$L_z$, which are essentially 2D runs, while for $L_z/\rho_L > 30$ the emitted power appears to become independent of~$L_z$.

\begin{figure*}
  \noindent\includegraphics[width=6.0in]{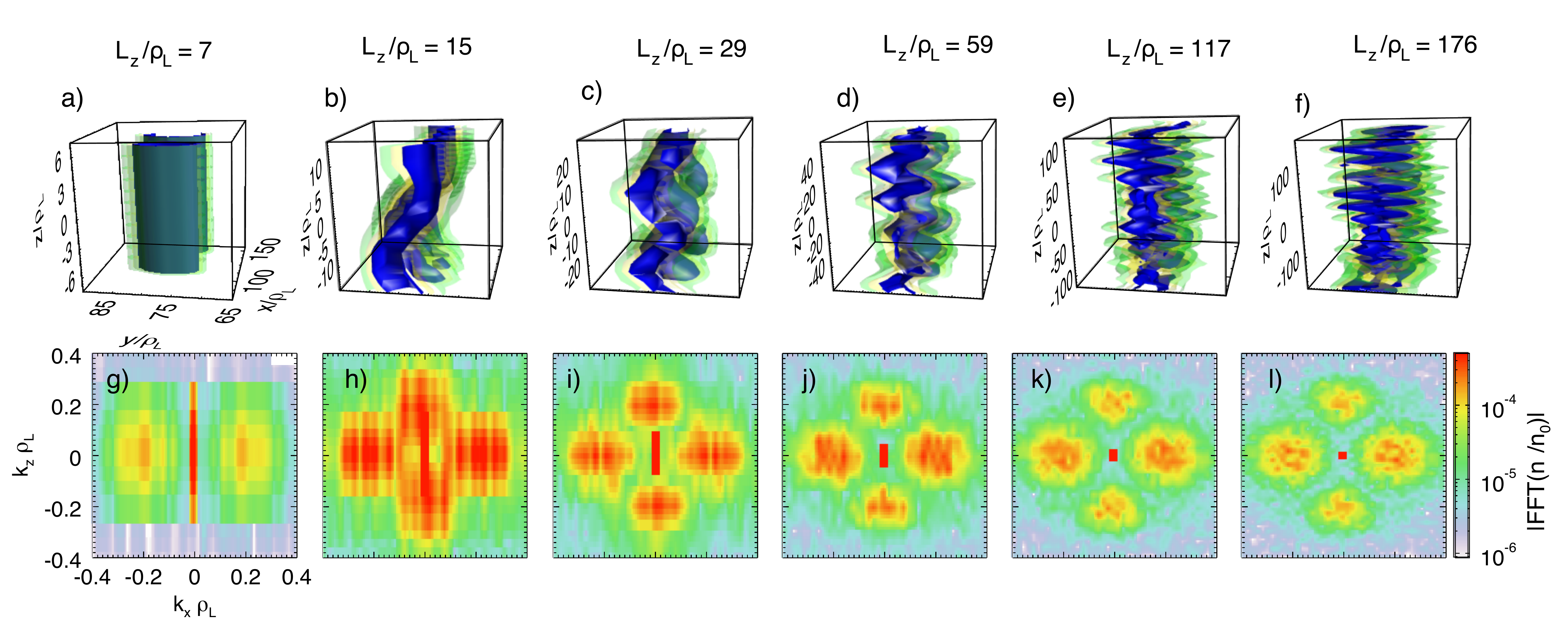}
  \caption{\label{kinkz}
	3D renderings of density contours (top row: panels a-f) at $tc/L_y = 0.566$ and the spatial Fourier decomposition of the density (bottom row: panels g-l) at $tc/L_y = 0.404$, for 3D radiative-case simulations with different~$L-z$: 
    from left to right, $L_z/\rho_L=7.3$, $14.7$, $29.3$, $58.6$, $117$, and~$176$. Bottom-row panels (g-l) are 2D cuts of the 3D FFT at $k_y\rho_L = 0.34$. 
	}
\end{figure*}

As in the previous Subsection~\ref{subsec-BG}, we show early-time 3D iso-contours of the density at $tc/L_y = 0.566$ in~\fig{kinkz}(a-f) and the FFT of the density in the $xz$ plane (at $k_y \rho_L = 0.34$) at $tc/L_y = 0.404$ in~\fig{kinkz}(g-l), now for several values of~$L_z/\rho_L$.  The initial kinking mode appears at the same value of $k_z\rho_L \simeq 0.2$ as in the previous subsection, independent of~$L_z$, for all the runs except for the quasi-2D case $L_z/\rho_L = 7.3$. The reason why there are no signs of the kinking mode in the $L_z/\rho_L = 7.3$ case is simply that the wavelength of this instability mode, $\lambda_z/\rho_L \sim 30$, is too long to fit in the box of a full $z$-extent of $2L_z\simeq 15\rho_L$. On the other hand, our fiducial choice of the box half-length $L_z= 58.6\rho_L$ captures almost 4 full wavelengths.
Besides this main RDKI kinking mode, we do not see any other significant variations along the $\hat{z}$ direction, e.g., any clear evidence of the MHD kink modes, at this relatively early time, even for the largest value of~$L_z/\rho_L= 176$ ($L_z/L_y = 1.12$, four times longer than our fiducial~$L_z$). Therefore, $L_z/\rho_L = 58.6$ ($L_z/L_y = 0.37$) appears to be sufficient to capture the 3D effects, at least at early times.
It is still not clear if 3D effects might become more important at higher values of $L_y/\rho_L$ or~$L_z/L_y$, which we did not simulate in this study. 
As we discussed in Section~\ref{subsubsec-nbound}, the peak density of the current filaments appears to be limited at late times to a certain region in $n\mbox{-}B$ space, governed by the marginal stability condition for the flux-rope kink mode in the filaments, and this may explain the lack of clear kinking modes.

We thus conclude that, although rough, order-of-magnitude predictions of compression and radiation are possible based on 2D simulations, an accurate prediction in 3D requires the domain's length in the third dimension to be large enough to capture at least the initial RDKI instability.

\subsection{Parameter scan: \texorpdfstring{$L_y$}{TEXT}}
\label{subsec-Ly}

Other parameters may lead to stronger compression, greater~$\chi_e$, and thus more powerful emission of gamma-rays, but are also more computationally difficult to study. 
In particular, in this subsection, we consider the dependence on the system size~$L_y/\rho_L$. Increasing the system size  leads to a longer time of evolution of plasmoids, and thus potentially to stronger compression of the magnetic field in plasmoid cores, and more pronounced nonthermal particle acceleration.

In order to see the effects of system size, we have performed, in addition to our fiducial $L_y/\rho_L = 314.4$ run, two more simulations, with both larger ($L_y/\rho_L = 471.6$) and smaller ($L_y/\rho_L = 157.2$) sizes, keeping $L_z/\rho_L=58.6$, $B_0/B_Q = 4.53 \times 10^{-3}$, $B_G/B_0=0.4$, and $\sigma_h = 25.76$ fixed. The simulation duration was $t_{\rm max}c/L_y = 3.16$ in all the runs.
\begin{figure}
  \noindent\includegraphics[width=0.5\textwidth]{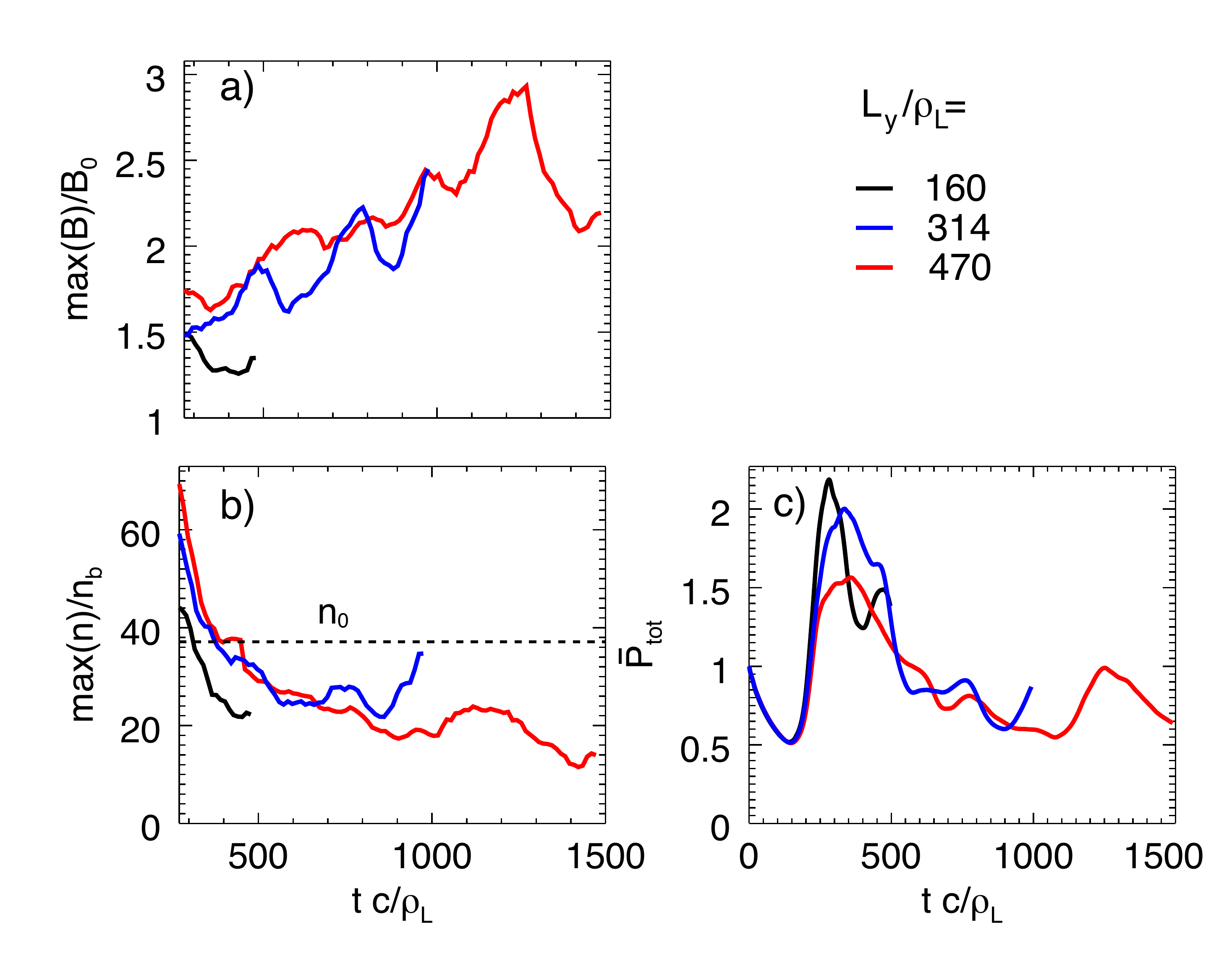}
  \caption{\label{Lyscan}
 	Peak (a) magnetic field and (b) plasma density at the upper right vertex of the best-fit polygon boundary (see Section~\ref{subsec-histogram}),
	and (c) total normalized emitted power~$\bar{P}_{\rm tot}$,  
	as functions of time for 3D radiative-case simulations with a range of system sizes: $L_y/\rho_L= 157.2$ (black), $314.4$ (blue), and $471.6$~(red).
    The dotted line in panel~(b) is the initial density at the center of the Harris current layer.
 }
\end{figure}

We find that the degree of compression (from the histogram diagnostic in Section~\ref{subsec-histogram}) scales with the time normalized to the (microscopic) cyclotron timescale $c/\rho_L = \gamma_T \Omega_c^{-1}$, rather than to the (macroscopic) system's light-crossing time~$c/L_y$.  Therefore, in~\fig{Lyscan}, we show the comparisons of the time histories of the magnetic field (panel~a) and density (panel~b) compression, as well as of the total normalized power (panel~c), using this microscopic time normalization, which indicates little dependence on the system size.  However, while the level of compression in~$n$ seems to reach an approximate steady state at late times that is weakly dependent on system size, the compression of $B$ appears to continue to grow roughly linearly with time.
Therefore, for simulations with a given duration in light-crossing times, magnetic compression can eventually reach larger values for larger system sizes. 
Note that for the smallest system size ($L_y/\rho_L = 157.2$) there is not enough time for significant compression, and hence the magnetic compression does not exhibit a clear linear trend like in the other two cases. 

Like the density compression, the total normalized emitted power $\bar{P}_{\rm tot}$ decreases as a function of time at intermediate times and approaches a steady state at late times [see \fig{Lyscan}(c)]. 
In general, more power is emitted at relatively early times, $tc/\rho_L \sim 300-500$.
However, for larger system sizes, there is more time for the
compressed magnetic field and increased temperature (not shown) to lead to more energy radiated at higher photon energies. 
Therefore, larger systems have greater potential for producing brighter gamma-ray emission and hence possibly more copious pair production.

\subsection{Parameter scan: \texorpdfstring{$\sigma_h$}{TEXT}}
\label{subsec-sigmah}

Although large-$\sigma_h$ simulations are computationally challenging, 
because it is numerically difficult to handle initial Harris equilibria with very large density contrasts~$n_0/n_b$
(a force-free initial equilibrium may be more amenable to simulation studies in this regime), these parameter regimes may be more relevant to astrophysical environments associated with gamma-ray flares.
We, therefore, look at the dependence of some of the key reconnection characteristics on the magnetization~$\sigma_h$.
As a reminder, this parameter quantifies the relative free energy in the upstream magnetic fields that can be converted by reconnection into plasma heating and the nonthermal acceleration of particles. For high~$\sigma_h$, the magnetic pressure dominates over the plasma pressure, potentially enabling stronger density compression. 
Therefore, as $\sigma_h$ is increased, we expect to find both greater heating and acceleration, and stronger compression of plasma density, leading to higher $\chi_e$ and thus brighter gamma-ray emission.

We have performed a parameter scan of $\sigma_h = 6.44, 12.88,$ and~$25.76$, keeping $T_b = 4 m_e c^2$, $B_0/B_Q = 4.53 \times 10^{-3}$, $B_G/B_0=0.4$, $L_z/\rho_L=58.6$, and $L_x/\rho_L =L_y/\rho_L = 314.4$ constant. 
The magnetization~$\sigma_h \sim 1/n_b$ is varied by changing~$n_b$ while keeping the other two basic background plasma parameters, $T_b$ and~$B_0$, constant. 
We also keep fixed most of the initial parameters of the current sheet, namely, its initial thickness $\delta = 2.55 \rho_L$ and temperature $T_0 = 6.92 m_e c^2$, as well as the drift velocity $v_d/c=0.56$ of current-carrying particles [see \eq{drift} in Appendix~\ref{sec-append-setup}]. This, in turn, implies that the difference in density $n_0-n_b$ is kept constant [\eq{pressurebalance1} in Appendix~\ref{sec-append-setup}] as $\sigma_h$ (and hence~$n_b$) is varied, and so $n_0$ is changed slightly. 
The respective density contrasts for increasing $\sigma_h$ are $n_0/n_b= 10, 19,$ and~$37$.

\begin{figure}
  \noindent\includegraphics[width=0.5\textwidth]{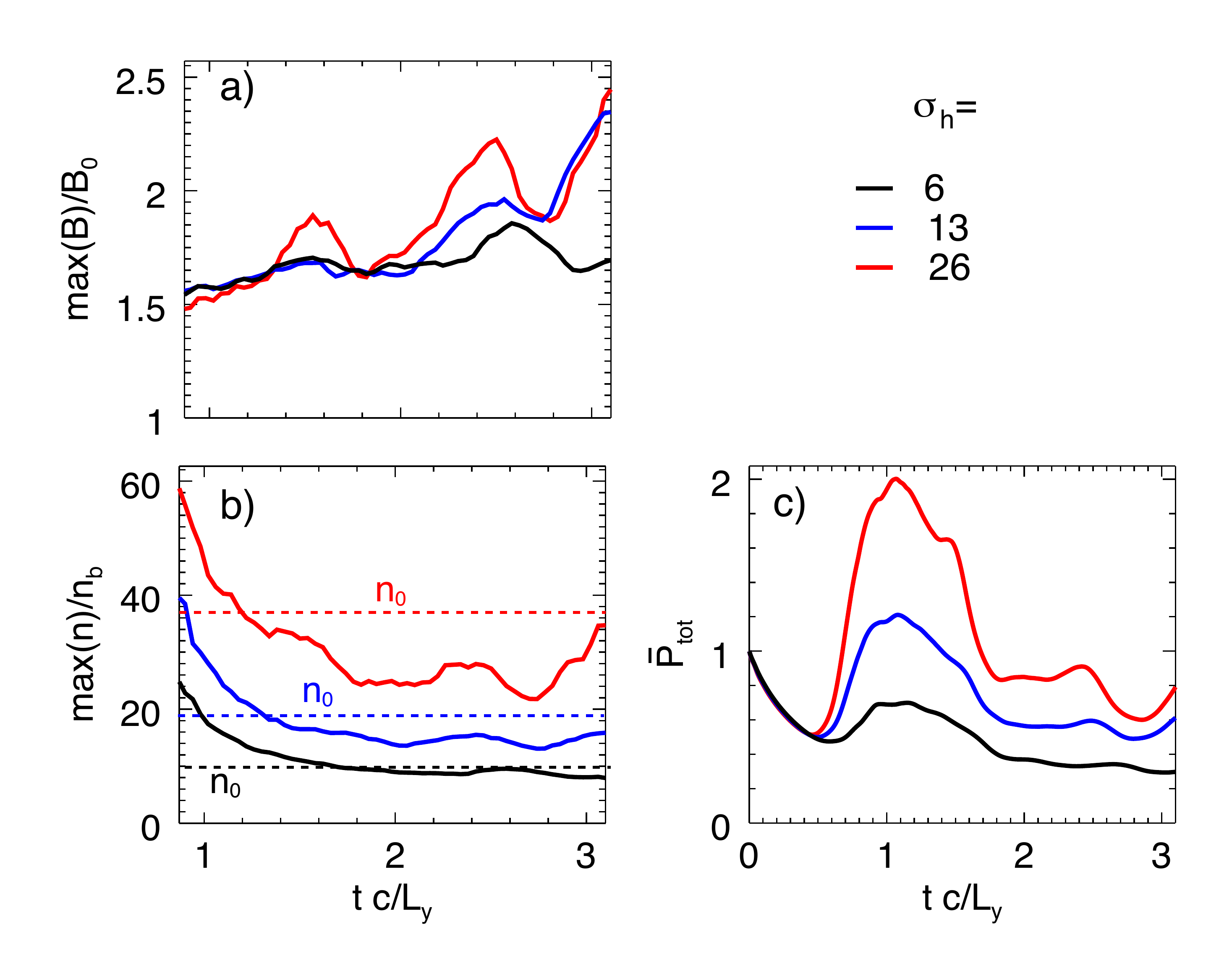}
  \caption{\label{sigmahscan}
 	Peak (a) magnetic field and (b) plasma density at the upper right vertex of the best-fit polygon boundary (see Section~\ref{subsec-histogram}),
	and (c) total normalized emitted power~$\bar{P}_{\rm tot}$,  
	as functions of time for 3D radiative-case simulations with a range of upstream magnetizations: $\sigma_h= 6.44$ (black), $12.88$ (blue), and $25.76$~(red).
    The dotted lines in panel~(b) are the initial densities at the center of the Harris current layer. 
    }
\end{figure}

\begin{figure*}
  \noindent\includegraphics[width=0.98\textwidth]{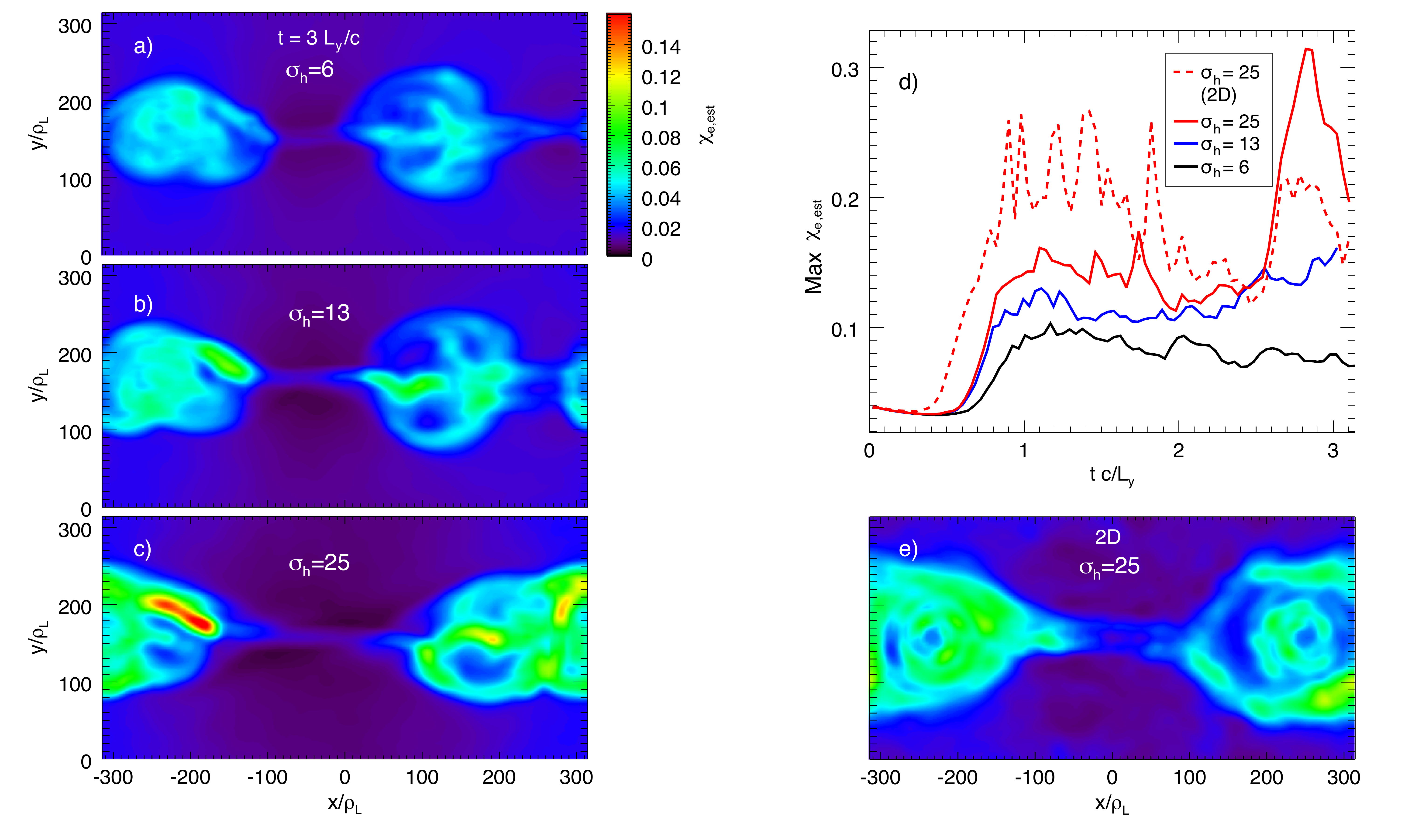}
  \caption{\label{chivssigma3D}
	2D maps of estimated local averaged $\chi_{e,{\rm est}} \equiv (2 T/m_e c^2)\, (B/B_Q)$ at $t = 3 L_y/c$ from a representative $z=0$ cut for 3D radiative-case simulations with (a) $\sigma_h = 6.44$, (b)~$12.88$, and (c)~$25.76$.
	Panel (d) shows the time evolution of the maximum $\chi_{e,{\rm est}}$ for each of the simulations.
	Panel (e) shows the $\chi_{e,{\rm est}}$-map from the 2D radiative case with $\sigma_h = 25.76$.
	}
\end{figure*}

Again, we examine the compression from the histogram diagnostic in Section~\ref{subsec-histogram}, after $t = L_y/c$, once the distribution has sufficiently mixed.
As shown in~\fig{sigmahscan}(a), the magnetic field compression for $\sigma_h\simeq 6.4$ stays nearly constant at a modest value of about~(1.6-1.8) throughout the simulation. However, for higher magnetization, $\sigma_h \simeq 13$ and~26, the magnetic compression exhibits an overall growth, up to $\sim 2.5$ at the end of the simulation, on top of strong fluctuations. Generally, higher $\sigma_h$ results in stronger magnetic compression.
The peak density compared to the initial background density~$n_b$, displayed in~\fig{sigmahscan}(b), is also higher for higher magnetization. However, this dependence does not actually reflect the degree of compression, simply because higher magnetization just corresponds to lower normalization~$n_b$ relative to the peak density $n_0$ in the initial Harris layer. Therefore, comparing the peak density of the compressed regions to $n_0$ would be a better measure.
Although a density compression has clearly occurred by the time the histogram diagnostic is first available [$\max(n)/n_0\sim 2$ at $t\approx L_y/c$], the peak density decreases with time and quickly drops below~$n_0$, especially for higher~$\sigma_h$. After that, during the second half of each simulation ($t \gtrsim 1.5\,L_y/c$), the peak density fluctuates below~$n_0$, around a saturated level that decreases, when normalized to~$n_0$, as $\sigma_h$ is increased, see \fig{sigmahscan}(b).
Since the late-time peak density remains below $n_0$ for all cases, any sustained compression can only be due to the background plasma, rather than the plasma from the initial current sheet.

Taking into account the magnetic field compression, we expect an overall increase in the values of~$\chi_e$ for typical particles in magnetic islands, which would be consistent with the clear enhancement of the normalized radiated power for increased values of $\sigma_h$ seen in~\fig{sigmahscan}(c).
Note that although there is a greater enhancement of normalized radiated power for higher~$\sigma_h$, the total energy radiated by the end of the simulations remains about $1/3$ of the initial energy in the reconnecting field for all~$\sigma_h$. The initial radiated power decreases with~$\sigma_h$, because there are fewer radiating particles in the background.

A very good estimation for the average $\chi_e$ of particles at a given location is given by $\chi_{e,{\rm est}} \equiv (2T/m_e c^2)\, (B/B_Q)$, shown in~\fig{chivssigma3D}. A comparison of the 2D maps of this quantity, corresponding to the slices of the simulation domain at $z=0$ at $t = 3\,L_y/c$, illustrates the expected increase in~$\chi_e$ with increased~$\sigma_h$, as seen in~\fig{chivssigma3D}(a-c).
In addition to these maps, the dependence of $\chi_{e,{\rm est}}$ on $\sigma_h$ is clearly visible in the time histories of the maximum (over the domain) values of $\chi_{e,{\rm est}}$, plotted in~\fig{chivssigma3D}(d).  
For the $\sigma_h = 25.76$ case (the radiative case in Section~\ref{sec-3D}), the maximum local average $\chi_{e,{\rm est}}$ grows rapidly before plateauing at about~$0.15$ (i.e., somewhat below the 0.2-0.25 range of variation found in the corresponding 2D simulation). It then fluctuates around this level throughout most of the active reconnection phase, before spiking suddenly near the end of the simulation to a value as high as~$0.3$, almost a factor of $10$ higher than the initial background value of~$0.038$ [see~\fig{chivssigma3D}(d)].  
Although the total radiated power in 3D is not a high as in~2D [the radiative case in Section~\ref{sec-2D}], the high values that $\chi_{e,{\rm est}}$ reaches in 3D at late times [see~\fig{chivssigma3D}(c)] exceed those seen in the 2D case with the same $\sigma_h \approx 26$ [see~\fig{chivssigma3D}(e) and the red dashed line in~\fig{chivssigma3D}(d)].

We have thus shown that the initial background magnetization $\sigma_h$ has a significant
effect on the magnetic field and density compression, the total emitted power, and the average $\chi_e$ parameter in the cores of plasmoids, all pointing to regimes where gamma-ray emission can be more efficient.

\section{Conclusions}
\label{sec-concl}

We have presented the results of a comparative 2D and 3D numerical study of collisionless relativistic reconnection of strong magnetic fields in an electron-positron pair plasma, self-consistently taking into account synchrotron radiation reaction. 
The main focus of our study was on investigating reconnection-powered sudden bursts of enhancement of the estimated local emissivity $\epsilon_{\rm est}$ and the total radiated power $P_{\rm tot}$, especially in the gamma-ray regime.
Our radiative-PIC simulations were conducted with the OSIRIS radiative-PIC code and were initialized with a self-consistent equilibrium relativistic, dense Harris sheet immersed in a lower-density ambient background pair plasma.  
We have investigated the effects of the relative strength of synchrotron cooling, controlled by the reconnecting magnetic field~$B_0/B_Q$, in both 2D (Section~\ref{sec-2D}) and 3D (Section~\ref{sec-3D}).
In addition, in~3D, we have performed an extensive study of the effects of several other key physical parameters (Section~\ref{sec-param}), namely, the relative strength of the non-reconnecting, guide magnetic field along the $\hat{z}$ direction~$B_G/B_0$ (Section~\ref{subsec-BG}), the length of the current sheet in the $\hat{z}$ direction~$L_z/\rho_L$ (Section~\ref{subsec-Lz}), which characterizes 3D effects, the system size in the perpendicular direction $L_y/\rho_L$ (Section~\ref{subsec-Ly}), and the upstream plasma magnetization~$\sigma_h$ (Section~\ref{subsec-sigmah}).

We have developed two novel diagnostic estimates of the total radiated power, which also help elucidate what causes the bursts in photon emission found in both 2D and 3D simulations. 
These diagnostics are based on fluid-level quantities obtained as  reductions of the PIC-simulation kinetic data.
The first one is a simple estimate of radiated power~$P_{\rm tot,est}$ [\eq{Pest}], which integrates the estimated local emissivity~$\epsilon_{\rm est}$, ignoring both bulk-flows and kinetic effects/non-Maxwellian distributions; it eventually overestimates the emission of radiation (for more radiative cases), but takes into account the important correlation between the magnetic energy density and plasma energy density. 
The second estimate, $P_{\rm tot,est2}$ [see~\eq{Pest2}], is even simpler, as it ignores this correlation, and thus underestimates the emission.
We have found that reconnection naturally leads to the development of inhomogeneities in the system, e.g., via plasma and magnetic-field compression inside plasmoids (magnetic islands; or flux ropes in~3D). The resulting increased values of the plasma density, magnetic field, and temperature are, in general, spatially correlated and found concentrated in plasmoid cores. This enhancement and correlation, which is taken into account by~$P_{\rm tot,est}$, increases the local emissivity and the total radiated power in comparison with the more naive estimate~$P_{\rm tot,est2}$, especially in~2D.
Although, as we had pointed out in our previous paper~\citep{Schoeffler2019}, radiative cooling can drive an even stronger compression and hence a further concentration of the radiative regions in the central cores of magnetic islands, making them more effective radiators, in the present study we have found that enhanced radiative cooling of the plasma caused by stronger magnetic fields actually reduces the appropriately normalized radiated power.

We have found that, for the most part, 2D simulations yield reasonable qualitative estimates (i.e., within factors $\sim 2$) for several key characteristics, such as the radiated power, particle spectra, magnetic field compression, etc., for the full 3D system with a moderate guide field $B_G/B_0 = 0.4$. However, for some other quantities, we have observed rather large differences between 2D and 3D results. For example, the localized compression of plasma density and the enhancement of local emissivity in 2D can reach peak values around a factor of $10$ greater than in~3D.
Such unphysically strong plasma compression does not occur in 3D because compression becomes disrupted by kinking instabilities, in particular, the~RDKI at early times, capturing which requires accessing modes with $k_z \delta \sim 0.5$.
 
To study the development and limits of compression, and to help highlight the correlations between density~$n$, magnetic field~$B$ (and sometimes temperature~$T$) that drive the enhancements of reconnection-powered radiation, we have designed and made use of a novel compression diagnostic based on 2D histograms in $n\mbox{-}B$ and $n\mbox{-}T$ spaces of our simulations (Section~\ref{subsec-histogram}). 
We have observed that the distributions of points on these histograms, especially in 3D simulations, tend to develop very clear, well-defined borders, described by power laws, corresponding to sharp limits on the compression. 
 
We have presented tentative theoretical explanations for two such compression limits seen on the histograms in our simulations: 
for the maximum $B/B_0 \sim (n/n_b)^{1/6}$ [\eq{Blimit}] (moderate-density part of the top boundary) for 2D and 3D radiative cases, which we attribute to radiative-resistive dissipation of magnetic fields in secondary plasmoid cores; and (in~3D only) for the maximum density, given by $n/n_b \sim (B/B_0)^1$ [\eq{nlimit}] (right boundary), which appears to be determined by the kink instability condition.

We have further explored the compression and the resulting power radiated using parameter-space scans employing 3D radiative simulations that are subject to both of these compression limits. We have found that the relative enhancement of the radiated power decreases with increased $B_G/B_0$ or~$B_0/B_Q$ but increases significantly with increased magnetization~$\sigma_h$. 
The density compression (compared to~$n_0$) does not change significantly for all parameters studied, while the compression of the magnetic fields, on the other hand, increases with $B_0/B_Q$, $L_y/\rho_L$, and~$\sigma_h$.

In light of the results of our study, we expect that magnetic reconnection in strongly-radiative,  strong-field astrophysical environments is capable of producing bright flares of gamma-rays and X-rays. This study should help in understanding to what degree these bursts of radiation can explain observations of gamma-ray and X-ray flares, e.g., from the magnetospheres of neutron stars, including magnetars.

In addition, this study paves the way for future numerical 3D investigations of even more extreme astrophysically relevant regimes, characterized by stronger magnetic fields~$B_0/B_Q$ and higher upstream magnetizations~$\sigma_h$. This will enable us to reach the $\chi_e \sim 1$ regime, where QED effects, including pair creation, can become dominant. Furthermore, extending this study to larger normalized system sizes~$L_y/\rho_L$, in combination with the strong magnetic field, will allow us to reach higher, more realistic values of magnetic compactness~$\ell_B$, a key parameter governing the importance of radiative and QED effects. 

\section*{Acknowledgements}
This work is supported by the European Research Council (ERC-2015-AdG Grant 695088), FCT (Portugal) Grants  SFRH/IF/01780/2013.  This work was also supported by NASA grants NNX17AK57G, 80NSSC20K0545, 80NSSC22K0828, and NSF grants AST-1411879 and AST-1806084.
DAU also thanks Greg Werner for fruitful discussions.
We acknowledge PRACE for awarding access to resource SuperMUC based in Germany at Leibniz research center.
Simulations were performed at the Accelerates cluster (Lisbon, Portugal), and SuperMUC (Germany).

\appendix
	
\section{Simulation Setup}
\label{sec-append-setup}

In our 2D (3D) simulations, we model a $2L_x \times 2L_y (\times 2L_z)$ domain with two oppositely
	directed thin current sheets located at $y = \pm L_y/2$.  The current
	is directed out of the $(x,y)$ simulation plane in the respective $\pm \hat{z}$ directions, which leads to an asymptotic magnetic field ${\bm B} =
	B_0 \hat{x}$, between $-L_y/2 > y > L_y/2$, and ${\bm B} = -B_0 \hat{x}$ on
	the outside of the two current sheets.  An initially uniform background
	Maxwell-J\"uttner population of relativistic electrons and positrons,
	each with density $n=n_b$ at temperature $T=T_b$, is included to
	represent the ambient (upstream) plasma.  This population is initially
	stationary and does not contribute to the current.  Furthermore, we
	include a uniform guide magnetic field $B_G$ along the
	$\hat{z}$ direction. 

The current and self-consistent magnetic field profiles are in pressure balance in a kinetic equilibrium, known as the relativistic Harris
	sheet~\citep{Harris1962,KirkHarris}.  
The current is carried by counter-drifting Maxwell-J\"uttner distributions of positrons and electrons with a uniform temperature~$T_0$, boosted into opposite $\pm \hat{z}$-directions with a uniform velocity~$v_d$.  
The lab-frame density profile (of both electrons and positrons) in the Harris current sheet at $y = \pm L_y/2$ is:
\begin{equation}
	n=\left(n_0 - n_b\right) {\rm sech}^2 \left(\frac{y \mp L_y/2}{\delta}\right),
\end{equation}
where $n_0$ is the total electron (or positron) density at the center of each current sheet.
The self-consistent initial reconnecting magnetic field is:
\begin{eqnarray}
	B_x &=&B_0\left[
	1 -\tanh\left(\frac{y - L_y/2}{\delta}\right)
	+\tanh\left(\frac{y + L_y/2}{\delta}\right)\right.\nonumber\\
	&+&\left.\tanh\left(\frac{y - 3L_y/2}{\delta}\right)
-\tanh\left(\frac{y + 3L_y/2}{\delta}\right)\right].
\end{eqnarray}
We conduct our simulations with periodic boundary conditions, so we also include the self-consistent magnetic field due to two more current sheets at $y = 3L_y/2$ and $y = -3L_y/2$ (outside of the simulation box).  This is a small correction due to the periodic boundary conditions introduced to account for the exponential tail that passes through the boundary.  
 
This current-sheet setup is unstable to the tearing instability, which grows naturally from the particle noise without externally imposed seed fields.
In order to facilitate the onset of magnetic reconnection, the initial thickness of the current sheet $\delta$ is chosen to be sufficiently small (of order the gyro-radius of the particles in the sheet), so that the tearing instability growth rate approaches the characteristic cyclotron period~\citep{DaughtonPOP}.  
We normalize all the length scales in our simulations to $\rho_L \equiv \gamma_T m_e c^2/eB_0 = \gamma_T c/\Omega_c$, defined as the Larmor radius of a background particle with a Lorentz factor corresponding to the peak of the initial upstream relativistic Maxwell-J\"uttner distribution, $\gamma_T \equiv 2T_b/m_e c^2$, and choose $\delta > \rho_L$, $\rho_{L0}$, where $\rho_{L0} = \rho_L T_0/T_b$ is the gyroradius of a typical particle in the current sheet.

The three main physical parameters that describe the upstream plasma conditions outside of the current sheets---$T_b$, $n_b$, and $B_0$---define two important dimensionless parameters: the magnetization $\sigma_h$ and the plasma-$\beta$ parameter, $\beta_{\rm up}$ (the ratio of the background plasma pressure to the magnetic pressure):
\begin{eqnarray}
	\label{eq-sigma_h}
    \sigma_h &\equiv &\frac{B_0^2}{4 \pi (2n_b)h_b}, \\
	\label{eq-beta_up}
    \beta_{\rm up} & \equiv & \frac{8 \pi (2n_b) T_b}{B_0^2} = \frac{2 T_b}{h_b} \frac{1}{\sigma_h}.  
\end{eqnarray}

The subscript $h$ refers to the ``hot" magnetization $\sigma_h$, defined with the upstream background relativistic enthalpy per particle~$h_b$~\citep{Melzani}. 
In the nonrelativistic limit ($T_b \ll m_e c^2$), the enthalpy $h_b \approx m_e c^2 +  5/2 T_{b}$ is dominated by the rest-mass $m_e c^2$  and so the ``hot" magnetization $\sigma_h$ approaches the so-called ``cold" magnetization 
\begin{equation}
	\label{sigma_c}
    \sigma_c \equiv \frac{B_0^2}{4 \pi (2n_b)\, m_e c^2}\, , 
\end{equation}
which is often used in the literature.  In the ultrarelativistic limit ($T_b \gg m_e c^2$), however, $h_b \approx 4T_{b}$, and then $\sigma_h = 1/(2\beta_{\rm up})$.

Using the $\beta_{\rm up}$ parameter allows us to cast the electron and positron drift speed inside the two Harris current layers, determined by Amp\`ere's law, in a convenient form as
\begin{equation}
	\label{amperes}
	\frac{v_d}{c} = \frac{1}{\beta_{\rm up}}\frac{\rho_L}{\delta}\frac{n_b}{n_0-n_b}. 
\end{equation}
In addition, the temperature $T_0$ of the drifting plasma in the layer, determined by the cross-layer pressure balance, can be written as 
\begin{equation}
	\label{pressurebalance1}
	\frac{T_0}{m_e c^2} = \frac{T_b}{m_e c^2}\frac{\gamma_d}{\beta_{\rm up}}\frac{n_b}{n_0- n_b}, 
\end{equation}
where $\gamma_d \equiv 1/\sqrt{1- v_d^2/c^2}$.
We can thus derive a convenient expression for the proper drift velocity $u_d = \gamma_d v_d/c$,
\begin{equation}
	\label{drift}
	u_d = \frac{\rho_L}{\delta}\frac{T_0}{T_b} 
    = \frac{\rho_{L0}}{\delta}\, .
\end{equation}
This shows that for constant values of $\delta/\rho_L$ and~$T_0/T_b$, the drift $u_d$ is also constant.

\section{Boundary from Kinking}
\label{sec-append-nbound}
As we pointed out in the histogram in~\fig{Histogram3D}, there exists a boundary in $n\mbox{-}B$ space in the 3D case, that corresponds to a maximum value of~$n$, or equivalently a minimum value of~$B$, following the
scaling
\be
\label{nlimit}
B_{\rm min} \propto n.
\ee
In this appendix, we will sketch a heuristic argument aimed at understanding the origin of this scaling and will discuss how the location of the limit is likely determined by unstable kinking modes that occur for regimes with large densities.

As we discussed in Section~\ref{subsec-histogram}, the upstream background plasma is frozen into the magnetic field and follows~\eq{frozenin}. In the inner non-ideal (diffusion) regions near X-points, where magnetic reconnection takes place, the frozen-in condition is broken, and the plasma density may move to new regions of $n\mbox{-}B$ space.  
Once the plasma and the associated reconnected magnetic flux escape from the X-point regions and join nearby circularized magnetic islands (magnetic flux ropes in~3D), the frozen-in condition holds again. In the outer regions of these flux ropes, assuming that the background guide magnetic field is weak, the magnetic field strength is dominated by the in-plane, reconnected component~$B_{xy}$, and the ideal-MHD evolution of a given fluid element in the $n\mbox{-}B$ space follows equation~(\ref{frozenin2}) (see Section~\ref{subsec-space}). Then, however, as the flux rope grows further on the outside by accumulating more and more reconnected flux and by merging with other flux ropes, the given fluid element gets buried deeper and deeper inside the flux rope and experiences compression. The magnetic flux also compresses, but this compression has a greater effect on the out-of-plane (guide) magnetic field component~$B_z$, which thus eventually comes to dominate over $B_{xy}$ deep inside the flux rope's core.
As long as ideal MHD holds in this region during this compression process (i.e., the rope's core radius is much greater than the typical particle gyro-radius, and the radiative resistive effects, discussed in the next two appendices, can be neglected), and before any 3D instabilities, such as the kink, get excited and cause mixing of plasma, the joint evolution of the plasma density and the guide magnetic field $B_z$ (which dominates in these regions) follows the scaling~\eq{nlimit}.
Since the central cores of these flux ropes/ current filaments are also the highest-density regions, their behavior determines the slope of the high-$n$ compression boundary. 

Up to this point, the above discussion was applicable to both 2D and 3D cases. However, in 2D, the density and magnetic field remain relatively constant at the centers of plasmoids (i.e., regions where $n/n_b > 10$). These regions, therefore, do not occupy much area in $n\mbox{-}B$ space, as was shown in~\fig{Histogram2D}, and thus do not result in a clear power-law high-$n$ boundary. In contrast, in~3D, these quantities evolve and fill the $n\mbox{-}B$ space, in part because of the freedom of motion of plasma along the third dimension. 

We believe that the slope and the location of this boundary in 3D are governed by the marginal stability condition of the compressed current-filament (flux-rope) cores to the kink instability. Approximating these flux ropes as simple cylindrical pinches, we can invoke the well-known Grad-Shafranov (GS) criterion for the instability onset, cast in terms of the safety factor~$q$:
\be
\frac{1}{q(r)} \equiv \frac{B_{xy}}{B_z} \, \frac{\ell_z}{r} > 1.
\label{eq-GS-1}
\ee
Here, $r$ is the cylindrical radius inside the flux rope's core, $\ell_z$ is its length in the $z$-direction, $B_{xy}(r)$ is the in-plane magnetic field, and $B_z$ is the out-of-plane (guide) magnetic field inside the flux rope.

In the following, we shall assume that inside each flux-rope core, the out-of-plane magnetic field $B_z$ and the current density $j_z$ are approximately uniform in~$r$.  
The in-plane magnetic field as a function of radius $r$ inside a given core can then be estimated using Ampere's law as 
\be
B_{xy}(r) \simeq  
\frac{2 \pi j_z}{c} \, r \, , 
\label{eq-B_xy}
\ee
i.e., increases linearly with the radius inside the flux rope. This is important because, once this expression for the in-plane magnetic field is plugged into the GS instability condition~(\ref{eq-GS-1}), the radius $r$ cancels, and the condition becomes
\be
\frac{1}{q} = 2\pi \ell_z \frac{j_z} {c}\, \frac{1}{B_z} > 1. 
\label{eq-GS-2}
\ee

Next, it is reasonable to assume that the guide magnetic field $B_z$ dominates over (or is at least comparable to) $B_{xy}$ inside plasmoid cores, and thus provides a good estimate for the total magnetic field strength $B$ there. 
Furthermore, we shall assume that counter-streaming (in $z$) electrons and positrons contribute equally to the current density in the $z$-direction, so that $j_z = 2 e  n v_d$, where $n$ is the density of the electrons (or positrons, which we will assume is equal) in the plasmoid core and $v_d = \beta_d c$ is the absolute value of the drift $z$-velocity of the current-carrying particles. 
For simplicity, we shall view both $n$ and $v_d$ as being uniform inside a given plasmoid core. We can then recast the marginal kink stability condition $q=1$ for a given flux-rope core in terms of a linear relationship between the magnetic field $B\approx B_z$ and density~$n$ inside of it as
\be
\frac{B}{n} \simeq 4\pi e\, \ell_z \, \beta_d \, .
\label{eq-kink_boundary-1}
\ee

We can go one step further and express this relationship in terms of the dimensionless, normalized density and magnetic field, $n/n_b$ and~$B/B_0$, that form the axes of our 2D $n\mbox{-}B$ histograms. We then get 
\be
\frac{B/B_0}{n/n_b} \simeq 
\frac{1}{2 \sigma_c}\, \frac{\ell_z}{\rho_0}\, \beta_d, 
\label{eq-kink_boundary-2}
\ee
where $\sigma_c \equiv B_0^2/(8\pi n_b m_e c^2)$ is the initial upstream ``cold" magnetization [corresponding to the total, electron plus positron, particle density~$2n_b$, see equation~(\ref{sigma_c})] and $\rho_0 \equiv m_e c^2/eB_0$ is the nominal relativistic Larmor radius. The cold magnetization provides the basic scale for the available upstream magnetic energy per particle, and then the combination $\rho_c \equiv \sigma_c \rho_0$ gives the corresponding characteristic Larmor radius of reconnection-energized particles. For a relativistically hot upstream plasma, 
$\sigma_c = 4 \sigma_h (T_b/m_e c^2) = 2 \gamma_T \sigma_h$, and hence 
\be
\rho_c \equiv  \sigma_c \rho_0 = 2 \sigma_h \rho_L \, .
\label{eq-rho_c}
\ee

Thus, the kink-based density boundary can be written as
\be
\frac{B/B_0}{n/n_b} \simeq 
\frac{\ell_z}{2\rho_c}\, \beta_d  =
\frac{1}{4\sigma_h}\, \frac{\ell_z}{\rho_L}\, \beta_d. 
\label{eq-kink_boundary-3}
\ee
Note that $\sigma_h$ and $\rho_L$ appearing in the expression on the right-hand side are just fixed parameters, defined in terms of the initial upstream plasma conditions; they are, therefore,  constant, by definition, within a given simulation. Thus, in order to see whether the $B\sim n$ scaling~(\ref{nlimit}) for the high-density histogram boundary holds, one just needs to examine $\beta_d$ and~$\ell_z$.

Empirically, in our simulations we see that different plasmoid cores reach roughly similar typical peak values of $\beta_d \simeq 0.2$, with relatively little variation.  

As for estimating the relevant values of~$\ell_z$, one can consider two arguments. First, the upper limit on $\ell_z$ is given by the $z$-extent of the computational box: $\ell_z = 2 L_z$. Then, all the quantities on the right-hand side of equation~(\ref{eq-kink_boundary-3}) have fixed (i.e., the same for all flux-rope cores) values for a given simulation, and we thus recover the high-$n$ histogram-boundary scaling(\ref{nlimit}), i.e., $B\sim n$. 
Quantitatively, for our fiducial simulations with $\sigma_h = 25.76$ and $2L_z = 117\rho_L$, we obtain 
\be
\frac{B/B_0}{n/n_b} \simeq 
 \beta_d \simeq 0.2\, , 
\label{eq-kink_boundary-4}
\ee
which agrees reasonably well with the location of this boundary for both radiative and nonradiative cases as can be seen in~\fig{Histogram3D}(a,b).

Alternatively, one can argue that kink modes that are particularly effective in disrupting the compression of a flux rope and causing efficient plasma mixing, are those with their $z$-wavelength, $\ell_z$, comparable to, but perhaps somewhat longer (but not much longer) than, the flux-rope core's diameter~$2r$. The GS condition equation~(\ref{eq-GS-1}) is then roughly equivalent to $B_{xy} \sim B_z/2$ \cite[c.f.,][]{Pritchett2004}. 
We observe that in our 3D simulations, compressed flux-rope cores have characteristic radii%
\footnote{See, e.g., Fig.~\ref{3Dmaps}; note that Fig.~\ref{3Dmaps} shows an earlier time, $t c/ L_y=1.5$, whereas the histogram in Fig.~\ref{Histogram3D} is at a later time, $t c/L_y = 3$. While the flux ropes do become larger over time, the relevant cores remain about the same size.} 
of $r\sim 20 \rho_L$, 
and hence further compression is disrupted by kink modes with $\ell_z \sim 40\rho_L$.  Substituting this estimate into equation~(\ref{eq-kink_boundary-3}), we get, for our fiducial $\sigma_h= 25.76$ case, 
\be
\frac{B/B_0}{n/n_b} 
\simeq \frac{10}{\sigma_h}\, \beta_d  
\simeq \beta_d/2.5 \simeq 0.08\, . 
\label{eq-kink_boundary-5}
\ee
This provides an excellent fit for the boundaries in~\fig{Histogram3D}(a,b) and in~\fig{wHistogram}(a).

After the kink instability gets excited and mixes high- and low-density regions, we expect the density to be limited to the stable regions in $n\mbox{-}B$ space, such that the boundary occurs at marginal stability given by the above condition, as we discussed in Section~\ref{subsec-hist3D}.

\section{Effective Radiative Resistivity Derivation}
\label{sec-append-raddiss}

In this appendix we derive an expression for the effective radiative resistivity~$\eta_{\rm eff}$, which acts in a manner similar to the standard collisional Spitzer resistivity
\citep{Spitzer1953,KrallandTrivelpiece} in the magnetic induction equation.
This radiative dissipation is caused by the synchrotron radiation reaction instead of binary particle collisions. 

Just like with the collisional resistivity, the radiative resistivity can be formulated for an arbitrary orientation of the electric field relative to the magnetic field. In the case when the two fields are not strictly aligned (or anti-aligned), the perpendicular component of the electric field drives an $\bm E\times \bm B$ drift of the magnetic field lines; in the case of a resistively decaying magnetic flux rope with an azimuthal magnetic field and an axial electric field, this drift is directed inwards, towards the rope's center.
However, if the electric force on the electron (or positron) fluid is balanced by the net radiation-reaction force~${\bm F}_{\rm rad}$, then the $\bm E\times \bm B$ drift of the particles is canceled by the oppositely directed drift due to the radiative friction force. The resulting resistive slippage of the plasma particles relative to the inward-drifting magnetic field lines allows the plasma in a flux-rope core to remain approximately static while the azimuthal magnetic flux moves inward and eventually gets destroyed at the flux rope's O-point. 
	
In order to obtain a simple estimation for the synchrotron radiative resistivity quantifying this resistive slippage, we will make a few assumptions. First, we shall assume that the electron and positron populations move in opposite directions in response to a super-imposed electric field~${\bm E} = E\hat{x}$, contributing equally to the resulting electric current (i.e., the net $e^+e^-$ flow is zero). We shall also assume for simplicity that the electron and positron distributions~$f_{e,p}(\bm u)$, where $\bm u$ is the normalized momentum (proper velocity normalized to~$c$), can be approximated by two drifting ultrarelativistic Maxwell-J\"uttner distribution functions with the same  density~$n$ and normalized temperature~$\Theta_e \equiv T_e/m_ec^2 \gg 1$. The distributions are boosted along the flow ($\hat{x}$) direction by a drift velocity $c\bm{\beta}_d = \pm c{\beta}_d \hat{x}$ (where the ``$+$" sign is for positrons and ``$-$" sign is for electrons), corresponding to a drift Lorentz factor $\gamma_d \equiv (1-\beta_d^2)^{-1/2} \sim 1 \ll \theta_e$.
In this Appendix, we ignore any possible spatial dependence of our quantities.

The standard way to formulate a resistivity is to calculate, in a steady state, how much electric current can be driven by an externally imposed electric field~${\bm E}$, taking into account the presence of friction on the charge-carrying particles.  The steady state is then determined by balancing the total volumetric electric force on one of the species, e.g., the electrons, with the volumetric friction force, which, in the case under consideration here, is the radiation-reaction force per unit volume, ${\bm F}_{\rm rad}$. Thus, for electrons,
\be
\label{Ebalance}
- e n {\bm E}  = 
- \, {\bm F}_{\rm rad} =
- \int {\rm d}^3 u \, f_e({\bm u}) \, 
{\bm f}_{\rm rad}({\bm u}), 
\ee
where the synchrotron radiation-reaction force on an ultrarelativistic particle is
\be
	\label{coolingexpr}
{\bm f}_{\rm rad} ({\bm u}) = 
 -\, \frac{2}{3}\, \bm \beta\, \frac{(\bm B \times \bm u)^2 e^4}{m_e^2 c^4}= 
 -\, 2\, \sigma_T\,  \frac{B^2}{8\pi} \, \bm\beta\, u^2 \sin^2\alpha\, .
\ee
Here $\bm \beta \equiv \bm u/\gamma$ is the particle's 3-velocity, $\alpha$ is its pitch angle with respect to the magnetic field~$\bm B$, and $\sigma_T \equiv (8\pi/3)\,r_e^2 = (8\pi/3)\, e^4/m_e^2 c^4$ is the classical Thomson cross-section, $r_e \equiv e^2/m_e c^2$ being the classical electron radius.	

Substituting \eq{coolingexpr} into \eq{Ebalance}, we get
 \be
	\label{Ewint}
	- e n \bm{E} = 2 \sigma_T\, \frac{B^2}{8\pi} \int {\rm d}^3 u \,\bm{\beta}\, \frac{(\bm B \times \bm u)^2}{B^2} \, f_e(\bm u) \, .
\ee
If we consider a set of coordinates where the applied electric field and the current are along the $\hat{x}$ direction and the magnetic field is in the $x\mbox{-}y$ plane, we can then express the $x$-component of the integral in \eq{Ewint} as 
\be
	\label{Ewintexp}
\int {\rm d}^3 u\, \beta_x \left[u_x^2\sin^2\theta + u_y^2 \cos^2\theta + u_z^2\right]f_e(\bm u) \, , 
\ee
where $\theta$ is the angle between the current and the magnetic field, and where we made use of the assumption that the distribution function $f_e(\bm u)$ is even with respect to~$u_y$.
For an ultrarelativistically hot plasma with $\Theta_e \gg 1$, we can evaluate the relevant integrals over the assumed boosted Maxwell-J\"uttner electron distribution as follows:
\be
	\label{dintegral}
	\int {\rm d}^3 u\, \beta_x u_x^2\, f_e(\bm u) 
 \simeq -12\, n \Theta_e^2\,\beta_d \left(1  + 2u_d^2 \right),
 \ee
 and
 \be
 	\int {\rm d}^3 u\, \beta_x u_y^2\, f_e(\bm u) = \int {\rm d}^3 u\, \beta_x u_z^2 \,f_e(\bm u) 
  \simeq - 4\, n \Theta_e^2\, \beta_d \, ,
\ee
where the negative sign appears because the electrons are boosted in the direction opposite to the current.
Then, \eq{Ewint} becomes:
\be
	\label{Ewintged}
	- e n \bm{E} \simeq -16\,n \sigma_T\, \frac{B^2}{8\pi}\, \Theta_e^2\,\beta_d \left[1 + (1+3u_d^2)\sin ^2\theta\right]\, \hat{x}.
\ee

One can now solve for the electric field in terms of the current density generated by both the electrons and positrons $\bm j =  2 e n \beta_d c \, \hat{x}$:

\be
	\bm{E} \simeq 8\, \frac{\sigma_T}{cn e^2}\, \frac{B^2}{8\pi}\,\Theta_e^2 
 \left[1 + (1+3u_d^2)\sin ^2\theta\right]\,\bm j\, .
\ee
 
Comparing with Ohm's law, $\bm E = \eta \bm j$, we thus find an effective radiative resistivity~$\eta_{\rm eff}$:
\begin{eqnarray}
\label{effres}
	\eta_{\rm eff} & \simeq & 
 \frac{\sigma_T}{cn e^2}\, \frac{B^2}{\pi}\,\Theta_e^2  
 \left[1 + (1+3u_d^2)\sin ^2\theta\right] \nonumber \\
 &= & \frac{64}{3}\pi\,\frac{r_e}{c} \, \sigma_{c}\,\Theta_e^2 \left[1 + (1+3u_d^2)\sin ^2\theta\right] \, , 
\end{eqnarray}
where   
$\sigma_{c}= B^2/8\pi n m_e c^2$ is the cold magnetization based on the local values of the magnetic field and total particle density~$2n$.
For simplicity and since the expression in \eq{effres} only varies by a factor of about $2$ when varying the angle~$\theta$ (assuming $\gamma_d \sim 1$), let us average over $\theta$ assuming a uniform, isotropic distribution of these angles in three-dimensional space.
This isotropic average leads to the estimate:
\begin{eqnarray}
\label{effressim}
	\eta_{\rm eff} &\simeq &
    \frac{\sigma_T}{cn e^2}\, \frac{B^2}{\pi}\,\Theta_e^2
\left(\frac{5}{3} + 2u_d^2\right) \nonumber \\
    &= & \frac{8}{3}\,\frac{e^2B^2}{n m_e^2 c^5}\,\Theta_e^2 \left(\frac{5}{3} + 2u_d^2\right) \nonumber \\
    &=& \frac{64}{3}\pi\,\frac{r_e}{c} \, \sigma_{c}\,\Theta_e^2 \left(\frac{5}{3} + 2u_d^2\right)\, .
\end{eqnarray}

This effective radiative resistivity can act as a dissipative term in the resistive-magnetohydrodynamics (MHD) magnetic induction equation,
\be
\label{induction}
	\frac{\partial \bm B}{\partial t} = 
 -\, c\, \bm \nabla \times \bm E = 
 \bm \nabla \times \left(\bm v \times \bm B\right)
-\bm \nabla \times \left[\frac{\eta_{\rm eff}c^2}{4\pi}\, \bm\nabla\times \bm B \right] \, ,
\ee
resulting in a simple expression for the radiative magnetic diffusivity: 
\be
\label{finaldiffus}
\frac{\eta_{\rm eff}c^2}{4\pi} \simeq 
\frac{16}{3}\, c r_e \, \sigma_{c}\, \Theta_e^2 \left(\frac{5}{3} + 2u_d^2\right) \approx \frac{80}{9}\, c r_e \, \sigma_{c}\, \Theta_e^2\, ,
\ee
where the last expression is valid for small~$u_d$.

Note that all the quantities appearing in the  expressions presented in this appendix, e.g., $B$, $n$, $\Theta_e$, $\beta_d$, $\gamma_d^2$, and~$\sigma_{c}$, are local, and so their values may, in general, be different from the globally defined initial system parameters that are used elsewhere in the paper.

\section{Radiative Dissipation Boundary}
\label{sec-append-Bbound}

In addition to the right boundary of the  $n\mbox{-}B$ histogram shown in~\fig{Histogram3D}, which we discussed in Appendix~\ref{sec-append-nbound}, we can offer an explanation for the upper limit in the $n\mbox{-}B$ space for the radiative cases, especially promising for the 3D radiative case.
We empirically find a very shallow power law of $B \sim n^{1/12}$ which can be seen in both classical and radiative 3D cases [see \fig{Histogram3D}(a,b)]. However, in the radiative case [\fig{Histogram3D}(b)],
we also observe a somewhat steeper upper boundary at lower densities, $n \lesssim 2 n_b$, consistent with the power-law scaling $B \sim n^{1/6}$. In this appendix, we will argue that this upper limit on the magnetic field strength is due to the radiative-resistive dissipation of the magnetic field in the central cores of secondary plasmoids filled with low-density background plasma from the upstream region.

To understand the radiative dissipation of magnetic flux ropes (current filaments), we first consider resistive decay described by the resistive-MHD magnetic induction equation~(\ref{induction}). 
In the collisionless relativistic plasmas under consideration in this study, the usual collisional resistivity can be  neglected, but the effective radiative resistivity~$\eta_{\rm eff}$ due to synchrotron radiation reaction, introduced in Appendix~\ref{sec-append-raddiss}, needs to be considered.
The dissipation time scale~$t_d$ can be obtained by comparing the left-hand side of \eq{induction} with the second (diffusive) term on the right-hand side, yielding 
\be
\frac{1}{t_d} \sim \frac{\eta_{\rm eff}c^2}{4\pi r_{\rm var}^2} \, , 
\ee
where $r_{\rm var}$ is the gradient length scale of the magnetic fields in current filaments.
Using our estimate~(\ref{effres}) for the effective radiative resistivity derived in Appendix~\ref{sec-append-raddiss}, we can express the corresponding radiative-resistive decay rate, normalized to the global light crossing time~$L_y/c$, assuming $u_{d} \ll 1$, as 
\be
\frac{L_y}{c t_d} \sim 
\frac{80}{9}\, \frac{L_y r_e}{r_{\rm var}^2}\, \sigma_{c,\rm loc}\, \Theta_{e,\rm loc}^2 \, ,
\ee
where we have added the additional subscript ``loc" in $\sigma_{c,\rm loc}$ and $ \Theta_{e, \rm loc}$ in order to distinguish these local quantities, describing a given flux-rope core, from the globally defined initial parameters describing the simulation setup such as~$\gamma_T = 2T_b/m_e c^2$.

This expression for the normalized radiative magnetic decay rate can be conveniently recast in terms of our simulation parameters as
\be
\label{dissipationrate}
\frac{L_y}{c t_d} \sim 
\frac{80}{9}\,\alpha_{\rm fs}\,\gamma_T^{-1}\, \frac{L_y}{\rho_L} \frac{\rho_L^2}{r_{\rm var}^2} \frac{B_0}{B_Q}\, \sigma_{c,\rm loc}\, \Theta_{e,\rm loc}^2 \propto \frac{B^2 T^2}{n}.
\ee
The last expression represents the scaling of the normalized radiative-resistive dissipation rate with the local quantities ($B$, $n$, and~$T$) used in our histograms from Section~\ref{subsec-histogram}.
Note that here we used 
\be
\sigma_{c,\rm loc} \equiv 
\frac{B^2}{8\pi n m_e c^2} = 
2 \sigma_h \gamma_T \, \frac{B^2}{B_0^2} \, \frac{n_b}{n} \propto \frac{B^2}{n}\, ,
\ee
where $\sigma_h$, $\gamma_T$, $B_0$, and $n_b$ are the initial upstream plasma parameters introduced in Section~\ref{sec-setup}.
 
Similarly, we define a radiative cooling time $t_{\rm cool}$ as the time for a typical (thermal) particle with energy $\gamma = 2\Theta_{e,\rm loc}$ (corresponding to the peak of the relativistic Maxwellian distribution), gyrating perpendicular to the local magnetic field~$B$,
to lose an order-unity fraction of its energy to synchrotron cooling.
The ratio of the light crossing time to the cooling time can be written as 
\be
\frac{L_y}{c t_{\rm cool}} =  \frac{4}{3}\, \alpha_{\rm fs} \, \frac{B_0}{B_Q} \frac{L_y}{\rho_L} \frac{B^2}{B_0^2} \, \Theta_{e,\rm loc} \, \gamma_{T} \propto B^2 T \, .
\ee
Again, the last expression is the scaling with respect to the space-dependent parameters used in our histograms from Section~\ref{subsec-histogram}.
In our radiative case, $L_y/c t_{\rm cool} = 0.44$ based on the background conditions $T_b$ and~$B_0$, 
and $L_y/c t_{\rm cool}= 0.77$ for the initial Harris sheet conditions $T_0$ and~$B_0$.
While these values correspond to only moderate cooling of the initial plasma throughout the simulation, for the hot plasma energized by the reconnection process, especially in plasmoid cores, the cooling rate can be quite significant. 

If the cooling rate $1/t_{\rm cool}$ in a given region is slow compared to the magnetic field resistive dissipation rate~$1/t_d$, then there is enough time for the magnetic field to decay before the plasma cools significantly. 
This ratio can be expressed as:
\be
\label{timeratio}
\frac{t_{\rm cool}}{t_d} = \frac{20}{3}\sigma_h\, \frac{T}{T_b}  \frac{n_b}{n} \frac{\rho_L^2}{r_{\rm var}^2} \, .
\ee

We believe that the magnetic field indeed suffers radiative-resistive dissipation in certain localized regions in the parameter regime of the present paper. As shown in Section~\ref{subsec-hist3D}, the magnetic field strength in regions with density $n/n_b \lesssim 2$ seems to be limited by the power-law boundary $B\sim n^{1/6}$. 
For the low-density secondary plasmoids in the 3D case, where $r_{\rm var}/\rho_L \simeq 7$, 
$n/n_b \simeq 2$, $B/B_0 \simeq 1.5$, $T/T_b \simeq 1$, and $\gamma_{d,\rm loc}\simeq 1$, we find the ratios $L_y/c t_d \simeq  1.8$ and $L_y/c t_{\rm cool} \simeq 1$ (i.e., $t_{\rm cool}/t_d \simeq 1.8$).
Therefore, there is marginally enough time for the magnetic dissipation to become important and limit the compression of the magnetic field.
In the 2D case, the parameters are about the same, and thus magnetic dissipation should also play a role in limiting magnetic field amplification. However, in this case, we do not observe the very clear boundaries in $n\mbox{-}B$ space found in 3D to help confirm this hypothesis.

When, for a given flux-rope core, the normalized resistive dissipation rate~(\ref{dissipationrate}) exceeds unity (and at the same time also exceeds the normalized radiative cooling rate, $t_d<t_{\rm cool}$), the magnetic field has sufficient time to dissipate within a light-crossing time (i.e., the flux rope's characteristic dynamical lifetime, enough time for a power-law boundary in $n\mbox{-}B$ space to develop).  
The magnetic field's amplification by compression is then checked by the effective radiative-resistive decay.
To evaluate the location of the corresponding histogram boundary in $n\mbox{-}B$ space, we will assume, based on the numerical observation from~\fig{Histogram3D}(d), that $T \sim n^{1/3}$, a scaling that is expected from simple adiabatic compression for a relativistic plasma.
As we have argued above, it is justifiable to ignore radiative cooling in these regions because the radiative-resistive decay of the magnetic field occurs faster than the cooling of the plasma.
This adiabatic temperature scaling was also confirmed by checking the average local temperature at the boundary in $n\mbox{-}B$ space for $n \lesssim n_b$ (not presented), giving a value of $\Theta_{e,loc} \approx 5~(n/n_b)^{1/3}$, just slightly hotter than a scaling based on the initial background temperature $T_b=4m_e c^2$ at $n=n_b$. Then, substituting the $T \sim n^{1/3}$ scaling into \eq{dissipationrate}, and setting the normalized resistive dissipation rate to be constant and of order unity, we obtain the scaling for the maximum magnetic field, 
\be
\label{Blimit}
B_{\rm max} \sim n^{1/6}.
\ee
This scaling provides a good match with the upper boundary of the $n\mbox{-}B$ histogram observed in \fig{Histogram3D}(b).

\bibliography{magnetpairs}

\bsp
\label{lastpage}
\end{document}